\documentclass{article}
\usepackage{bm,latexsym,amsmath,amssymb,amsfonts,fancyhdr,color,graphicx,multirow}
\usepackage[a4paper,bottom=2cm,top=2.5cm,head=5mm,width=18cm,dvipdfm]{geometry}
\usepackage[usenames,dvipsnames,svgnames,table]{xcolor}
\usepackage[ps2pdf,
            colorlinks=true,
            linkcolor=blue,
            urlcolor=blue,
            citecolor=green,          
						bookmarks=true,
						bookmarksnumbered=true,
						breaklinks=true,
						pdfpagemode=Fullscreen,
						pdfstartview=FitBH]{hyperref}
\usepackage[dotinlabels]{titletoc}
\usepackage{titlesec}
\usepackage{authblk}
\usepackage[normalem]{ulem}

\graphicspath{{code/}}

\numberwithin{equation}{section}
\allowdisplaybreaks[4]

\titlelabel{\thetitle.\quad \hspace{-0.8em}}
\titlecontents{section}
              [1.5em]
              {\vspace{4mm} \large \bf}
              {\contentslabel{1em}}
              {\hspace*{-1em}}
              {\titlerule*[.5pc]{.}\contentspage}
\titlecontents{subsection}
              [3.5em]
              {\vspace{2mm}}
              {\contentslabel{1.8em}}
              {\hspace*{.3em}}
              {\titlerule*[.5pc]{.}\contentspage}
\titlecontents{subsubsection}
              [5.5em]
              {\vspace{2mm}}
              {\contentslabel{2.5em}}
              {\hspace*{.3em}}
              {\titlerule*[.5pc]{.}\contentspage}

\newcommand{\titledef}{Non-standard interactions and the CP 
phase measurements in neutrino oscillations at low energies} % Insert Title here!!!
\hypersetup{ pdfauthor = {Shao-Feng Ge},
	     pdftitle = {\titledef}, % Insert title here!!!
	     pdfsubject = {}, % Insert subject here!!!
             pdfkeywords = {}, % Insert keywords here!!!
	     pdfcreator = {LaTeX with hyperref package},
	     pdfproducer = {dvips + ps2pdf} }

\newcommand{\be}{\begin{equation}}
\newcommand{\ee}{\end{equation}}
\newcommand{\ba}{\begin{array}}
\newcommand{\ea}{\end{array}}
\newcommand{\bea}{\begin{eqnarray}}
\newcommand{\eea}{\end{eqnarray}}
\newcommand{\balg}{\begin{align}}
\newcommand{\ealg}{\end{align}}
\newcommand{\bit}{\begin{itemize}}
\newcommand{\eit}{\end{itemize}}

\definecolor{gesfpurple}{rgb}{0.47,0.19,0.42}
\newcommand{\gpurple}[1]{{\color{gesfpurple} #1}}
\definecolor{gesflanse}{rgb}{0.00,0.50,0.50}

\definecolor{gesfblue}{rgb}{0.08,0.42,0.76}
\newcommand{\gblue}[1]{{\color{gesfblue} #1}}
\definecolor{gesfred}{rgb}{1,0,0}
\newcommand{\gred}[1]{{\color{gesfred} #1}}
\definecolor{gesfwhite}{rgb}{1,1,1}

\definecolor{gesfblack}{rgb}{0,0,0}

\newcommand{\gsec}[1]{{\hypersetup{linkcolor=red}Sec.~\ref{#1}\hypersetup{linkcolor=blue}}}

\newcommand{\geqn}[1]{\hypersetup{linkcolor=blue}(\ref{#1})\hypersetup{linkcolor=blue}}
\newcommand{\gfig}[1]{{\hypersetup{linkcolor=violet}Fig.~\ref{#1}\hypersetup{linkcolor=blue}}}
\newcommand{\gtab}[1]{{\hypersetup{linkcolor=gesflanse}Table~\ref{#1}\hypersetup{linkcolor=blue}}}

\newcommand{\Dma}{{\Delta m^2_{31}}}
\newcommand{\Dms}{{\Delta m^2_{21}}}

\newcommand{\dD}{\delta_D}

\newcommand{\Ta}{\theta_{23}}
\newcommand{\Ca}{c_{23}}
\newcommand{\Sa}{s_{23}}
\newcommand{\Tr}{\theta_{13}}

\newcommand{\Cr}{c_{13}}
\newcommand{\Sr}{s_{13}}

\newcommand{\Ts}{\theta_{12}}

\newcommand{\Cs}{c_{12}}
\newcommand{\Ss}{s_{12}}

\newcommand{\te}{\epsilon'}

\newcommand{\rD}{r_\Delta}
\newcommand{\rV}{r_V}
\newcommand{\tk}{T2(H)K }

\begin{document}

\fontsize{12pt}{14pt}\selectfont

\title{\textbf{\Large \titledef}} % Insert title here!!!
\author[]{{\large Shao-Feng Ge}\footnote{gesf02@gmail.com} }
\author[]{{\large Alexei Yu. Smirnov}\footnote{smirnov@mpi-hd.mpg.de}}
\affil[]{\small Max-Planck-Institut f\"{u}r Kernphysik,
Saupfercheckweg 1,
Heidelberg 69117, Germany}
\date{\today}

\maketitle

\begin{abstract}

We study the effects of non-standard interactions (NSI) and the genuine CP phase $\dD$ in neutrino oscillations
at low, $E_\nu \lesssim 1\,\mbox{GeV}$, and very low,  $E_\nu \lesssim 0.1\,\mbox{GeV}$, energies.
For experimental setup with baseline and neutrino energy tuned to the first
1-3 oscillation maximum, we develop a simple analytic formalism to show the effects of NSI. 
The vacuum mimicking and its violation as well as the use of the separation basis play a central
role in our formalism. The NSI corrections that affect the CP phase measurement
mainly come from the violation of vacuum mimicking as well as from the corrections
to the 1-3 mixing angle and mass-squared difference.
We find that the total NSI correction to the $\nu_\mu - \nu_e$ probability $P_{\mu e}$ 
can reach $20\% - 30\%$ ($1 \sigma$) at T2(H)K. 
Correspondingly, the correction to the CP phase can be as large as $50^\circ$ and hence
significantly deteriorates the CP sensitivity at T2(H)K. The proposed TNT2K experiment,
a combination of T2(H)K and the short baseline experiment $\mu$Kam that uses the
Super-K/Hyper-K detector at Kamioka to measure the oscillation of the antineutrinos from muon decay
at rest ($\mu$DAR), can substantially reduce the degeneracy between NSI and the genuine
CP phase $\dD$ to provide high CP sensitivity. The NSI correction to
$P_{\mu e}$ is only $2\%$ ($1 \sigma$) for $\mu$DAR neutrinos.

\end{abstract}

\section{Introduction}
%%%%%%%%%%%%%%%%%%%%%%%%%%%%%%%%%%%%%%%%%%%%%%%%%%%%%%%%%%%%%%%%%%%%%%%%%%%%

The determination of the Dirac CP phase, $\delta_D$, and searches for new (non-standard) 
neutrino interactions \cite{Wolfenstein:1977ue}, \cite{Guzzo:1991hi,Roulet:1991sm}
are among the main objectives of present-day neutrino physics  
(see \cite{Ohlsson:2012kf,Miranda:2015dra} for reviews).  
There are certain connections between the two unknowns: 

- NSI can be additional source of CP violation on top of the PMNS matrix \cite{GonzalezGarcia:2001mp,Gago:2009ij}.

- The effects of NSI and CP violation are entangled in oscillation experiments, 
thus creating the $\delta_D$-NSI degeneracy problem \cite{Gago:2009ij, Coloma:2011rq}.

Within the standard $3\nu$ paradigm (no NSI, no sterile neutrinos, {\it etc.}), 
the global fit of oscillation data gives more than $2\sigma$ indication of CP violation 
with the best-fit value $\dD \approx 3\pi/2$.
This indication follows mainly from the results of 
T2K \cite{T2K} and NO$\nu$A \cite{NOvA} 
in combination with the results of reactor and some other experiments. 
The confirmation of this hint and measurement of the CP phase 
are associated to future experiments with neutrinos of different origins.   
The current T2K experiment can be extended to T2K-II \cite{T2KII,T2KII2} by upgrading the
flux and eventually to T2HK with a much larger Hyper-K detector \cite{HK}, or even
T2KK/T2KO \cite{T2K04,T2KK05a,T2KK05,T2KK06,T2KK06b,T2KK09,T2KK12,T2KK16} with longer baseline. 
The DUNE \cite{LBNE, DUNEvol1, DUNEvol2} experiment should provide precision measurement of $\dD$.  
In addition to accelerator experiments, the CP phase $\dD$ can also have large observable effect
in the oscillation of atmospheric neutrinos below $1\,\mbox{GeV}$. 
This can be explored using the future possible upgrades of PINGU and ORCA detectors
with $3 \sim 4$ times denser instrumentation \cite{SuperPINGU1}. 
At even lower energy, $E_\nu \lesssim 0.1~\mbox{GeV}$, one can also employ neutrinos from the muon 
decay at rest ($\mu$DAR) in proposed experiments like 
DAE$\delta$ALUS \cite{DAEdALUS} and CI-ADS \cite{ADS},
or in-flight in experiments like MOMENT \cite{MOMENT, MOMENT-CP} to measure $\dD$. 
Measuring the antineutrino oscillation $\bar \nu_\mu \rightarrow \bar \nu_e$ from the $\mu$DAR source 
with the detector of JUNO/RENO-50, which is originally designed to establish the 
neutrino  mass hierarchy,
can also determine the CP phase \cite{Ciuffoli:2014ika}.

Interpretation of the present results on CP violation can be affected by the possible presence of
 non-standard interactions \cite{Forero:2016cmb} or sterile neutrinos
\cite{Klop:2014ima,Palazzo:2015gja,Gandhi:2015xza}.
The effects of NSI substantially reduce the sensitivity to $\delta_D$ at the ongoing T2K and NO$\nu$A
experiments \cite{Liao:2016hsa,Forero:2016cmb,Masud:2016bvp}, and 
the future long-baseline experiments,  in particular, DUNE  
\cite{Masud:2015xva,deGouvea:2015ndi,Coloma:2015kiu,Liao:2016hsa,Forero:2016cmb,Farzan1602,Masud:2016bvp,deGouvea:2016pom, 
Blennow:2016etl, Bakhti:2016gic}.
So, the problem is to disentangle the genuine CP violation and the NSI effects in a way that can
guarantee high sensitivity to the CP phase. To achieve this, one can use the fact that the effects of
NSI, in particular the effects on CP violation,
are proportional to the neutrino energy $E_\nu$. 
To disentangle the genuine CP phase from NSI, one can use
two or more experiments with different neutrino energies.
The effects of NSI can be as large as $O(1)$ at accelerator neutrino experiments
with $E_\nu > 500\,\mbox{MeV}$ like DUNE. The neutrino spectrum from the MOMENT
source peaks around $(200 \sim 300)\,\mbox{MeV}$ \cite{MOMENT}, being 10 times
smaller than the peak energy at DUNE. Hence, a combination of the future MOMENT and 
DUNE results could reduce the degeneracy between $\dD$ and the NSI parameters \cite{Farzan1602}.
Still, the NSI effects at the MOMENT energy are large. Rescaled from the
estimation for T2K, the relative correction of NSI to the oscillation probability $P_{\mu e}$
is roughly $10\%$ at MOMENT. 
For neutrinos from $\mu$DAR with energies $\sim (30 - 50)$ MeV the NSI corrections are only $2\%$. 

Various possible experiments with $\mu$DAR sources have been explored. 
In particular, it was proposed \cite{TNT2K} to use the Super-K detector (and later Hyper-K)
to detect the antineutrino oscillation $\bar \nu_\mu \rightarrow \bar \nu_e$ from $\mu$DAR source
to Kamioka (dubbed by $\mu$Kam, which can be either $\mu$SK with the Super-K detector or
$\mu$HK with the Hyper-K detector).
%Using the same detector for both neutrino sources will reduce some systematic errors. 
The combination of T2K running solely
in the neutrino ($\nu$T2K) and $\mu$SK in the antineutrino mode,
referred as TNT2K, provides even higher sensitivity to the CP phase $\delta_D$ \cite{TNT2K}. 

In most papers, the NSI effects have been computed numerically although several analytical studies
can be found in \cite{GonzalezGarcia:2001mp,Kopp:2007ne,Kikuchi:2008vq,Liao:2016hsa}.
These earlier studies provide analytical expressions of the oscillation probabilities
in the $\bar \nu_e \rightarrow \bar \nu_e$ \cite{Kopp:2007ne,Kikuchi:2008vq},
$\nu_e \rightarrow \nu_\mu$ \cite{GonzalezGarcia:2001mp,Kikuchi:2008vq}, 
$\nu_\mu \rightarrow \nu_e$ \cite{Kopp:2007ne,Kikuchi:2008vq,Liao:2016hsa}, and
$\nu_\mu \rightarrow \nu_\mu$ \cite{Kopp:2007ne,Kikuchi:2008vq} channels.
In particular, \cite{Kikuchi:2008vq} provides
the probability formula for all oscillation channels.  
In this paper, we present an analytical formalism   
which allows us to analyze the effects of NSI in a simple way. 
The vacuum mimicking \footnote{The effect of vacuum mimicking was noticed in 
\cite{Wolfenstein:1977ue,DeRujula:1998umv,Freund:1999gy}, 
explained in \cite{Akhmedov:2000cs}, further studied in 
\cite{Lipari:2001ds,Minakata:2000ee,Minakata:2000wm} and dubbed as such in
\cite{Yasuda:2001va}.} in the 1-2 sector and its violation
plays a central role in our formalism. 
We show that NSI can  induce large correction via the violation of vacuum mimicking and
apply our results to the CP measurement at TNT2K.  

%%%%%%%%%%%%%%%%%%%%%%%%%%%%%%%%%%%%%%%%%%%%%%%%%%%%%%%%%%%

The paper is organized as follows. In \gsec{sec:matter} we present a general formalism of matter 
effects and discuss the role of vacuum mimicking in the sector of 1-2 mass splitting and mixing.
We then apply this formalism to the case of standard interactions in  
\gsec{sec:SI} and non-standard interactions in \gsec{sec:NSI}. 
The vacuum mimicking can be significantly violated by NSI and the Dirac CP phase $\dD$ can
receive $\mathcal O(1)$ correction. 
In \gsec{sec:CP} we explore the sensitivity of T2K and $\mu$SK experiments 
to the genuine CP phase in the presence of NSI. Especially,
the improvement on the sensitivity due to the $\mu$SK component of TNT2K is shown.
Our conclusions are given in \gsec{sec:conclusion}.

\section{Neutrino Oscillation in Matter at Low Energies}
\label{sec:matter}
%%%%%%%%%%%%%%%%%%%%%%%%%%%%%%%%%%%%%%%%%%%%%%%%%%%%%%%%%%%%%%%%%%%%%%%%%%%

\subsection{Generalities and physics setup}
%%%%%%%%%%%%%%%%%%%%%%%%%%%%%%%%%%%%%%%%%%%%%%%%%%%%%%%%%%%%%

In the flavor basis, the Hamiltonian $\mathcal H$ that describes the neutrino propagation 
is a sum of the vacuum term and the matrix of matter potential $\mathbb V$: 
\begin{equation}
  \mathcal H
\equiv
  \frac 1 {2 E_\nu}
  U_{PMNS}
\left\lgroup
\begin{matrix}
  0 \\ 
& \Dms \\
& & \Dma
\end{matrix}
\right\rgroup
  U_{PMNS}^\dagger
+
  \mathbb V \,,
\label{eq:H}
\end{equation}
where $\Dms \equiv m^2_2 - m^2_1$, $\Dma \equiv m^2_3 - m^2_1$,
and $E_\nu$ is the neutrino energy. The matrix of matter potential $\mathbb V$
in general has contributions from both standard and non-standard interactions.
For the mixing matrix in vacuum $U_{PMNS}$ 
we use the standard parametrization
$$
U_{PMNS} \equiv U_{23}(\Ta) \Gamma_{\dD} U_{13}(\Tr) \Gamma^\dagger_{\dD} U_{12}(\Ts), ~~~ 
 \Gamma_{\dD} = {\rm diag}(1, 1, e^{i\delta_D}). 
$$
Here $\theta_{ij}$ are the vacuum mixing angles and $\delta_D$ is the Dirac CP phase. 
In the case of standard interaction only, the matrix of matter potential takes the form as
\begin{equation}
  \mathbb V = \mbox{diag}\{V, 0, 0\}
\quad \mbox{and} \quad
  V = \sqrt{2} G_F n_e,
\label{eq:V}
\end{equation}
where $G_F$ is the Fermi coupling and $n_e$ is the electron number density.
For the standard interaction with a constant density $3\mbox{g/cm}^3$ of isotopically
neutral matter and $\sin^2 \theta_{12} \approx 0.31$,
the resonance neutrino energy due to the 1-2 mass splitting equals 
\begin{equation}
E_R = \cos 2 \Ts \frac{\Dms}{2V} = 122\,\mbox{MeV}. 
\end{equation}

At low energy, $E_\nu \lesssim 1\,\mbox{GeV}$, the three mass-energy scales 
($\Delta m^2_{21}$, $\Delta m^2_{31}$, $2 E_\nu V$) in the Hamiltonian (\ref{eq:H})
satisfy the following conditions: 
\begin{equation}
\frac{\Dms}{2 E_\nu} \sim V \ll \frac{\Dma}{2 E_\nu}. 
\label{eq:conditions}
\end{equation}
This allows us to introduce two small parameters
\begin{equation}
\rD \equiv \frac{\Dms}{\Dma} \approx 0.03, ~~~~ 
x_{31} \equiv \frac{2 EV}{\Dma} = 0.01 \left(\frac{E_\nu}{ 100 {\rm MeV}}\right)
\end{equation} 
which will be used
for perturbative diagonalization of the Hamiltonian.
For typical energies $E_\nu = 50$ MeV at $\mu$DAR and
$E_\nu = 600$ MeV at T2K, we obtain 
$x_{31} = 5 \cdot 10^{-3}$ and $x_{31} = 0.06$, respectively.

At low energies the matter corrections to the 1-3 mixing angle
and mass splitting are small. It is more convenient to consider 
the neutrino oscillation in a different basis $\nu'$ that is related to the flavor
basis $\nu_f$ as, 
\cite{Peres:2003wd,Blennow:2013rca},
\begin{equation}
\nu_f = \mathcal R \nu', 
\end{equation}
where 
\begin{equation}
  \mathcal R
\equiv U_{23}(\Ta) \Gamma_\delta U_{13}(\Tr) \,.
\label{eq:rmat}
\end{equation}
The matrix $\mathcal R$ depends only on the vacuum mixing angles ($\theta_{23}$, $\theta_{13}$)
and the phase $\delta_D$. 
Using \geqn{eq:H} and \geqn{eq:rmat}, we obtain the Hamiltonian in the $\nu'$ basis
\begin{equation}
  \mathcal H'
\equiv
  \mathcal R^\dagger \mathcal H \mathcal R
=
  \frac {\Dma}{2 E_\nu}
\left\lgroup
\begin{matrix}
  0 \\
& 0 \\
& & 1
\end{matrix}
\right\rgroup
+
\frac {\Dms} {2 E_\nu}
\left\lgroup
\begin{matrix}
  \Ss^2 & \Cs \Ss \\
  \Cs \Ss & \Cs^2 \\
& & 0
\end{matrix}
\right\rgroup
+
  \mathbb V_{sep} \,,
\label{eq:H'}
\end{equation}
where
\begin{equation}
  \mathbb V_{sep}
\equiv
  U^T_{13} \Gamma^\dagger_\delta U_{23}^T \mathbb V U_{23} \Gamma_\delta U_{13} \,, 
\end{equation} 
and we use the notation $(c_{ij}, s_{ij}) \equiv (\cos \theta_{ij}, \sin \theta_{ij})$. In this basis,
the first term with the largest mass scale $\Delta m^2_{31}$ is diagonal and separated from
the 1-2 sector in vacuum. 
The second term in (\ref{eq:H'}) depends on the vacuum parameters of the 1-2 sector only while
the last term is the matter term that provides all other (in particular 1-3) mixings.  
For convenience, we call this basis the {\it separation basis}. 
It is related to the usual propagation basis 
\cite{Akhmedov:1998xq,Yokomakura:2002av,Ge:2013zua} 
by an additional rotation $\Gamma_\delta U_{13}(\Tr)$ \cite{Peres:2003wd,Blennow:2013rca}.

Let us introduce a unitary mixing matrix $U'$ to diagonalize the Hamiltonian $\mathcal H'$ (\ref{eq:H'}): 
\begin{equation}
U'^{\dagger} \mathcal H' U' = {\rm diag} (H_1, H_2, H_3), 
\label{eq:eigenval}
\end{equation}
where $H_i$  are the eigenvalues of $\mathcal H'$. 
Then the  total mixing matrix in matter, $U^m$,  which connects 
the flavor states and the eigenstates of the Hamiltonian $\mathcal H$, 
$\nu_f = U^m \nu^m$,  is given by the product 
\begin{equation}
U^m = {\mathcal R} U'.  
\label{eq:Um}
\end{equation}
In \geqn{eq:Um}, the vacuum and matter parts are factorized. 
Without matter effect, we obtain $U' = U_{12}(\theta_{12})$.

In matter with constant density, solving the evolution 
equation is straightforward. For the eigenstates of the 
Hamiltonian, $\nu^m$, we obtain according to (\ref{eq:eigenval})
the evolution matrix 
\begin{equation}
S^d = 
\left\lgroup
\begin{matrix}
  1 \\
& e^{- i \Delta H_{21} L } \\
& & e^{- i \Delta H_{31} L}
\end{matrix}
\right\rgroup,
\label{eq:ssdd}
\end{equation}
where $\Delta H_{ij} \equiv H_i - H_j$ and $L$ is the distance (baseline).  
Then the matrix of amplitude in the flavor basis is  
\begin{equation}
S^f = U^m S^d U^{m \dagger}, 
\label{eq:sfmat}
\end{equation}
and the 
$\nu_\alpha \rightarrow \nu_\beta$ ($\alpha, ~ \beta  = e, ~\mu, ~\tau$) oscillation probability
is given by  $P_{\alpha \beta} \equiv |S_{\beta \alpha}^f|^2$. 

In what follows we will elaborate on the analytical description 
of the neutrino oscillation effects for the following physics setup:

(i) Neutrino energies are from a few tenths to a few hundreds of MeV. 

(ii) The baselines are about half of the oscillation length 
for the 1-3 mass splitting.
\begin{equation}
L \sim \frac{1}{2} l_{31} = \frac{2 \pi E_\nu}{\Dma}.   
\end{equation}
Correspondingly, the (half) oscillation phase equals $\phi_{31} \approx \pi/2$. 

(iii) The phase associated with the 2-1 splitting is small: 
\begin{equation}
\phi_{21} = r_\Delta \phi_{31} \sim \frac{\pi r_\Delta }{2} 
\sim  5 \cdot 10^{-2}~~~ (\sim 3^{\circ}). 
\label{eq:phases}
\end{equation}
Also the matter phase 
\begin{equation}
\phi_V = VL \sim \phi_{31} r_\Delta \cos 2 \theta_{12} \frac{E_\nu}{E_R}
\end{equation}
is small. The setup is realized in accelerator experiments such as T2(H)K, NO$\nu$A,
MOMENT, and  experiments based on the  K-meson, pion and muon decays at rest:  
KDAR, $\mu$DAR, {\it etc.} \cite{Wongjirad}.

%%%%%%%%%%%%%%%%%%%%%%%%%%%%%%%%%%%%%%%%%%%%%%%%%%%%%%%%%%%%%%%%%%%%%%%%%%%%%%%%%
\subsection{Oscillation probability}
%%%%%%%%%%%%%%%%%%%%%%%%%%%%%%%%%%%%%%%%%%%%%%%%%%%%%%%%%%%%%%%%%%%%%%%%%%%%%%%%

We consider the $\nu_\mu \rightarrow \nu_e$ transition in matter. 
Using (\ref{eq:ssdd}) and (\ref{eq:sfmat}), the probability $P_{\mu e}$ can be
generally presented as  
\begin{equation}
  P_{\mu e}
=
\left|
  2 U_{e2}^m U_{\mu 2}^{m*} \sin \phi_{21}^m
+ 2 U_{e3}^m U_{\mu 3}^{m*} \sin \phi_{31}^m e^{-i \phi_{32}^m}
\right|^2,  
\label{eq:me-prob}
\end{equation}
where $U_{\alpha j}^m$ is the $\alpha j$-element of the mixing matrix $U^m$ in matter,
and the half-oscillation phases equal 
\begin{equation}
\phi_{21}^m = \frac{1}{2} \Delta H_{21} L, ~~~ 
\phi_{31}^m = \frac{1}{2} \Delta H_{31} L, ~~~ 
\phi_{32}^m = \phi_{31}^m - \phi_{21}^m. 
\end{equation}
We can rewrite the probability $P_{\mu e}$ using the explicit expressions of
$U_{\alpha j}^m$ in the standard parametrization.
In particular, the first term in (\ref{eq:me-prob}) becomes 
\begin{equation}
  2 c_{13}^m s_{12}^m
\left(
  c_{12}^m c_{23}^m
- s_{13}^m s_{12}^m s_{23}^m e^{- i \delta_D^m}
\right) 
\sin \phi_{21}^m \,,
\label{eq:combine}
\end{equation}
where $(c^m_{ij}, s^m_{ij}) \equiv (\cos \theta^m_{ij}, \sin \theta^m_{ij})$.
It is convenient to  combine  the second term in (\ref{eq:combine}) 
with the second term in (\ref{eq:me-prob}) to get   
\begin{equation}
P_{\mu e} \equiv \left|A_{e\mu}^S +   A_{e\mu}^A \right|^2.
\end{equation}
Here, we have introduced the ``solar'' amplitude
\begin{equation}
A_{e\mu}^S =  c_{13}^m c_{23}^m A_{12}^m 
\end{equation}
with 
\begin{equation}
A_{12}^m \equiv \sin 2\theta_{12}^m \sin \phi_{21}^m, 
\label{eq:sol-amp}
\end{equation}
and the ``atmospheric'' amplitude
\begin{equation}
	A_{e\mu}^A
\equiv
  s_{23}^m \sin 2\theta_{13}^m e^{- i \delta_D^m}
\left[
  \sin \phi_{31}^m e^{-i\phi_{32}^m} 
- s_{12}^{m2} \sin \phi_{21}^{m}
\right].
\label{eq:atm-amp1}
\end{equation}
The atmospheric amplitude contains the term with oscillation phase 
related to the small 2-1 mass splitting and it can be rewritten as 
\begin{equation}
  A_{e\mu}^A
=
  s_{23}^m A_{13}^m e^{i ( - \delta_D^m - \phi_{32}^m)} 
\left[1 - (s_{12}^{m})^2 e^{i  \phi_{32}^m} \frac{\sin \phi_{21}^m}{\sin \phi_{31}^m} \right],  
\label{eq:atm-a33}
\end{equation}
where 
\begin{equation}
A_{13}^m \equiv  \sin 2\theta_{13}^m \sin \phi_{31}^m. 
\end{equation}
The amplitudes  $A_{13}^m$  and $A_{12}^m$ 
coincide with the standard $2\nu$ oscillation amplitudes.  

The 2-1 splitting correction to the atmospheric amplitude (expression in the 
brackets of (\ref{eq:atm-a33})) 
can be parametrized as 
\begin{equation}
\left[
  1
- y e^{i  \phi_{32}^m}  
\right]
\equiv
  \kappa e^{-i\phi_\kappa},
\label{eq:kappa2}
\end{equation}
where
\begin{equation} 
  y
\equiv
  (s_{12}^m)^2 \left(\frac{\sin \phi_{21}^m}{\sin \phi_{31}^m}\right). 
\end{equation}
For the experimental setup under consideration $y \approx (s_{12}^m)^2 r_\Delta$.  
Below the 1-2 resonance ($E_\nu < E_R$) we have  $(s_{12}^m)^2 \approx s_{12}^{2} = 0.31$,   
in the resonance $(s_{12}^m)^2 = 0.5$, and $(s_{12}^m)^2$ approaches 1 with 
$E_\nu$ above the resonance. 
So, typically $y = (1 - 3) \times 10^{-2}$ is a small quantity.
The quantities in (\ref{eq:kappa2}) equal
\begin{subequations}
\begin{eqnarray}
  \kappa
& = &
  \sqrt{1 - 2y \cos \phi_{32}^m   + y^2} \approx 1 - y \cos \phi_{32}^m, 
\label{eq:kappa}
\\
  \tan \phi_\kappa
& = &
  \frac{y \sin \phi_{32}^m}{ 1 - y \cos \phi_{32}^m} \approx  y \sin \phi_{32}^m.
\end{eqnarray}
\end{subequations}
Since $y \ll 1$, the phase $\phi_\kappa$ is much smaller than $\phi_{32}^m$.

Thus, the total probability  (\ref{eq:me-prob}) can be written as 
\begin{equation}
  P_{\mu e}
=
\left|
  c_{13}^m c_{23}^m A_{12}^m
+ s_{23}^m A_{13}^m  \kappa e^{i(- \delta_D^m -  \phi_{32}^m - \phi_\kappa)}
\right|^2. 
\label{eq:probemu2}
\end{equation}
Here $\phi_\kappa = \phi_\kappa(\phi_{32}^m)$ is small correction to the oscillation phase $\phi_{32}^m$. 
If $\kappa = 1$, Eq.  (\ref{eq:probemu2}) reproduces the standard expression for $P_{\mu e}$.

In the case of standard interactions, the vacuum mimicking for the 1-2 amplitude 
is realized due to smallness of the 2-1 phase $\phi^m_{21} \ll 1$,
\cite{Wolfenstein:1977ue,DeRujula:1998umv,Freund:1999gy,Akhmedov:2000cs,Lipari:2001ds,Minakata:2000ee,Minakata:2000wm,Yasuda:2001va} 
 \begin{equation}
A_{12}^m \equiv   \sin 2 \theta_{12}^m \sin \phi_{21}^m 
\approx \sin 2 \theta_{12} \sin \phi_{21} = A_{12} \,,
\label{eq:vacuum-mimicking}
\end{equation}
For $\phi^m_{21} \ll 1$, \geqn{eq:vacuum-mimicking} gives 
$A^m_{12} \simeq \sin 2 \theta_{12} \frac {\Delta m^2_{21} L}{2 E_\nu}$.
The condition $\phi^m_{21} \ll 1$ for vacuum
mimicking is fulfilled in the majority of existing and proposed experiments.

In general, vacuum mimicking is broken in the presence of NSI
\cite{Akhmedov:2000cs}. Introducing parameter  
\begin{equation}
  r_V \equiv \frac{A_{12}^m}{A_{12}} 
\label{eq:rv-param}
\end{equation}
to quantify the deviation from vacuum mimicking,
we can rewrite the ptobability $P_{\mu e}$ as
\begin{equation}
  P_{\mu e}
\approx
\left|
  c_{13}^m c_{23}^m r_V A_{12}
+ s_{23}^m A_{13}^m  \kappa e^{i (- \delta_D^m -  \phi_{32}^m - \phi_\kappa)}
\right|^2.  
\label{eq:probemu2}
\end{equation}
This parametrization of the probability $P_{\mu e}$ is convenient for 
understanding the matter effects, including both standard and non-standard interactions.

The probability $P_{\mu e}$ in \geqn{eq:probemu2} can also be rewritten 
as sum of the atmospheric, solar and interference terms: 
\begin{eqnarray}
  P_{\mu e} 
\approx
  P^A_{\mu e}
+ P^S_{\mu e}
+ P^I_{\mu e}
  \cos \left( \dD^m  + \phi_{32}^m  +  \phi_\kappa \right) \,. 
\label{eq:Pme-SI}
\end{eqnarray}
Here 
\begin{eqnarray}
  P^A_{\mu e} & = & |A^A_{\mu e}|^2  
\equiv 
\kappa^2 \Sa^{m2} \sin^2 2 \Tr^m \sin^2 \phi_{31}^m \,,
\\
  P^S_{\mu e} & = & |A^S_{\mu e}|^2 \approx  
  r_V^2 \Ca^{m 2} \Cr^{m2} \sin^2 2 \Ts \sin^2 \phi_{21} \,,
\\
 P^I_{\mu e}  
& \equiv &
 \kappa  r_V \sin 2 \Ta^m  c_{13}^{m} \sin 2 \Tr^m \sin\phi_{31}^m \sin 2 \Ts \sin \phi_{21} \,. 
\label{eq:Pme-coefficients}
\end{eqnarray}
For constant density, the problem then reduces to finding the mixing angles and mass splittings in matter. 
The exact vacuum mimicking corresponds to $r_V = 1$. If in addition $\kappa = 1$ and $\phi_\kappa = 0$,
Eqs.(\ref{eq:Pme-SI}-\ref{eq:Pme-coefficients}) reproduce the usual approximate 
expression for $P_{\mu e}$ \cite{PDG12}. 

It is easy to see that the 1-2 correction to the atmospheric amplitude 
\begin{equation}
\kappa - 1 \propto  \sin^2 \theta_{12}^m \sin \phi_{21}^m = A \frac{\Delta m_{21}^2 L }{4E_\nu}
\label{eq:kappa-1}
\end{equation}
does not show vacuum mimicking and has strong dependence on the matter potential 
$V$.  For $V \rightarrow 0$: $A \rightarrow \sin^2 \theta_{12}$, in the resonance 
$A \rightarrow \sin 2\theta_{12}$  and with further increase of $V$  
the coefficient $A$ increases as $\propto V$.

\subsection{On vacuum mimicking}
%%%%%%%%%%%%%%%%%%%%%%%%%%%%%%%%%%%%%%%%%%%%%%%%%%%%%%%%%%%%%%%%%%%

Let us present vacuum mimicking in a general form for both standard and non-standard interactions. 
Consider for simplicity the $2\nu$ Hamiltonian with moduli of  the off-diagonal 
elements $|\bar{\mathcal H}|$ and difference of the eigenvalues 
$\Delta \mathcal H$. The mixing angle is given by 
\begin{equation}
  \sin 2 \theta_{12}^m
=
  \frac{2 |\bar{\mathcal H}|}{ \Delta \mathcal H}, 
\label{eq:s2sm}
\end{equation}
and the (half) oscillation phase equals 
\begin{equation}
\phi_{21}^m = \frac{1}{2} \Delta \mathcal H L. 
\end{equation}
Then 
\begin{equation}
A_{21}^m 
\approx \sin 2 \theta_{12}^m ~\phi_{21}^m  = 
\frac{2 |\bar{\mathcal H}|}{\Delta \mathcal H} \frac{1}{2} \Delta \mathcal H L = 
|\bar{\mathcal H}| L. 
\label{eq:vacmim3}
\end{equation}
So, the oscillation amplitude $A^m_{21}$ in the first order is given by 
the off-diagonal element of the $2\nu$ Hamiltonian multiplied by distance. 
Since for standard interaction the matter potential appears only in the diagonal elements
of the Hamiltonian in the flavor basis, $\bar{\mathcal H} =  \bar{\mathcal H}_V$, we 
obtain $A_{21}^m = A_{21}$. The deviation from vacuum mimicking appears when 
the off-diagonal element $\bar{\mathcal H}$ depends on matter potential.  

Even for the standard interactions, matter effect appears in $A^m_{12}$ when
$\sin \phi^m_{21}$ is expanded to higher order,
\begin{equation}
A_{12}^m \equiv   \sin 2 \theta_{12} 
\frac{\Dms L}{2 E_\nu} \left[1 - 
\frac{1}{6}\left(\frac{\Delta H_{21}L}{2} \right)^2 \right]. 
\label{eq:mim2}
\end{equation}
For the vacuum amplitude $A_{12}$ we have the same expression with substitution 
$\Delta H_{21} \rightarrow {\Dms}/{2 E_\nu}$. Then the ratio of the matter 
to vacuum amplitudes equals
\begin{equation}
r_V  = 1 + \frac{1}{3} (\phi_{21})^2  
\cos^2 2\theta_{12} c_{13}^2 
\left(1 - \frac{c_{13}^2}{2} \frac{E}{E_R} \right).
\end{equation}
Using (\ref{eq:phases}) we can estimate the matter corrections as
\begin{equation}
r_V - 1 \approx   \frac{1}{3} 
\left(  \frac{\pi r_\Delta }{2}\right)^2
\cos^2 2\theta_{12} 
\left(1 - \frac{E}{2 E_R} \right) \approx 10^{-4}. 
\label{eq:rV-1}
\end{equation}
According to \geqn{eq:rV-1},
tn addition to the small oscillation phase squared,
the correction $r_V - 1$ is suppressed by $\cos^2 2\theta_{12}/3 \approx 0.03$.  
Note that $r_V - 1$ is positive for neutrinos because $\Delta H_{21} < {\Dms}/{2 E_\nu}$  
and increases with energy. For antineutrinos it is negative.

One comment is in order. Recently there was a discussion on why the 
formulas for the $3\nu$-oscillation probabilities in matter derived 
for $E \gg E_{21}^R$ ({\it i.e.},  far above the 1-2 resonance) 
work well at $E \sim E_{21}^R$ \cite{Asano:2011nj}.
Rather complicated explanation  has been proposed in \cite{Xu:2015kma}.
In fact, the reason is simple. For small values of the 2-1 phase,
which is true at all proposed long-baseline experiments, the vacuum mimicking is realized. 
In spite of large matter corrections to the oscillation phase and mixing angle, 
the corrections cancel with each other in the oscillation probabilities.   
The same expression for the oscillation probability applies at all energies
as long as the matter phase is small and the presence of the 1-2 resonance or not is irrelevant. 

Corrections due to $\kappa$ \geqn{eq:kappa} and \geqn{eq:kappa-1} depend on the 1-2 sector parameters,
$\kappa = \kappa(\theta_{21}, \Delta_{21}, V)$, and break vacuum breaking. 
In the $3 \nu$ case the vacuum mimicking does not work exactly even for the
standard interaction. As can  be seen from \geqn{eq:Pme-SI}
the matter effect associated with the 1-2 splitting, $\phi_{21}^m$, 
appears in the phase of the interference term. 
To realize the vacuum mimicking, one needs to reduce $3\nu$ evolution to $2\nu$ evolution
associated with small mass splitting.
In the $3\nu$ case, the violation of vacuum mimicking can be induced 
even by the diagonal elements of the matter potential matrix (see below).

%%%%%%%%%%%%%%%%%%%%%%%%%%%%%%%%%%%%%%%%%%%%%%%%%%%%%%%%%%%%%%%%
\section{Standard Interaction and Vacuum Mimicking}
\label{sec:SI}
%%%%%%%%%%%%%%%%%%%%%%%%%%%%%%%%%%%%%%%%%%%%%%%%%%%%%%%%%%%%%%%%%%

\subsection{Oscillation parameters}
%%%%%%%%%%%%%%%%%%%%%%%%%%%%%%%%%%%%%%%%%%%%%%%%%%%%%%%%%%%%%%%%%%%%%%%%

Using the expression \geqn{eq:V} of $\mathbb V$ for the standard interactions in the flavor basis, 
we obtain the matrix of matter potential in the separation basis,
\begin{equation}
\mathbb V_{sep}  = U_{13}^T \mathbb V U_{13} = 
V
\left\lgroup
\begin{matrix}
  \Cr^2 & & \Cr \Sr \\
        & 0 & \\
  \Cr \Sr & & \Sr^2
\end{matrix}
\right\rgroup \,.
\end{equation}
Consequently, the total Hamiltonian (\ref{eq:H'}) can be written explicitly as
\begin{equation}
  \mathcal H'
=
  \frac {\Dma}{2 E_\nu}
\left\lgroup
\begin{matrix}
  0 \\
& 0 \\
& & 1
\end{matrix}
\right\rgroup
+
\frac {\Dms} {2 E_\nu}
\left\lgroup
\begin{matrix}
  \Ss^2 & \Cs \Ss \\
  \Cs \Ss & \Cs^2 \\
& & 0
\end{matrix}
\right\rgroup
+
V
\left\lgroup
\begin{matrix}
  \Cr^2 & & \Cr \Sr \\
        & 0 & \\
  \Cr \Sr & & \Sr^2
\end{matrix}
\right\rgroup \,.
\label{eq:H''-SI}
\end{equation}
Here the 1-3 mixing (being proportional to $\Tr$) is generated by
matter potential. 

An additional 1-3 rotation
\begin{equation}
  \delta \Tr \approx \tan \delta \Tr \approx 
  \Sr \Cr \frac {2 E_\nu V}{\Dma} = \Sr \Cr x_{31}
\label{eq:add13}
\end{equation}
eliminates the 1-3 and 3-1 elements of the Hamiltonian 
(\ref{eq:H''-SI}). This rotation, in turn, generates non-zero 2-3 and 3-2 elements 
which have next order of smallness and can be neglected. 
Consequently, $\theta_{23}^m \approx \theta_{23}$ 
(see \cite{Blennow:2013rca} for details). 

After the rotation (\ref{eq:add13}) the third state decouples and
for the rest of the system we obtain the effective $2 \nu$ Hamiltonian 
\begin{equation}
{\mathcal H'}_{2\nu} \approx  \frac {\Dms} {2 E_\nu}
\left\lgroup
\begin{matrix}
  \Ss^2 + c_{13}^2 x_{21} & \Cs \Ss \\
  \Cs \Ss & \Cs^2 
\end{matrix}
\right\rgroup, 
\label{eq:2nuham}  
\end{equation}
where 
\begin{equation}
  x_{21}
\equiv
  \frac{2 E_\nu V}{ \Dms} =  0.28 \left(\frac{E_\nu}{ 100 ~{\rm MeV}}\right) . 
\end{equation}
For typical energies of $\mu$DAR and T2K,
we have $x_{21} = 0.14$ and $x_{21} = 1.5$. 
The correction to the 1-1 element generated by decoupling is
of the order $r_\Delta s_{13}^2$, and therefore has been neglected in (\ref{eq:2nuham}). 
Notice that the Hamiltonian (\ref{eq:2nuham}) 
can be obtained by block-diagonalization 
(\ref{eq:H''-SI}) and decoupling of the third state. 

The diagonalization of (\ref{eq:2nuham}) gives the effective mass splitting
\begin{equation}
\Delta H_{21} = 
  \frac{\Dms}{2E}
  \sqrt{\sin^2 2 \Ts + (\cos 2 \Ts - c _{13}^2 x_{21})^2} \,,
\label{eq:12spl}
\end{equation}
and the 1-2 mixing angle in matter  
\begin{equation}
\sin 2 \Ts^m =
  \frac {\Dms}{2E \Delta H_{21}} \sin 2 \Ts \,. 
\label{eq:sangle}
\end{equation}
The 1-3 splitting (upper sign) and the 2-3 spitting (lower sign) are
\begin{equation}
\Delta H_{31} (\Delta H_{32})
\approx
  \frac{\Dma}{2E} 
-
  \frac{1}{2} 
\left[
  \frac{\Dms}{2E}
\mp \Delta H_{21}
+ (\Cr^2 - 2 \Sr^2 )  V 
\right].
\label{eq:1323spl}
\end{equation}
Thus, the diagonalization matrix in the separation basis is given by 
\begin{equation}
U' = U_{13}(\delta \theta_{13}) U_{12}(\theta_{12}^m), 
\end{equation}
and the total mixing matrix in matter becomes 
\begin{equation}
U^m = \mathcal R U' = U_{23}(\theta_{23}) \Gamma_{\delta_D}   
U_{13} (\theta_{13}^m) U_{12}(\theta_{12}^m), 
\label{eq:tot-mix}
\end{equation}
where 
\begin{equation}
\theta_{13}^m = \theta_{13} + \delta \theta_{13}. 
\label{eq:13corr}
\end{equation}
The matter correction to the CP phase is absent: $\delta_D^m = \delta_D$.

\subsection{The oscillation probability} 
%%%%%%%%%%%%%%%%%%%%%%%%%%%%%%%%%%%%%%%%%%%%%%%%%%%%%%%%%%%%%%%%%%%

%Let us consider consequences of the  vacuum mimicking. 
%For $\phi_{21}^m \ll 1$ expansion of sine in $A_{12}^m$ gives 
%in the first approximation in $\phi_{21}^m$:  $\Delta H_{21}L/2$, so that using 
%(\ref{eq:sangle}) we have 
%\begin{equation}
%  \sin 2 \theta_{12}^m ~\sin \phi_{21}^m
%\approx
%  \sin 2 \theta_{12} ~\phi_{21}
%=
%  \sin 2 \theta_{12} \frac{\Dms L}{2 E_\nu}  \,,
%\label{eq:vacmim2}
%\end{equation}
%which precisely coincides with the result in vacuum. The reason is that  
%the standard matter potential appears in the diagonal elements of the 
%Hamiltonian $\mathcal H$ in the flavor basis. For small distances $L$ 
%the oscillation amplitudes (S-matrix) are given simply by 
%$\mathcal H L $. In particular,  the transition amplitude $\nu_\mu \rightarrow \nu_e$ 
%equals $(\mathcal H)_{\mu e} L $ and according to (\ref{eq:2nuham})
%this reduces immediately to (\ref{eq:vacmim2}). 

The probability of $\nu_\mu \rightarrow \nu_e$ transition
is given in (\ref{eq:Pme-SI} - \ref{eq:Pme-coefficients}).
Taking $r_V \approx 1$,   
$c_{23}^m \approx c_{23}$, $s_{23}^m \approx s_{23}$ and 
$\delta_D^m = \delta_D$, it becomes
\begin{eqnarray}
  P_{\mu e}
& = &
  \kappa^2 s^2_{23} \sin^2 2 \theta^m_{13} \sin^2 \phi^m_{31}
+ c^2_{23} c^{m2}_{13} \sin^2 2 \theta_{12} \sin^2 \phi_{21}
\nonumber
\\
& + &
  \kappa \sin 2 \theta_{23} c^m_{13} \sin 2 \theta^m_{13} \sin \phi^m_{31} \sin 2 \theta_{12} \sin \phi_{21}
  \cos (\delta_D + \phi^m_{32} + \phi_\kappa) \,.
\label{eq:Pmett}
\end{eqnarray}
The correction to the atmospheric amplitude given by 
$(\kappa - 1) \sim r_\Delta$ can also be neglected. Then for $\kappa = 1$
and $\phi_\kappa = 0$, \geqn{eq:Pmett} reproduces
the commonly used formula of $P_{\mu e}$ \cite{PDG12}. The oscillation parameters in matter
$\theta^m_{13}$, $\phi^m_{31}$, and $\phi^m_{32}$ that enter this formula
have been obtained in Eqs. (\ref{eq:12spl}),  
(\ref{eq:sangle}),  (\ref{eq:1323spl}), (\ref{eq:13corr}). 

\begin{figure}[h!]
\centering
\includegraphics[height=0.48\textwidth,angle=-90]{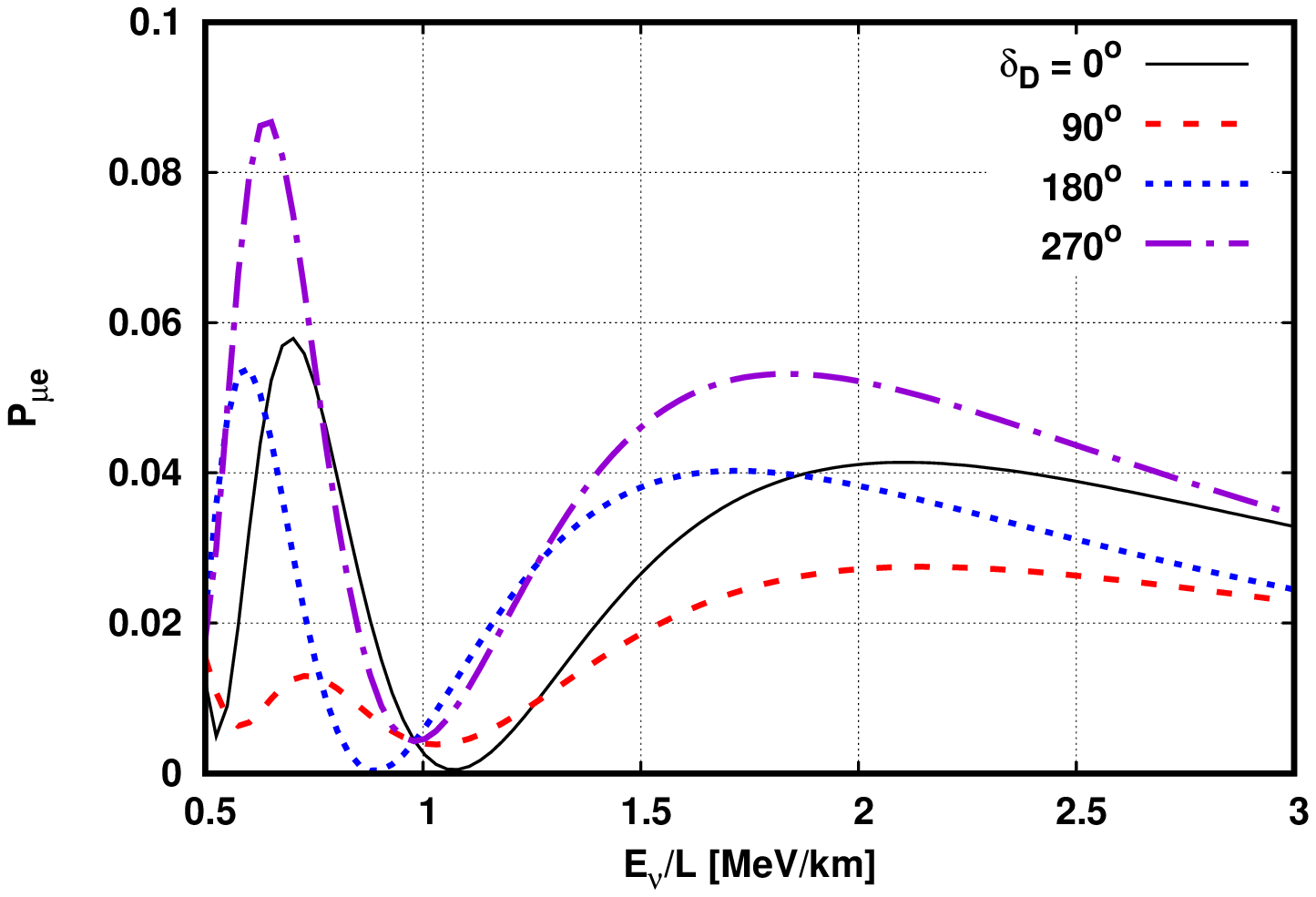}
\includegraphics[height=0.48\textwidth,angle=-90]{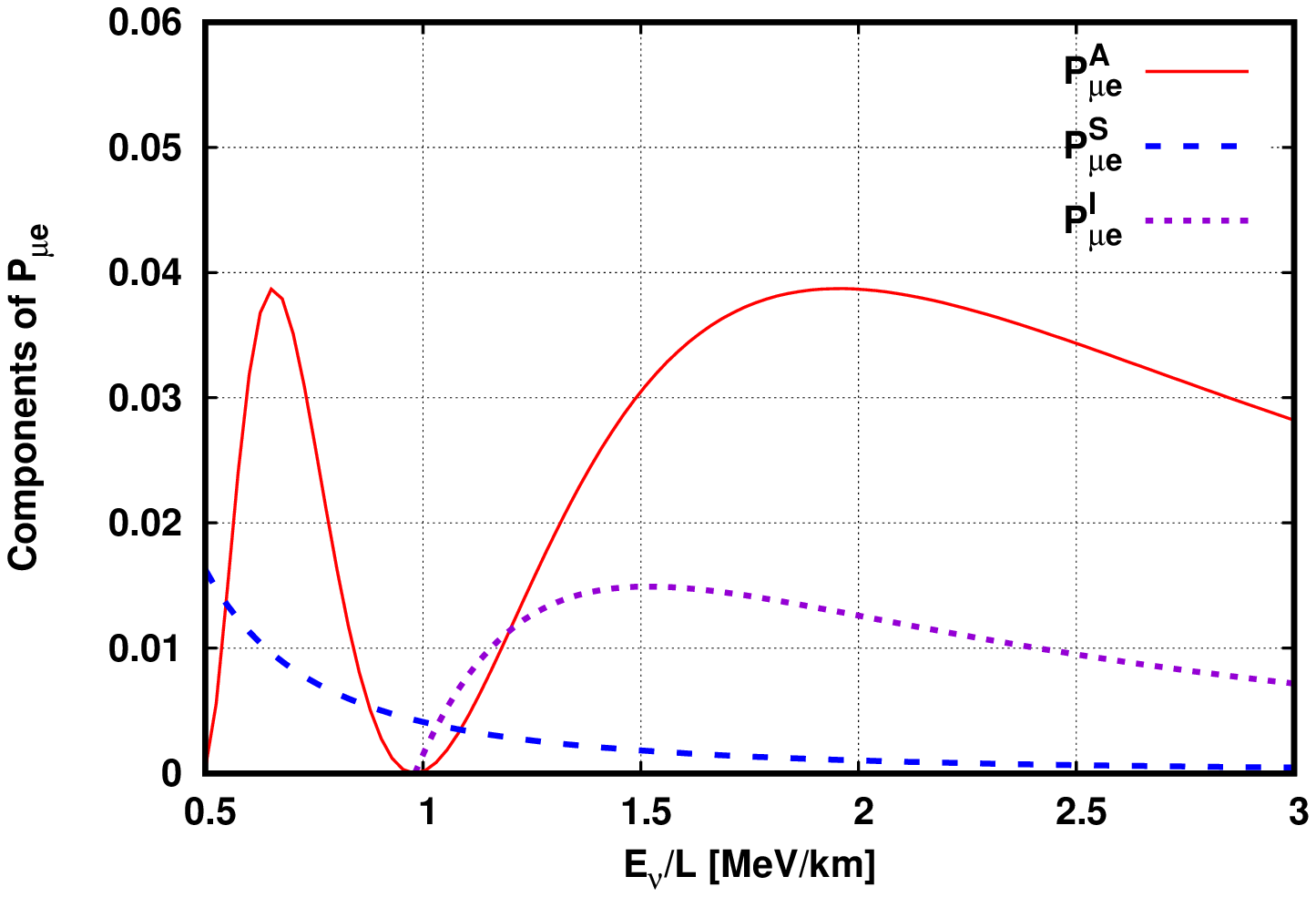}
\caption{The oscillation probability $P_{\mu e}$ as a function of $E_\nu/L$ 
for different values of the CP phase $\delta_D$
(left). Different components of   $P_{\mu e}$    
($P^A_{\mu e}$, $P^I_{\mu e}$, $P^S_{\mu e}$) as functions of $E_\nu/L$ for $\delta_{CP} = 3\pi/2$.  
We use the \tk baseline $L = 295\,\mbox{km}$.  
}
\label{fig:Pme}
\end{figure}

In \gfig{fig:Pme} (left panel) we show the dependence of the probability $P_{\mu e}$
on $E_\nu/L$ for the T2K setup and different values of $\delta_D$.
At the first oscillation maximum, $E_\nu/L \approx 1.8$, we have 
$P_{\mu e}(0) = P_{\mu e}(\pi)$ and $P_{\mu e}(\pm \pi/2)  
= (1 \mp 1/3) P_{\mu e}(0)$ where
the plus sign corresponds to the presently favored $\delta_D = 3\pi/2$.
The spread of the values of $P_{\mu e}(\delta_D)$ with varying $\delta_D$ is more than $60\%$.

According to  (\ref{eq:Pme-coefficients}) 
$P^A_{\mu e} \propto \Sr^2$, $P^S_{\mu e}$ is suppressed 
by $\sin^2 \phi_{21} \propto r_\Delta^2$, and
$P^I_{\mu e} \propto \Sr r_\Delta$. 
So, the following  hierarchy 
\begin{subequations}
\begin{eqnarray}
  \frac {P^S_{\mu e}}{P^A_{\mu e}}
& \approx & 
  \sin^2 2 \Ts \frac {\phi_{21}^2}{4 \Sr^2}
\approx
  \frac 2 9 \left( \frac{\pi r_\Delta} {2 \Sr}  \right)^2
\approx
  2\% , 
\\
  \frac {P^I_{\mu e}}{P^A_{\mu e}}
& \approx & 
  \frac {\sin 2 \Ts}{\Sr} \phi_{21}
\approx
  \frac {\pi r_\Delta}{2\Sr}
\approx
  30\% \,,
\end{eqnarray}
\end{subequations}
is realized  between the three components at the first oscillation maximum.
Roughly, $P^A_{\mu e} : P^I_{\mu e} : P^S_{\mu e} = 50 : 15 : 1$
which holds also above the resonance (see \gfig{fig:Pme}) but changes 
significantly 
with decrease of energy: at $E_\nu/L \approx 1.2$ we have  $P^A_{\mu e} =  P^I_{\mu e}$ 
and $P^S_{\mu e} = P^A_{\mu e}/4$.  
Similar relations between the different components of $P_{\mu e}$ are also realized at $\mu$SK.

%%%%%%%%%%%ffff1%%%%%%%%%%%%%%%%%%%%%%%%%%%%%%%%%%%%%%%%%%%%%%%%%%%%%%%%%%
%%%%%%%%%%%%%%%%%%%%%%%%%%%%%%%%%%%%%%%%%%%%%%%%%%%%%%%%%%

\begin{figure}[h]
\centering
\includegraphics[height=0.48\textwidth,angle=-90]{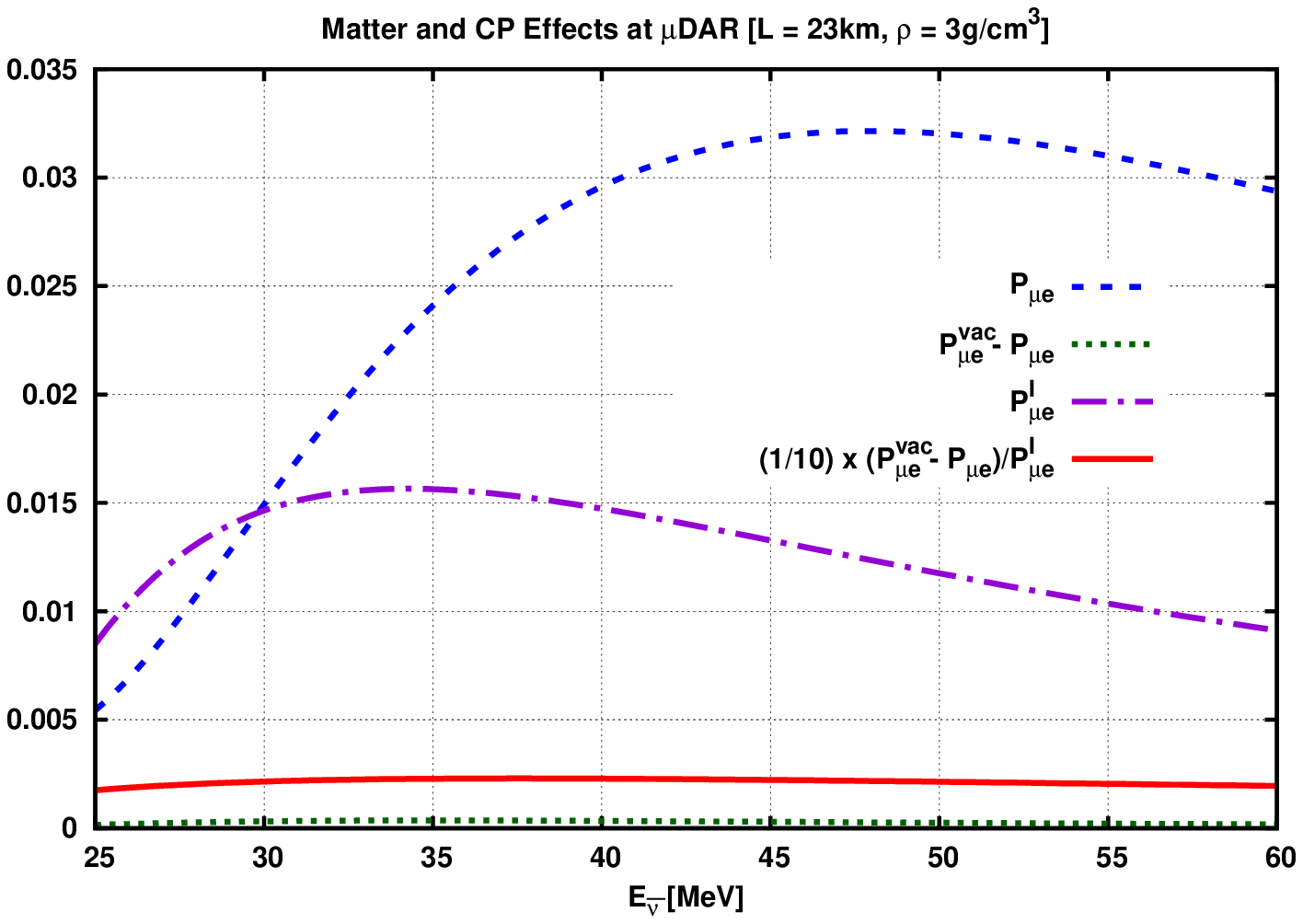}
\includegraphics[height=0.48\textwidth,angle=-90]{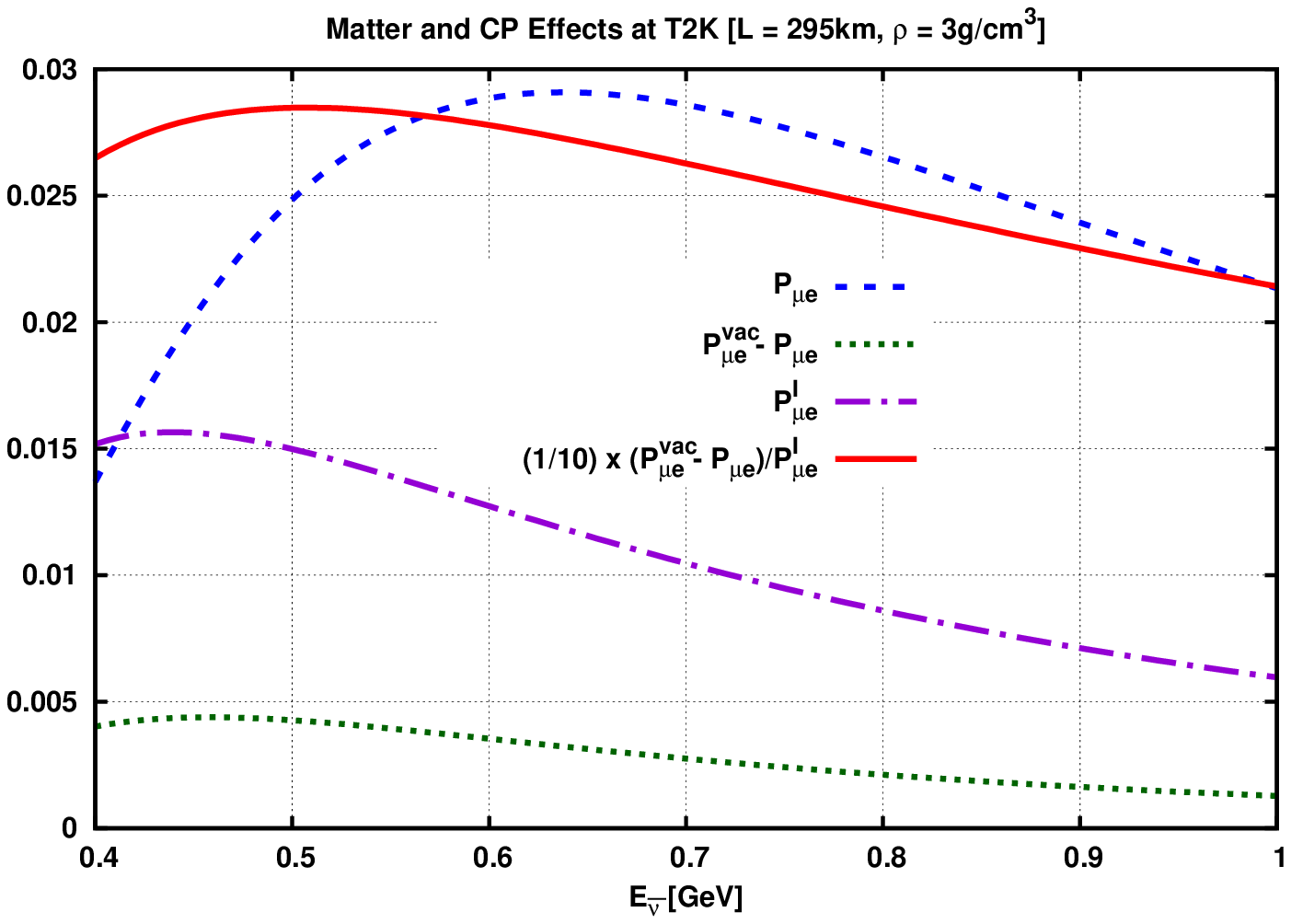}
\caption{The matter effect on probabilities 
($P^{vac}_{\mu e} - P_{\mu e}$) 
and the CP term coefficient
         $P^I_{\mu e}$ at $\mu$DAR (left) and \tk (right) 
for $\dD = 3\pi/2$.}
\label{fig:SI}
\end{figure}

The matter correction to the whole oscillation probability $P_{\mu e}$ mainly comes from
the angle $\Tr^m$ and the  phase $\phi_{31}^m$. Using $P_{\mu e}^A$ only we find 
\begin{eqnarray}
  V \frac {\partial P_{\mu e}^A}{\partial V}
& \approx & 
\frac{1}{2}  \Sa^2 \sin^2 2 \Tr \sin \phi_{31}
\left[4 \sin \phi_{31} \frac {2 E_\nu V}{\Dma}
-   \Cr^2 \cos \phi_{31}V L \right]  
\nonumber \\
& \approx & \Sa^2 \sin^2 2 \Tr \sin^2 \phi_{31} \frac{4 E_\nu V}{\Dma} 
\left[ 1 - c_{13}^2 \cot \phi_{31} \frac{\phi_{31}}{2} \right] 
\nonumber \\
& \approx &
\sin^2 2 \theta_{13} x_{31} (1 - 0.5  c_{13}^2 \phi_{31} \cot \phi_{31}).   
\label{eq:dPme-SI}
\end{eqnarray}
The first term in \geqn{eq:dPme-SI} comes from $\Tr^m$ 
while the second from $\phi_{31}^m$. 
At the first  oscillation maximum, the second  
term is suppressed by $\cot \phi_{31} \approx 0$. However,
if the neutrino energy spectrum is wide enough, $\cos \phi_{31}$ can become sizable 
out of the peak, as shown in \gfig{fig:SI}. According to \gfig{fig:SI},
$P_{\mu e}$  and $P_{\mu e}^I$ at T2K and $\mu$SK 
are very similar when expressed as functions of $E_\nu/ L$. 
At the first oscillation maximum $P_{\mu e}^I / P_{\mu e} \approx 0.3$ and
the matter effect increases with energy as 
$$ 
P_{\mu e} -  P^{vac}_{\mu e} \propto E_\nu
$$
according to (\ref{eq:dPme-SI}).
Therefore the matter effect at $\mu$SK is typically $12$ times smaller than at T2K. 
With respect to the interference (CP) term, the matter effect is about $(25 - 30)\%$ at T2K and 
$(2.0 - 2.2)\%$ at $\mu$SK.

%%%%%%%ffff2%%%%%%%%%%%%%%%%%%%%%%%%%%%%%%%%%%%%%%%%%%%%%%%%%%%
%%%%%%%%%%%%%%%%%%%%%%%%%%%%%%%%%%%%%%%%%%%%%%%%%%%%%%%%%%%%

For standard interactions, the matter corrections mainly come from the 1-3 
mixing $s_{13}^m$, being 
$\sim 13\%$ at T2K and $\sim 1\%$ at $\mu$DAR, at the first oscillation peak and for $\dD = 3\pi/2$.
The matter effect from the 1-2 sector is strongly suppressed by the vacuum mimicking, 
producing $\mathcal O(r_\Delta)$ correction to the atmospheric 
amplitude $A^A_{\mu e}$. The interference term is about 1/3 of the total probability.

%%%%%%%%%%%%%%%%%%%%%%%%%%%%%%%%%%%%%%%%%%%%%%%%%%%%%%%%%%%%%%%%%%%%%%%%%%%
\section{Effects of Non-Standard Interactions}
\label{sec:NSI}
%%%%%%%%%%%%%%%%%%%%%%%%%%%%%%%%%%%%%%%%%%%%%%%%%%%%%%%%%%%%%%%%%%%%

\subsection{Oscillation parameters in the presence of NSI}
\label{sec:NSI-param}
%%%%%%%%%%%%%%%%%%%%%%%%%%%%%%%%%%%%%%%%%%%%%%%%%%%%%%%%%%%%%%%%

In the presence of  NSI, the matter potential matrix  
$\mathbb V$ in the flavor basis can be parametrized 
as 
\begin{equation}
  \mathbb V
=
\left\lgroup
\begin{matrix}
  V \\
& 0 \\
& & 0
\end{matrix}
\right\rgroup
+
  V
\left\lgroup
\begin{matrix}
  \epsilon_{ee}      & \epsilon_{e \mu}      & \epsilon_{e \tau}   \\
  \epsilon^*_{e\mu}  & \epsilon_{\mu \mu}    & \epsilon_{\mu \tau} \\
  \epsilon^*_{e\tau} & \epsilon^*_{\mu \tau} & \epsilon_{\tau \tau}
\end{matrix}
\right\rgroup \,.
\label{eq:nsi-pot}
\end{equation}
(We do not separate here scattering on different components of matter: 
electrons, u- and d- quarks, summing up the effect.)  
The $1\sigma$ constraints on the NSI parameters $\epsilon_{\alpha \beta}$ 
from the global fit \cite{globalfit}  are 
\begin{subequations}
\begin{eqnarray}
&&
  |\epsilon_{e \mu}| < 0.16,
\qquad
  |\epsilon_{e \tau}| < 0.26,
\qquad
  |\epsilon_{\mu \tau}| < 0.02,
\\
&&
- 0.018 < \epsilon_{\tau \tau} - \epsilon_{\mu \mu} < 0.054,
\qquad
      0 < \epsilon_{ee}        - \epsilon_{\mu \mu} < 0.93.
\end{eqnarray}
\label{eq:globalfit}
\end{subequations}
The hierarchy among these constraints can be described by powers 
of small parameter $\lambda = 0.15$  (notice that $\Sr \approx \lambda$):     
\begin{equation}
  \epsilon_{ee} - \epsilon_{\mu \mu}
<
  \mathcal O(1) \,,
\qquad
\epsilon_{e \mu}, ~~ \epsilon_{e \tau} 
<
  \mathcal O(\lambda) \,,
\qquad
\epsilon_{\mu \tau}, ~~\epsilon_{\tau \tau} - \epsilon_{\mu \mu} 
<
  \mathcal O(\lambda^2) \,.
\label{eq:epsilon-sr}
\end{equation}
Consequently, the matrix of matter potential in the flavor basis can have the
maximal allowed $1 \sigma$ values,
\begin{equation}
  \mathbb V
\sim 
  V
\left\lgroup
\begin{matrix}
  1       & \lambda     &  \lambda     \\
 \lambda  &  \lambda^2  &  \lambda^2\\
 \lambda  &  \lambda^2  & \lambda^2 
\end{matrix}
\right\rgroup \,.
\label{eq:V-hierarchy}
\end{equation} 
In the separation basis, the matrix of matter potentials 
is given by transformation of (\ref{eq:nsi-pot}): 
\begin{equation}
  \mathbb V_{sep}
= U^T_{13} \Gamma^\dagger_\delta U_{23}^T \mathbb V U_{23} \Gamma_\delta U_{13} 
=   V
\left\lgroup
\begin{matrix}
  \te_{11} + c_{13}^2   & \te_{12}   & \te_{13} + s_{13}c_{13} \\
  \te^*_{12} & \te_{22}   & \te_{23} \\
  \te^*_{13} + s_{13}c_{13} & \te^*_{23} & \te_{33} + s_{13}^2
\end{matrix}
\right\rgroup , 
\label{eq:H'ns}
\end{equation}
where 
\begin{eqnarray}
\te_{11}
& = &
  \Cr^2 (\epsilon_{ee} - \epsilon_{\mu \mu})
+ \Sr^2
\left[
  \Ca^2 (\epsilon_{\tau \tau} - \epsilon_{\mu \mu})
+ 2 \Ca \Sa \mathbb R(\epsilon_{\mu \tau})
\right]
- 2 \Cr \Sr \mathbb R
\left[
(
  \Sa \epsilon_{e \mu}
+ \Ca \epsilon_{e \tau}
) e^{i \dD}
\right] \,,
\nonumber\\
%%%
\te_{12}
& = &
  \Cr (\Ca \epsilon_{e \mu} - \Sa \epsilon_{e \tau})
+ \Sr e^{- i \dD}
\left[
  \Ca \Sa (\epsilon_{\tau \tau} - \epsilon_{\mu \mu})
- (\cos 2 \Ta \mathbb R - i \mathbb I)(\epsilon_{\mu \tau})
\right] \,,
\nonumber\\
\te_{13}
& = &
  \Cr \Sr
\left[
  (\epsilon_{ee} - \epsilon_{\mu \mu})
- \Ca^2 (\epsilon_{\tau \tau} - \epsilon_{\mu \mu})
- 2 \Ca \Sa \mathbb R(\epsilon_{\mu \tau})
\right]
+
  (\cos 2 \Tr \mathbb R + i \mathbb I)
\left[
  ( \Sa \epsilon_{e \mu} + \Ca \epsilon_{e \tau} ) e^{i \dD}
\right] \,,
\qquad
\nonumber\\
%%%
\te_{22}
& = &
- 2 \Ca \Sa \mathbb R(\epsilon_{\mu \tau})
+ \Sa^2 (\epsilon_{\tau \tau} - \epsilon_{\mu \mu}) \,,
\nonumber\\
%%%
\te_{23}
& = &
  \Sr (\Ca \epsilon^*_{e \mu} - \Sa \epsilon^*_{e \tau})
- \Cr e^{i \dD}
\left[
  \Ca \Sa (\epsilon_{\tau \tau} - \epsilon_{\mu \mu})
- (\cos 2 \Ta \mathbb R + i \mathbb I)(\epsilon_{\mu \tau})
\right] \,,
\nonumber\\
\te_{33}
& = &
  \Sr^2 (\epsilon_{ee} - \epsilon_{\mu \mu})
+ \Cr^2
\left[
  \Ca^2 (\epsilon_{\tau \tau} - \epsilon_{\mu \mu})
+ 2 \Ca \Sa \mathbb R(\epsilon_{\mu \tau})
\right]
+ 2 \Cr \Sr \mathbb R
\left[
(
  \Sa \epsilon_{e \mu}
+ \Ca \epsilon_{e \tau}
) e^{i \dD}
\right] \,. 
\label{eq:tilde-epsilon-explicit}
\end{eqnarray}
The operators $\mathbb R$ and $\mathbb I$ extract the real and imaginary parts from
the attached quantities, e.g. $\mathbb R(\epsilon_{\mu \tau}) \equiv \mbox{Re}(\epsilon_{\mu \tau})$.
For convenience, we have subtracted from $\mathbb V$ 
a diagonal term $\epsilon_{\mu \mu}$
proportional to the unit matrix $\mathbb I_{3 \times 3}$ which
does not affect oscillation probabilities.
Notice that the elements of $ \mathbb V_{sep}$
in the separation basis  
have the same hierarchy of values as \geqn{eq:V-hierarchy} 
in the flavor basis due to hierarchical values of the mixing angles
$s_{13} = \mathcal O(\lambda)$ and $s_{23} = \mathcal O (1)$ in \geqn{eq:H'ns}.

We can diagonalize the total Hamiltonian in the same way as in the case of standard interaction. 
First, block-diagonalization gives the effective $2\nu$ Hamiltonian 
\begin{equation}
{\mathcal H'}_{2\nu} \approx  \frac {\Dms} {2 E_\nu}
\left\lgroup
\begin{matrix}
  \Ss^2 + x_{21} (c_{13}^2 +  \te_{11}) & \Ss \Cs   + x_{21} \te_{12} \\
  \Ss \Cs   + x_{21} \te^*_{12}  & \Cs^2 + x_{21} \te_{22}
\end{matrix} 
\right\rgroup
\label{eq:2nuham-ns}
\end{equation}
for the first two states.
The decoupled state has the eigenvalue 
\begin{equation}
H_3 \approx  \frac{\Dma}{2E} + V (\Sr^2  + \te_{33}). 
\label{eq:eig33}
\end{equation} 

In the first approximation, this block-diagonalization is equivalent 
to an additional 1-3 rotation 
$$
\Gamma_{\delta_3} U_{13}(\delta \Tr)
$$
which removes the 1-3 and 3-1 elements of the total Hamiltonian. 
Here 
\begin{equation}
\Gamma_{\delta_3} \equiv  diag (1, 1, e^{i\delta_3}), ~~~~  
\delta_3 = {\rm Arg} [s_{13}c_{13}  + \te_{13}], 
\end{equation}
and the rotation angle in $U_{13}(\delta \Tr)$ is determined approximately  by 
\begin{equation}
 \delta \Tr \approx \tan \delta \Tr \approx
x_{31} \frac{|\Sr \Cr + \te_{13}|}{1 + x_{31}(s_{13}^2 + \te_{33})}
\approx x_{31}|\Sr \Cr + \te_{13}|. 
\label{eq:add13ns}
\end{equation}
Numerically, $\delta \Tr \sim {\mathcal O}(\lambda^3)$, but the phase $\delta_3$
can be large, $\delta_3 =  {\mathcal O}(1)$, since $\te_{13} \sim s_{13} \sim \lambda$.

In this approximation, we have neglected the 2-3 and 3-2 elements 
of the Hamiltonian. They  can be eliminated by an additional 2-3  rotation on the angle 
\begin{equation}
 \delta \theta_{23} \approx x_{31} |\te_{23}| 
\end{equation}
which has the next order of smallness: $\delta \theta_{23} = {\mathcal O}(\lambda^4)$.  

Diagonalization of the matrix (\ref{eq:2nuham-ns}) gives 
the effective mass splitting
\begin{equation}
\Delta H_{21} 
= \frac{\Dms}{2E}
  \sqrt{
[\cos 2 \Ts  - x_{21}(\Cr^2 + \te_{11} - \te_{22})]^2 
+ (\sin 2 \theta_{12}  + 2 x_{21} \te_{12})^2},   
\label{eq:dM2s-NSI}
\end{equation}
and mixing $\Gamma_{\delta_2} U_{12}(\Ts^m)$ where
\begin{equation}
\sin 2 \Ts^m = 
\frac{\Dms}{2E \Delta H_{21}} |\sin 2 \theta_{12} + 2 x_{21} \te_{12}|, 
\label{eq:ts-NSI}
\end{equation}
\begin{equation}
\Gamma_{\delta_2} = {\rm diag}(1, e^{i\delta_2}), ~~~~\delta_2 = 
- {\rm Arg} [\sin 2 \theta_{12} + 2 x_{21} \te_{12}]. 
\label{eq:delta2}
\end{equation}
Notice that with NSI the off-diagonal element of the Hamiltonian $\mathcal H'_{2 \nu}$
are complex and so additional rephasing $\Gamma_{\delta_2}$ is needed.  
In fact, the phase $\delta_2$ also originates from the violation of vacuum mimicking, being
the phase of the off-diagonal element $\mathcal H_{12}$ (or equivalently
 $\bar{\mathcal H}$ in \geqn{eq:s2sm} for the $2 \nu$ Hamiltonian).

Using the trace of the Hamiltonian (\ref{eq:2nuham-ns})
as well as the equalities (\ref{eq:eig33}) and (\ref{eq:dM2s-NSI}),
we find the 1-3 mass splitting
\begin{equation}
\Delta H_{31} 
\approx
  \frac{\Dma}{2E} + V(s_{13}^2  + \te_{33}) 
- \frac{1}{2} \left[
  \frac{\Dms}{2E} - \Delta H_{21}
+ (\Cr^2 + \te_{11} + \te_{22}) V
\right]. 
\label{eq:13splitting}
\end{equation}

Combining all the rotations we obtain the total mixing matrix in matter 
\begin{equation}
U_m = \mathcal R U' = U_{23}(\theta_{23}) \Gamma_{\delta_D}
U_{13} (\theta_{13}) \Gamma_{\delta_3} U_{13} (\delta \theta_{13})  
\Gamma_{\delta_2} U_{12}(\theta_{12}^m). 
\label{eq:tot-mix2}
\end{equation}
Notice that if 
$\delta_3 = 0$, the correction $\delta \Tr$ can be combined with $\Tr$   
as in the standard interaction case (\ref{eq:13corr}).

To find the effective mixing angles in matter, the matrix (\ref{eq:tot-mix2}) should be reduced 
to the standard parametrization.
Since $\Gamma_{\delta_2}$ commutes with the 1-3 rotations on the left-hand side,  
it can be combined with $\Gamma_{\delta_D}$:
\begin{equation}
\Gamma_{\delta_D} \Gamma_{\delta_2} = {\rm diag}\left[1, e^{i\delta_2}, e^{i\delta_2} \right]
{\rm diag} \left[1, 1, e^{i(\delta_D - \delta_2)}\right]. 
\label{eq:rephasing-l}
\end{equation}
The first matrix in (\ref{eq:rephasing-l}) can be omitted since it commutes with
the 2-3 rotations in (\ref{eq:tot-mix2}), and therefore can be absorbed in the rephasing of 
the charged lepton states.

It is straightforward to show that the product of the 1-3 transformations
$U_{13} (\theta_{13}) \Gamma_{\delta_3} U_{13} (\delta \theta_{13})$
can be written as 
\begin{equation}
U_{13} (\theta_{13}) \Gamma_{\delta_3} U_{13} (\delta \theta_{13}) = 
\Gamma_3 \Gamma_C \Gamma_\alpha U_{13} (\theta_{13}^m) \Gamma_\beta. 
\label{eq:product}
\end{equation}
Here
\begin{eqnarray}
\Gamma_3 & = & {\rm diag} \left(e^{i\delta_3/2},~ 1 ,~ e^{i\delta_3/2}\right), \\
\Gamma_C & = & {\rm diag} \left(e^{i\phi_C}, ~1 , ~e^{i\phi_C} \right), \\
\Gamma_\alpha & = & {\rm diag} \left(1, ~1 , ~ e^{- i(\phi_S + \phi_C)}\right), \\
\Gamma_\beta & = & {\rm diag} \left(1,~ 1 ,~ e^{i(\phi_S - \phi_C)}\right),  
\end{eqnarray}
and the phase $\phi_S$ is determined from
\begin{equation}
\tan \phi_S = \tan \left(\frac{\delta_3}{2}\right)~  
\frac{\sin(\theta_{13} - \delta \theta_{13})}{\sin(\theta_{13} + \delta \theta_{13})},
\label{eq:expr-ps}
\end{equation}
and the phase $\phi_C$ has similar expression with substitution $\sin \rightarrow  \cos$ and overall
minus sign. Then, the 1-3 mixing angle in matter is given by
\begin{equation}
\cos \theta_{13}^m = \sqrt{\cos^2 (\delta_3/2)  
\cos^2(\theta_{13} + \delta \theta_{13}) + \sin^2 (\delta_3 /2)  
\cos^2(\theta_{13} - \delta \theta_{13})}. 
\end{equation}

After inserting (\ref{eq:product}) into (\ref{eq:tot-mix2}),
we find a number of simplifications. The matrix $\Gamma_\beta$ commutes with the
1-2 rotation and therefore can be absorbed into the rephasing of the eigenstates of the Hamiltonian.
The product $\Gamma_3 \Gamma_C \Gamma_\alpha$ can be written as 
\begin{equation}
\Gamma_3 \Gamma_C \Gamma_\alpha = {\rm diag} 
\left[e^{i(\delta_3/2 + \phi_C) },~ 1, ~1 \right] 
\times {\rm diag} \left[1, ~1, ~ e^{i(\delta_3/2 - \phi_S)}\right]. 
\end{equation}
The first matrix commutes with the 2-3 rotation of
(\ref{eq:tot-mix2}) and therefore can be absorbed into the rephasing of the flavor states. 
The second matrix can be combined with the matrix in (\ref{eq:rephasing-l}).  
As a result,  we obtain the standard expression for the mixing matrix 
\begin{equation}
U_m = \mathcal R U' = U_{23}(\theta_{23}) \Gamma_{\delta_D^m}
U_{13} (\theta_{13}^m) U_{12}(\theta_{12}^m)  
\label{eq:tot-mix4}
\end{equation}
with  
\begin{equation}
\Gamma_{\delta_D^m} = (1, 1, e^{i\delta_D^m}),  ~~~~~\delta_D^m \equiv \delta_D - \delta_2  
+ \frac{\delta_3}{2} - \phi_S.  
\end{equation}
Here  $\delta_D^m$ is 
the effective CP phase in matter which includes corrections from NSI.

The obtained expressions for the effective mixing parameters in matter can be simplified 
using the smallness of $\delta \theta_{13}$ (\ref{eq:add13ns}). 
From  (\ref{eq:expr-ps})  we find 
\begin{equation}
  \frac{\delta_3}{2} - \phi_S
\approx
  x_{31} c^2_{13} \left| 1 + \frac {\epsilon'_{13}}{c_{13} s_{13}} \right| \sin \delta_3 \, , 
\end{equation}
so that the effective CP phase in matter becomes 
\begin{equation}
  \delta_D^m
\approx
  \delta_D - \delta_2
+ x_{31} c^2_{13} \left| 1 + \frac {\epsilon'_{13}}{c_{13} s_{13}} \right| \sin \delta_3  \,.
\label{eq:dDm}
\end{equation}
For the effective 1-3 mixing angle in matter we have 
\begin{equation}
\theta_{13}^m \approx \theta_{13} + 
\cos \delta_3 \delta\theta_{13}.   
\label{eq:delta13mix}
\end{equation}
Recall that $\delta_2 = \delta_2(\te_{12})$ is a function of $\te_{12}$, 
whereas $\delta_3 = \delta_3(\te_{13})$ is a function of $\te_{13}$.

According to our consideration here, the matter potential (including NSI) influences 
the mixing angles, the CP phase, and the splittings of the eigenvalues of the Hamiltonian in rather specific ways 
which we will discuss in the next two subsections.

%%%%%%%%%%%%%%%%%%%%%%%%%%%%%%%%%%%%%%%%%%%%%%%%%%%%%%%%%%%
\subsection{Matter corrections and the violation of vacuum mimicking}
\label{sec:leading}
%%%%%%%%%%%%%%%%%%%%%%%%%%%%%%%%%%%%%%%%%%%%%%%%%%%%%%%%%%%%%%%%%%%%%%%

The matter potential influences the 1-2 mixing and splitting in a particular form 
that can be described with high accuracy as the violation of vacuum mimicking. 
The non-diagonal element of the Hamiltonian (\ref{eq:2nuham-ns}),  
\begin{equation} 
{\mathcal  H'}_{12} = s_{12} c_{12}  + x_{21} \te_{12},   
\end{equation}
depends on the NSI parameter $\te_{12}$
and therefore vacuum mimicking is broken \cite{Akhmedov:2000cs}
already at the lowest order, according to the discussions related to \geqn{eq:vacmim3}. 
The violation of vacuum mimicking is characterized by 
\begin{equation}
\rV  = \frac{|{\mathcal  H'}_{12}|}{{\mathcal  H}_{12}} = 
\left|1 + \frac{2 x_{21} \te^*_{12}}{\sin 2 \theta_{12}}\right| 
= \left|1 + \xi_{21} \te^*_{12}\right| ,  
\label{eq:rV}
\end{equation}
where we have denoted 
\begin{equation}
\xi_{21} \equiv \frac{2 x_{21}}{\sin 2 \theta_{12}}. 
\label{eq:def-xi}
\end{equation}
For T2K peak energy,  $E_\nu \sim 600$ MeV, the deviation
$\rV  - 1 \approx 2.30 \te_{12}$ can be as large as ${\mathcal O}(1)$.  
For $\mu$DAR ($E_\nu = 50$ MeV) we have  $(\rV  - 1) \sim 0.215~ \te_{12}$.  
Here the deviation also scales linearly with energy. Consequently, 
the deviation at T2K is about 12 times larger than the one at $\mu$DAR.

The violation parameter $r_V$ (\ref{eq:rv-param}) of vacuum mimicking equals
\begin{equation}
r_V = \sqrt{1 + 2\xi_{21}|\te_{12}| \cos \phi_\epsilon + \xi_{21}^2 |\te_{12}|^2 }
\end{equation}
with $\phi_\epsilon \equiv {\rm Arg} [\te_{12}]$.  

There are two effects of the violation of vacuum mimicking: 

1. Being attached to the solar amplitude \geqn{eq:rv-param},
the phase factor $e^{i\delta_2}$
contributes to the phase of the interference term (\ref{eq:Pme-SI}) directly: 
\begin{equation}
\delta_{D}  - \delta_2  + \phi_{32}^m.
\end{equation}
Consequently, the CP phase in matter equals
$$
\delta_{D}^m \approx \delta_{D} - \delta_2
$$
in agreement with (\ref{eq:dDm}) which differs by the small correction due to the
matter effect on the 1-3 mixing. Thus, the violation of vacuum mimicking gives the main correction to 
the CP phase in matter.

2. The interference term is modified by a factor $r_V$ and the solar term by a factor $r_V^2$. 
The deviation from the standard case equals 
\begin{equation} 
r_V^2 - 1 =  \xi_{21} |\te_{12}| (2 \cos \phi_\epsilon + \xi_{21} |\te_{12}| ). 
\end{equation}
Consequently, the correction vanishes when
\begin{equation}
|\te_{12}| = - \frac{2 \cos \phi_\epsilon}{\xi_{21}}. 
\end{equation}
For $\phi_\epsilon = 0$ and $\pi$, the factor $r_V$ itself can be zero if 
\begin{equation}
|\te_{12}| = \frac{1}{\xi_{21}} = \frac{\sin 2 \theta_{12}}{2 x_{21}} = 
\sin 2 \theta_{12}\frac{\Delta m_{21}^2}{4 EV}. 
\label{eq:te-zero}
\end{equation}
 
Let us consider the effects of the NSI parameters  
$\epsilon_{ee} - \epsilon_{\mu \mu}$,
$\epsilon_{\tau \tau} - \epsilon_{\mu \mu}$, $\epsilon_{e \mu}$, $\epsilon_{e \tau}$ 
and $\epsilon_{\mu \tau}$ separately. 
They modify the $\nu_\mu-\nu_e$ oscillation probability $P_{\mu}$ via $\te_{12}$,
see the second equation of (\ref{eq:tilde-epsilon-explicit}). 
Notice that the largest possible NSI parameter  
$\epsilon_{ee} - \epsilon_{\mu \mu}$ does not contribute to  $\te_{12}$. 

From (\ref{eq:tilde-epsilon-explicit}) we  
find the parameter of the violation of vacuum mimicking in terms of the NSI parameters: 
\begin{equation}
\rV  = \left|1 + \frac{2 x_{21} F}{\sin 2 \theta_{12}}\right|, ~~~ 
F =  
\begin{cases}
  0 \,    ~~~~~~~~~~~~~~~~~~~~~~~~~~~~~~~~~~~~~~~~~ \leq {\mathcal O}(\lambda^5)
\\[2mm]
\Ca \Sa \Sr e^{i \delta_D} (\epsilon_{\tau \tau} - \epsilon_{\mu \mu}) \, 
~~~~~~~~~~~~~\leq {\mathcal O}(\lambda^3)
\\[2mm]
\Ca \Cr \epsilon^*_{e \mu} \,
~~~~~~~~~~~~~~~~~~~~~~~~~~~~~~~~\leq {\mathcal O}(\lambda)
\\[2mm]
- \Sa \Cr \epsilon^*_{e \tau} \,
~~~~~~~~~~~~~~~~~~~~~~~~~~~~~~\leq {\mathcal O}(\lambda)
\\[2mm]
- \Sr e^{i \delta_D} (\cos 2 \Ta \mathbb R + i \mathbb I)(\epsilon_{\mu \tau}) \, 
~~~~~~\leq {\mathcal O}(\lambda^3). 
\end{cases}
\label{eq:vacuum-NSI}
\end{equation}
The presence of several non-zero 
$\epsilon$ can be easily taken into account by summing 
up the contributions to $F$.

Let us comment on the effect of individual NSI parameter (\ref{eq:vacuum-NSI})
when all others are zero. 
\begin{itemize}

\item 

$\epsilon_{ee} - \epsilon_{\mu \mu} \neq 0$.   
Being the biggest allowed parameter, it generates
$\te_{11}$, $\te_{13}$ and $\te_{33}$ while other $\te_{ij}$ parameters vanish. 
According to \geqn{eq:tilde-epsilon-explicit},
the parameter $\te_{11}$ is  the diagonal element  of the 
Hamiltonian ${\mathcal H'}_{2\nu}$ while $\te_{13}$ and $\te_{33}$ cannot contribute
to ${\mathcal H'}_{2\nu}$. Therefore, vacuum mimicking is realized:
$F = 0$ and
 $\rV =  1$ in the leading order.

\item 

$\epsilon_{\tau \tau} - \epsilon_{\mu \mu} \neq 0$.   
This parameter  contributes to all $\te_{ij}$. 
In $\te_{12}$ it appears with suppression factor $s_{13}$, so that  
$\te_{12} = {\mathcal O}(\lambda^3)$. 
Although $\epsilon_{\tau \tau}$ and  $\epsilon_{\mu \mu}$
are the diagonal elements of the matter potential matrix in the flavor basis,
they violate vacuum mimicking via $\te_{12}$. This happens due to the $3\nu$ mixing 
and large oscillation phase associated with the third state (otherwise mimicking would 
exist for all mass splittings \cite{Akhmedov:2000cs}). 
We can call such a violation {\it the induced} violation of vacuum mimicking due to the $3 \nu$ mixing. 
The corresponding contribution to the oscillation probability is proportional to $s_{23} s_{13}$.      

\item 
 
$\epsilon_{e \mu} \neq 0$  produces all
$\te_{ij}$ but $\te_{22}$. It appears in  $\te_{12}$ without suppression, 
and therefore provides the largest violation of vacuum mimicking. 
Similar statement applies for  $\epsilon_{e \tau} \neq 0$.

\item

 $\epsilon_{\mu \tau} \neq 0$  generates all $\te_{ij}$.   
Being small  it appears with $\Sr$  in $\te_{12}$.

\end{itemize}

%%%%%%%%%%%%%%%%%%%%%%%%%%%%%%%%%%%%%%%%%%%%%%%%%%%%%%%%%%%%%%%%%%%%
\begin{figure}[h!]
\centering
\includegraphics[width=5cm,height=8cm,angle=-90]{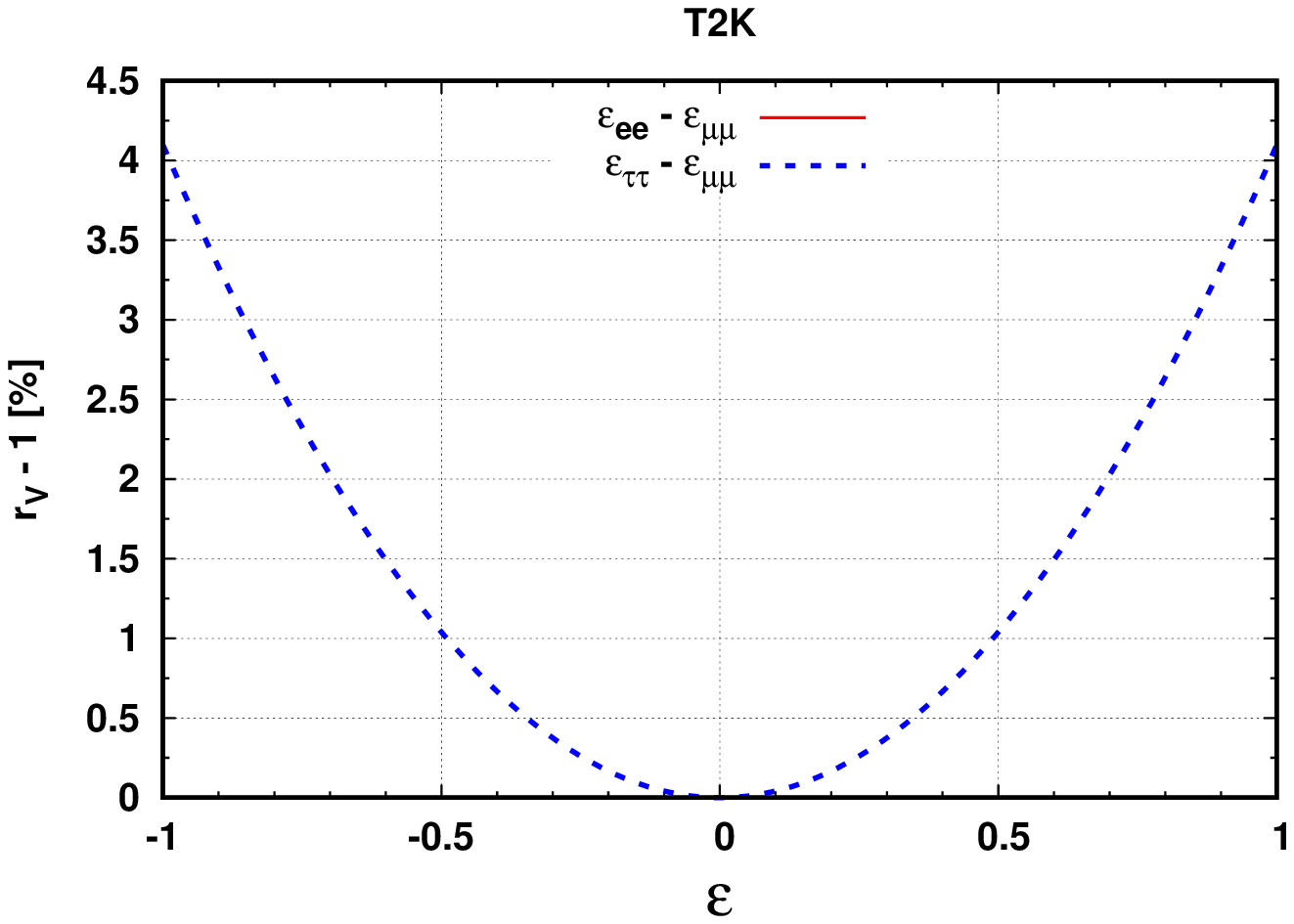}
\includegraphics[width=5cm,height=8cm,angle=-90]{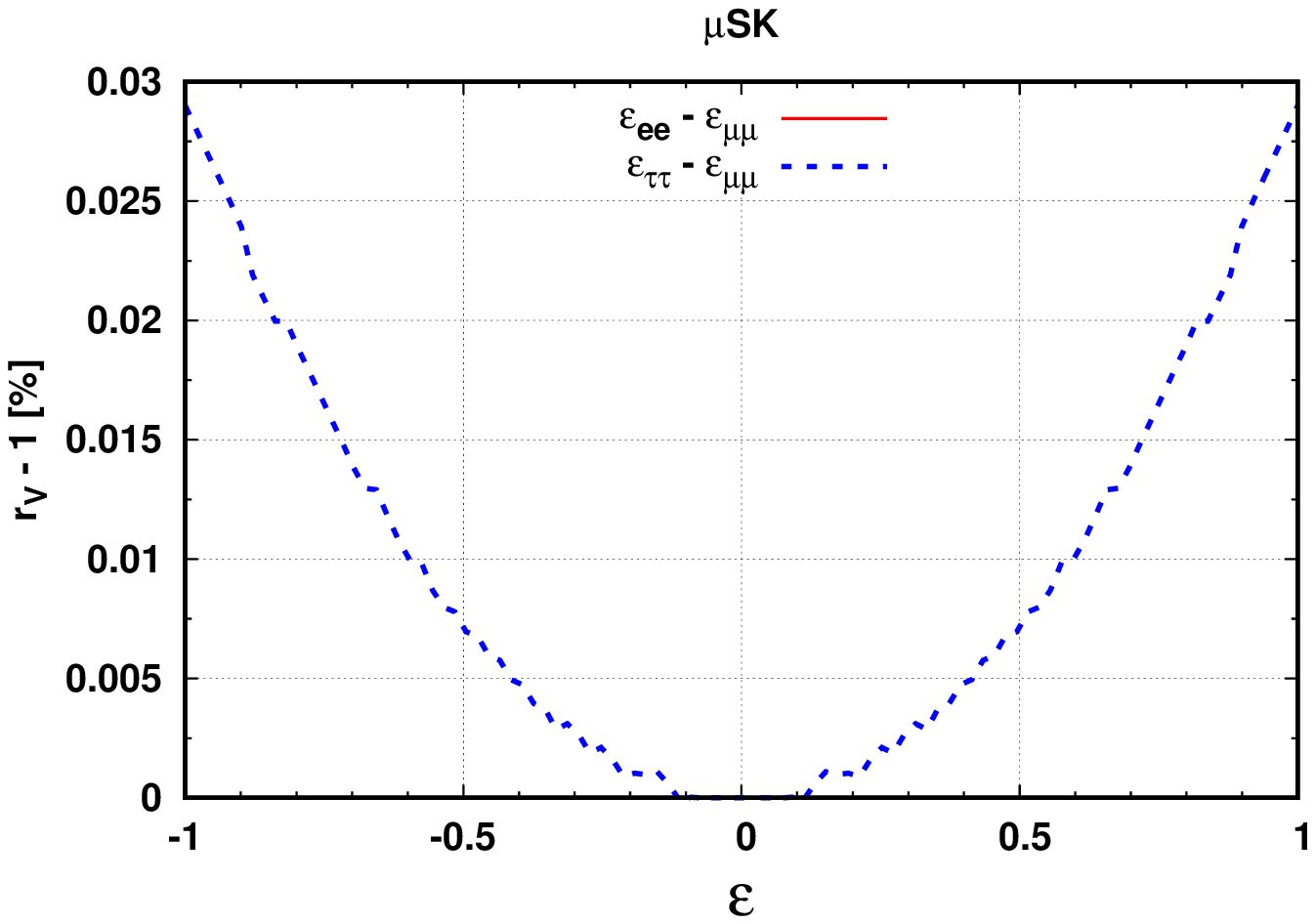}
\includegraphics[width=5cm,height=8cm,angle=-90]{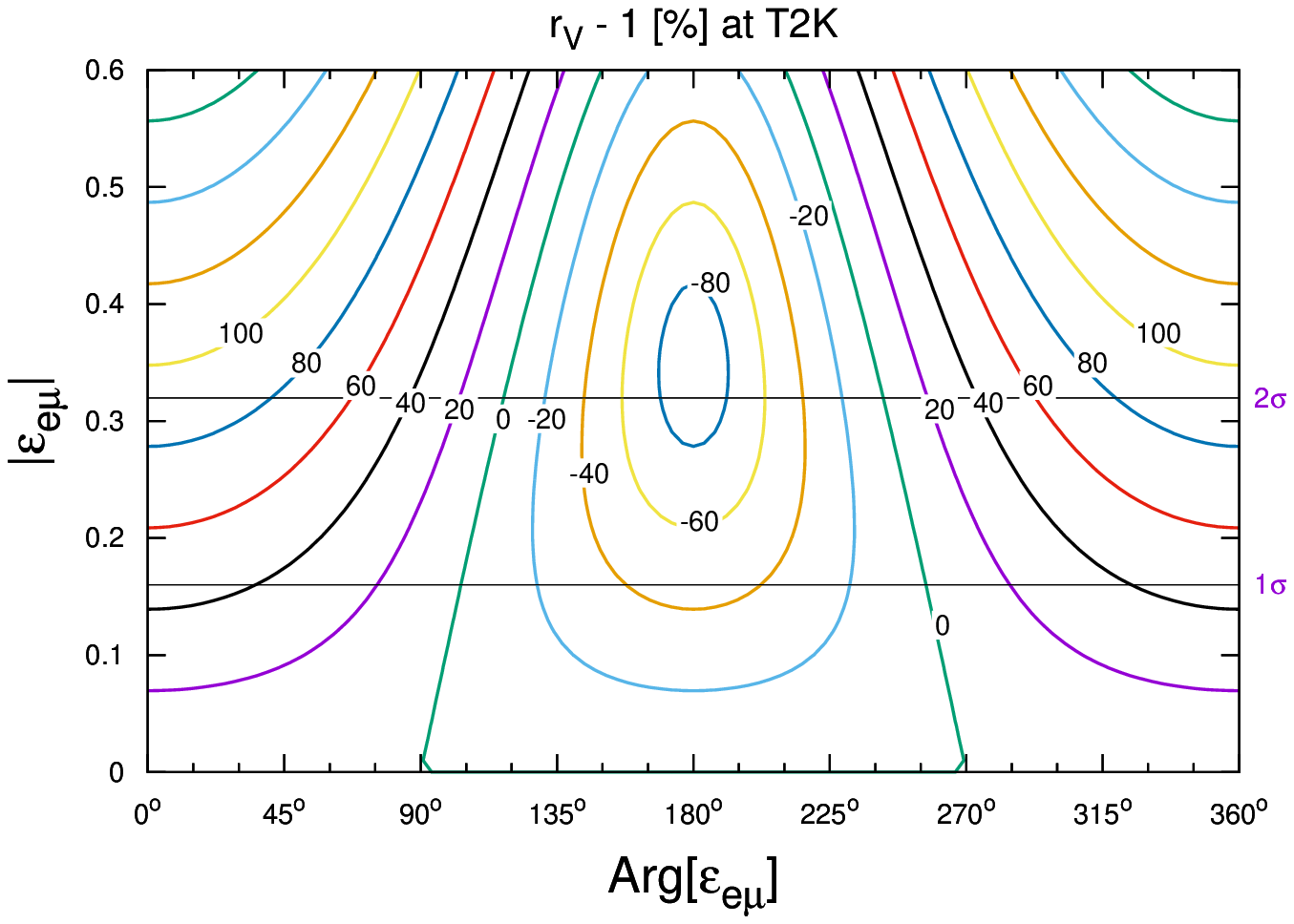}
\includegraphics[width=5cm,height=8cm,angle=-90]{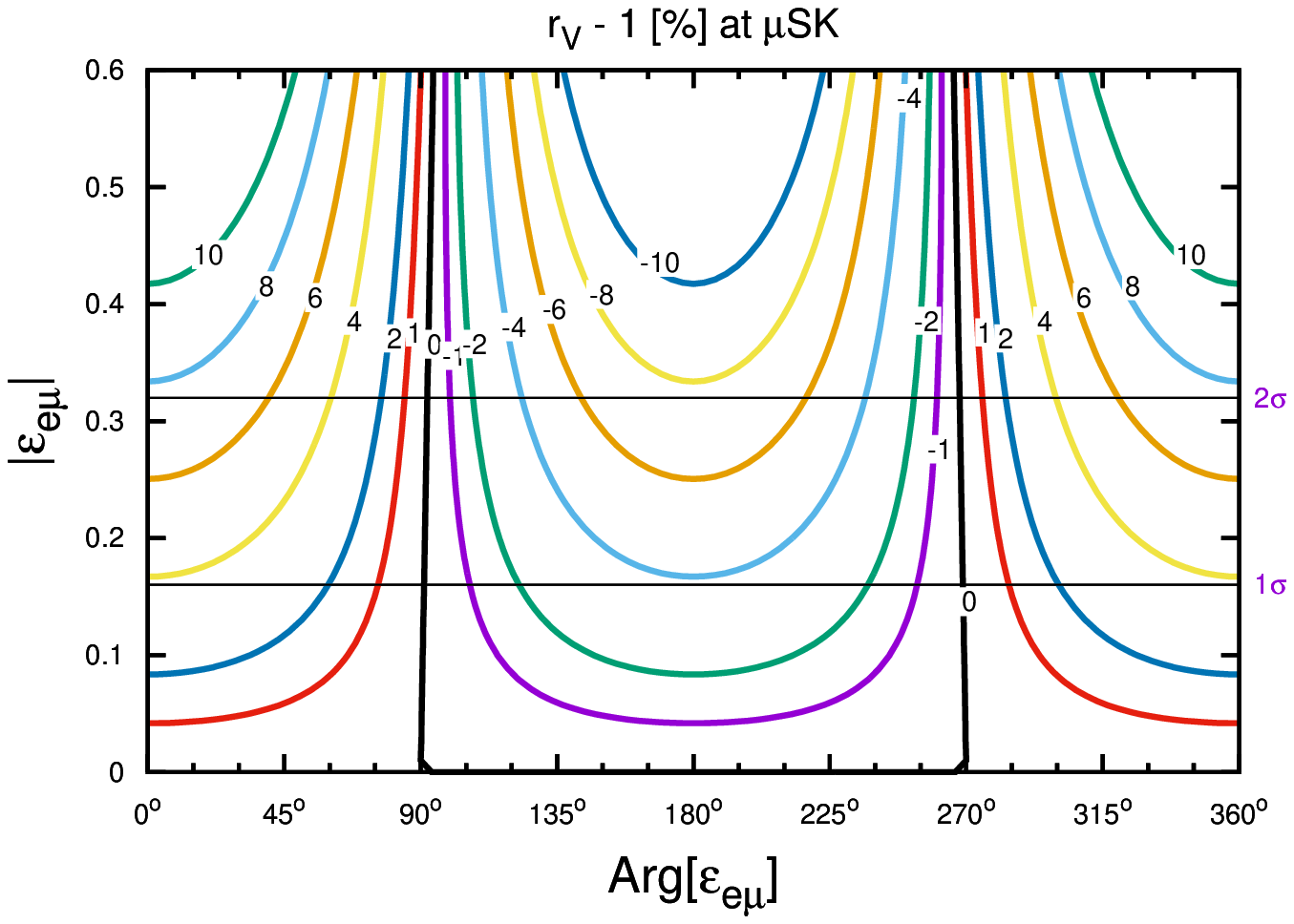}
\includegraphics[width=5cm,height=8cm,angle=-90]{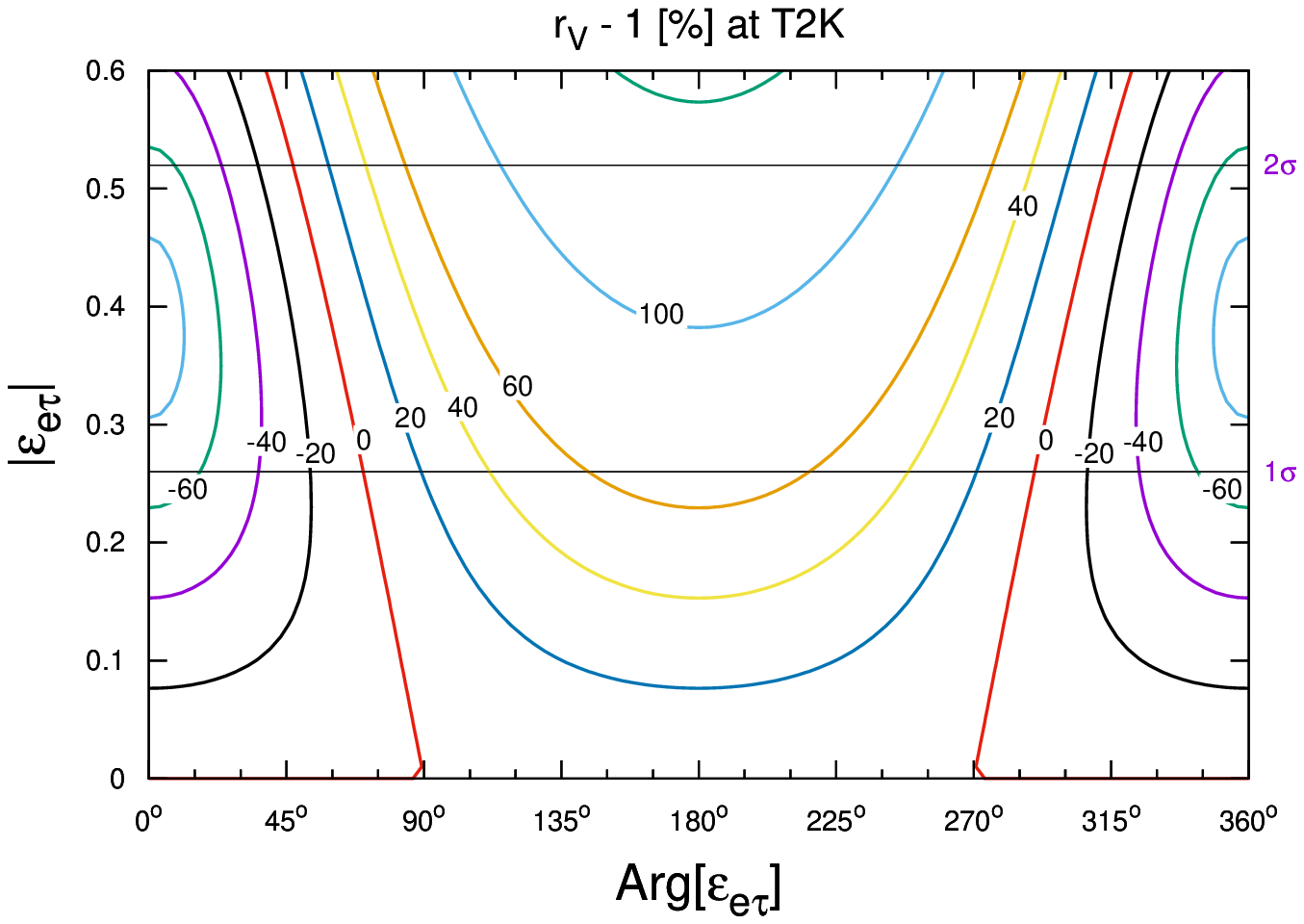}
\includegraphics[width=5cm,height=8cm,angle=-90]{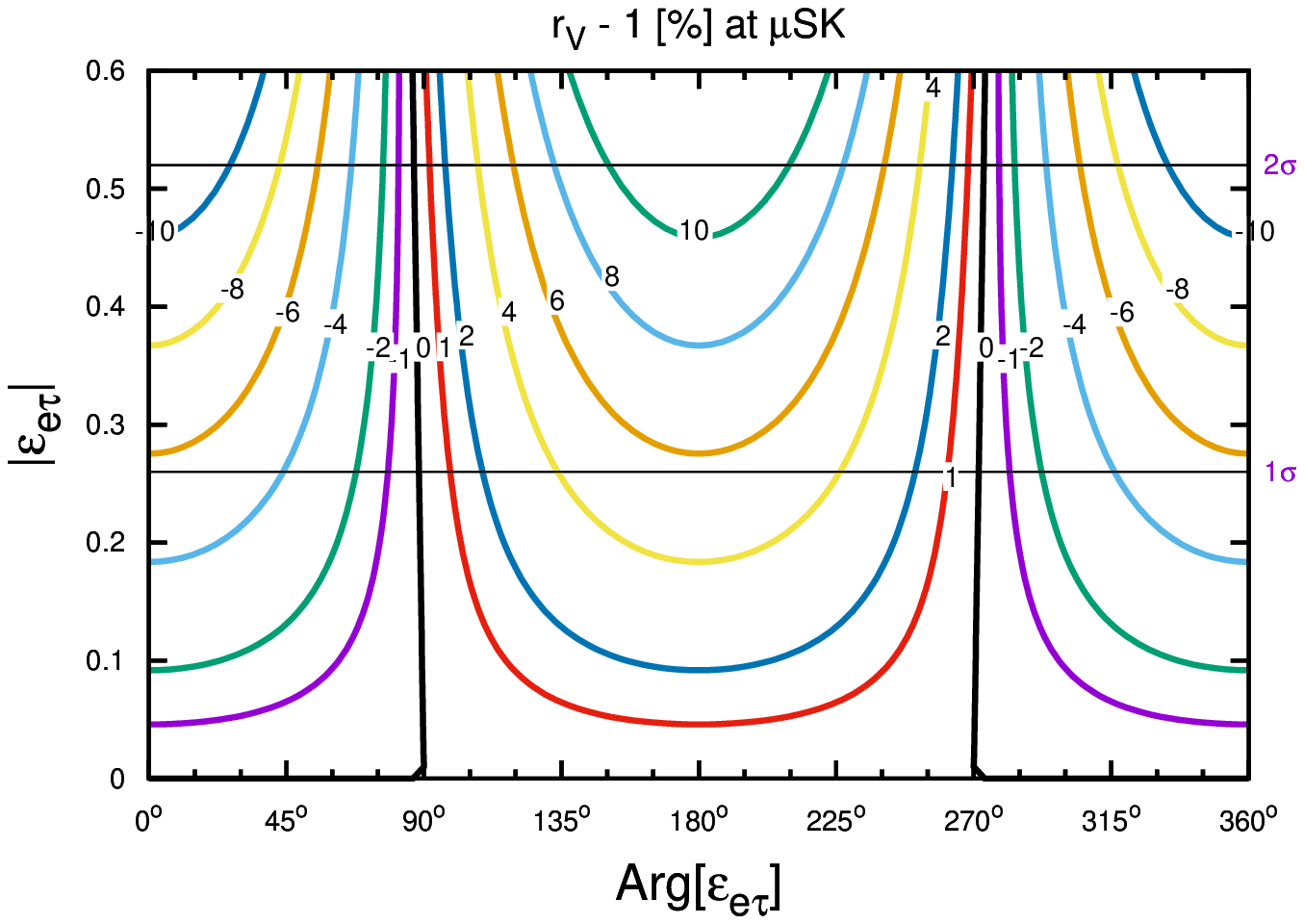}
\includegraphics[width=5cm,height=8cm,angle=-90]{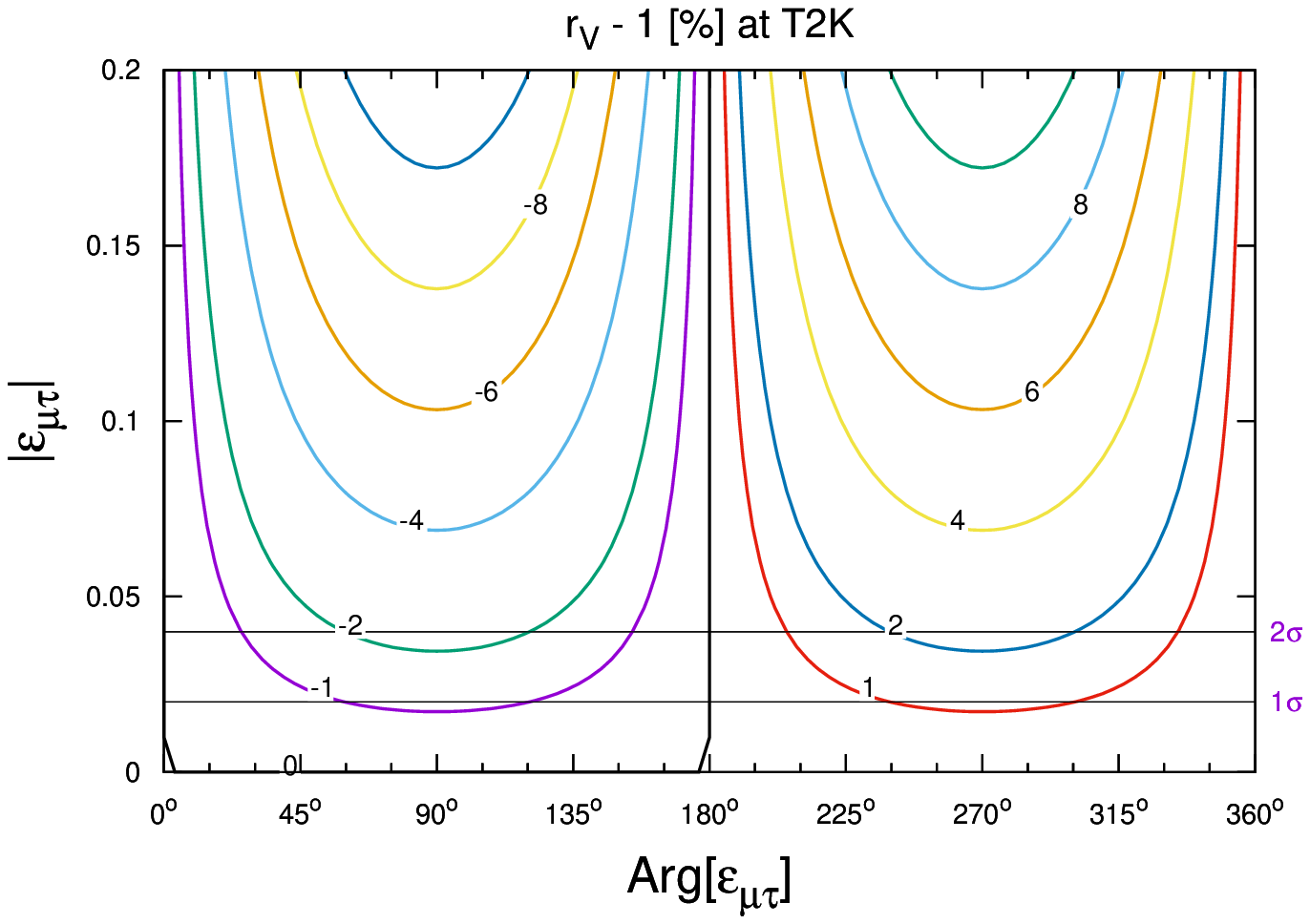}
\includegraphics[width=5cm,height=8cm,angle=-90]{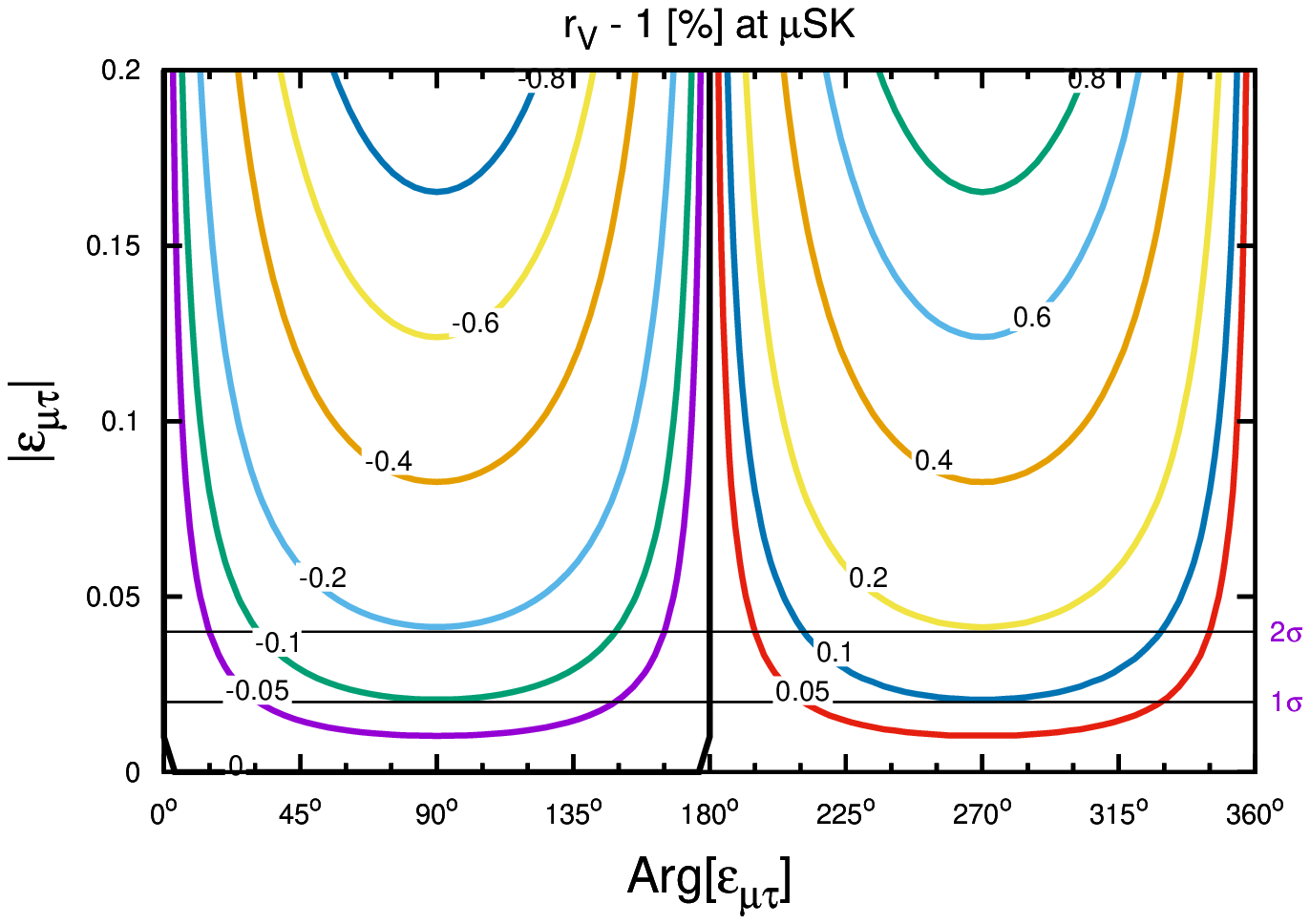}
\caption{Dependence of the deviation from vacuum mimicking,  $r_V - 1$,  
on the absolute values and phases of the NSI parameters:
         (a) $\epsilon_{ee} - \epsilon_{\mu \mu}$ and $\epsilon_{\tau \tau} - \epsilon_{\mu \mu}$;
         (b) $\epsilon_{e \mu}$;
         (c) $\epsilon_{e \tau}$;
         (d) $\epsilon_{\mu \tau}$. 
We use $\delta_D = 3\pi/2$.  
         The left panels are for $E_\nu = 600 \, \mbox{MeV}$  (\tk ) and the right
 panels are  for $E_\nu = 50 \, \mbox{MeV}$ ($\mu$SK). The horizontal lines show 
$1\sigma$ and $2\sigma$  upper bounds on the  NSI parameters. }
\label{fig:rVacuum-NSI}
\end{figure}
%%%%%%%%%%%%%%%%%%%%%%%%%%%%%%%%%%%%%%%%%%%%%%%%%%%%%%%%%%%%%%%%%%%%%%%%%%%%%%%%%%%%%

In \gfig{fig:rVacuum-NSI} we show the dependence of $ \rV -1$ on the NSI parameters. According
to \gfig{fig:rVacuum-NSI},
in the $1 \sigma$ allowed interval of $\epsilon_{e \mu}$, the correction is between $ \pm 45\% $ 
at T2K, with maxima at 
${\rm Arg} [\epsilon_{e \mu}] = 0$ and minimum at ${\rm Arg} (\epsilon_{e \mu}) = \pi$. 
At $\mu$DAR, the correction is 12 times smaller: $ \pm 4\%$. 
For $\epsilon_{e \tau}$ the correction at T2K is slightly larger: $ \pm 65\% $  
with maximum at ${\rm Arg} [\epsilon_{e \tau}] = \pi$. 

The correction from $\epsilon_{\mu \tau}$ depends on the phase $\delta_D$. 
For $\delta_D = 3\pi/2$ we have from (\ref{eq:vacuum-NSI})
$$
\xi_{21} \te_{12} =  - \xi_{21} \Sr |\epsilon_{\mu \tau}|  
(-i \cos 2 \Ta \cos \phi_{\mu \tau} +  \sin \phi_{\mu \tau}) \approx 
 \xi_{21} \Sr |\epsilon_{\mu \tau}|  \sin \phi_{\mu \tau}\
$$
which gives a correction between $\pm 1.2 \%$. 
The parameter $(\epsilon_{\tau \tau} - \epsilon_{\mu \mu})$ 
is real but its correction depends on $\dD$.  
For $\dD = 3\pi/2$, we have $r_V = |1  - 
i s_{13} s_{23} c_{23} (\epsilon_{\tau \tau} - \epsilon_{\mu \mu})| $
and consequently $r_V - 1 \approx 
[s_{13} s_{23} c_{23} (\epsilon_{\tau \tau} - \epsilon_{\mu \mu})]^2$
which is negligible being quadratic in $s_{13}$.

%%%%%%%%%%%%%%%%%%%%%%%%%%%%%%%%%%%%%%%%%%%%%%%%%%%%%%%%%%%%%%%%%%%%%%
\begin{figure}[h!]
\centering
\includegraphics[width=5cm,height=8cm,angle=-90]{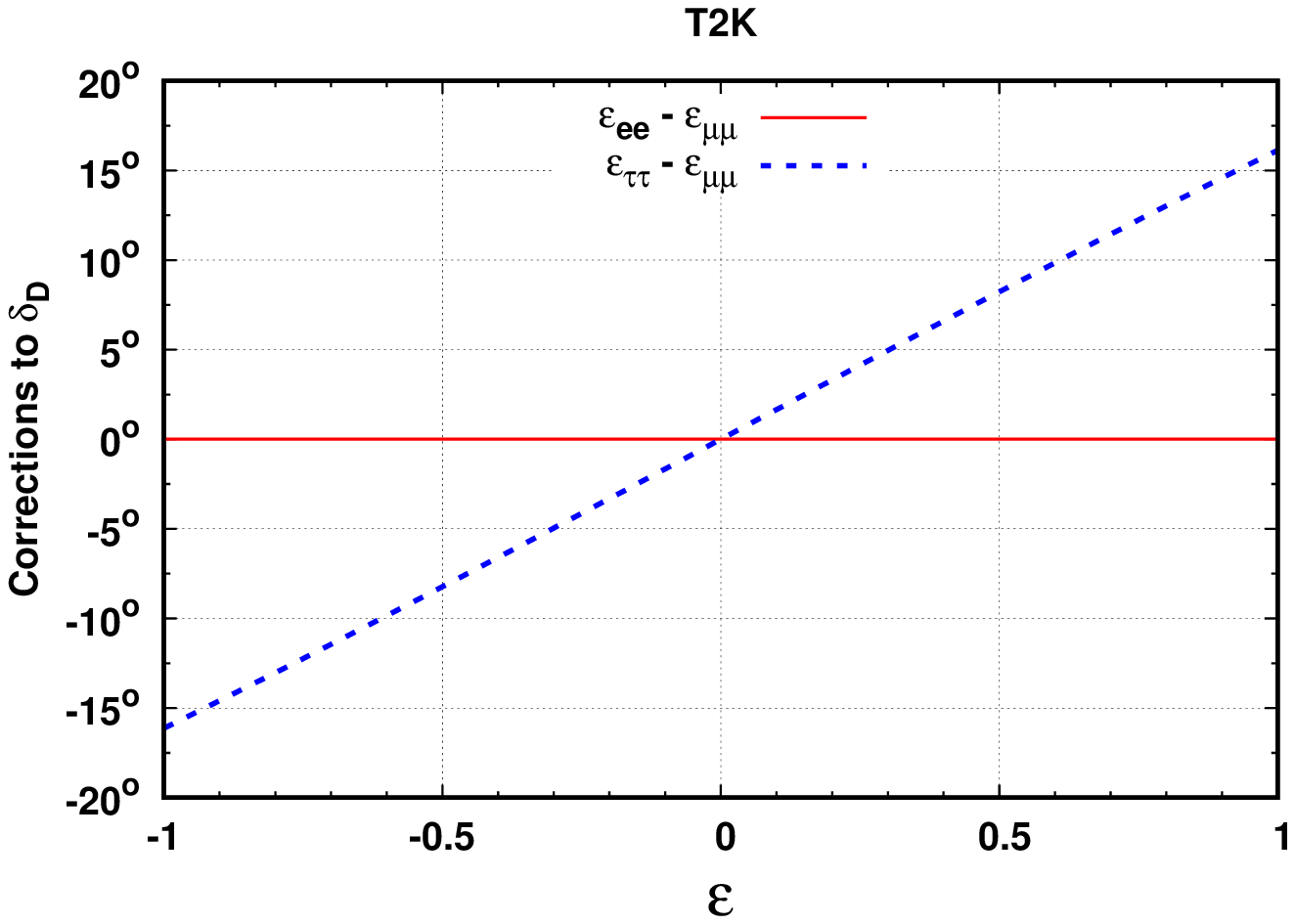}
\includegraphics[width=5cm,height=8cm,angle=-90]{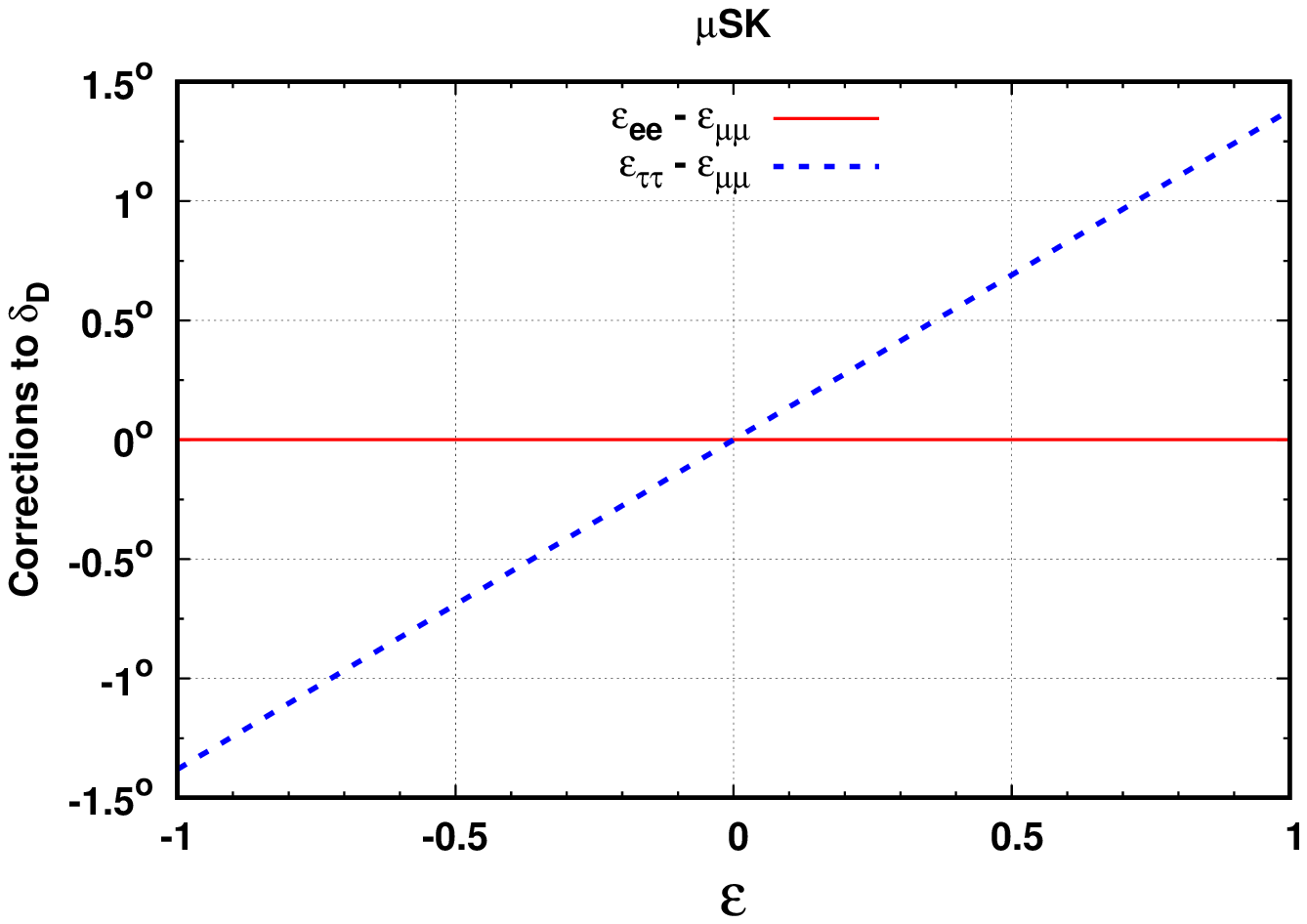}
\includegraphics[width=5cm,height=8cm,angle=-90]{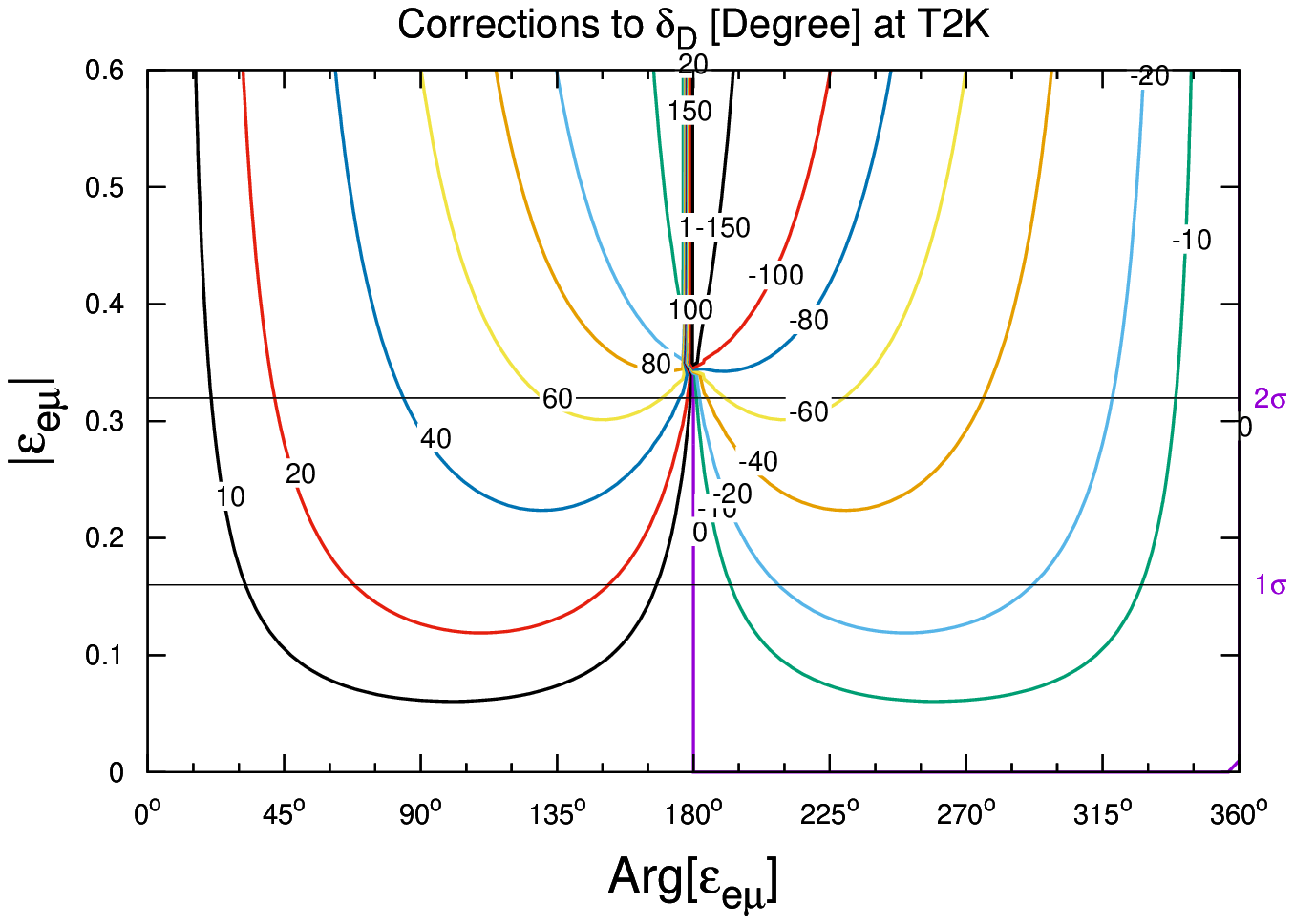}
\includegraphics[width=5cm,height=8cm,angle=-90]{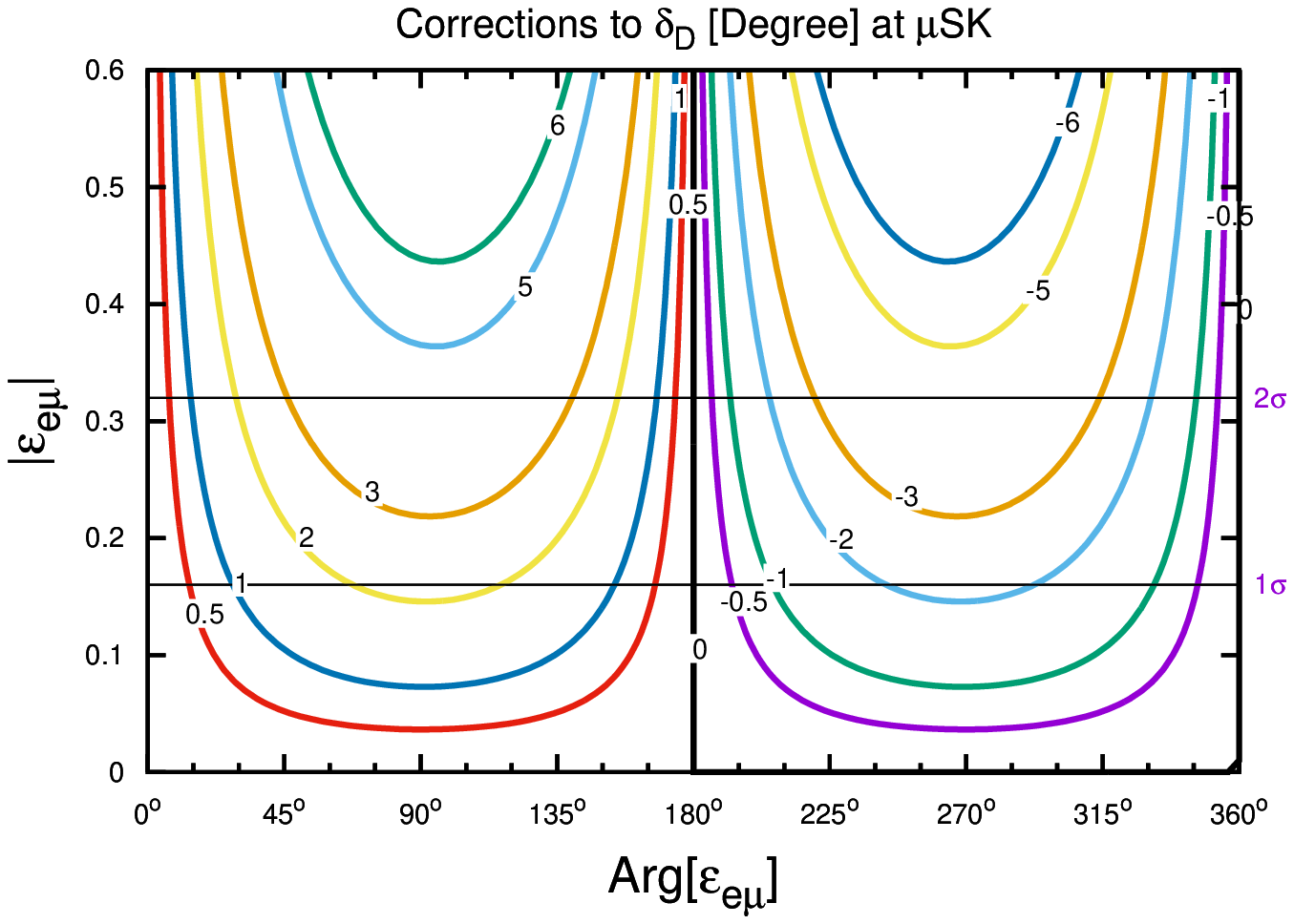}
\includegraphics[width=5cm,height=8cm,angle=-90]{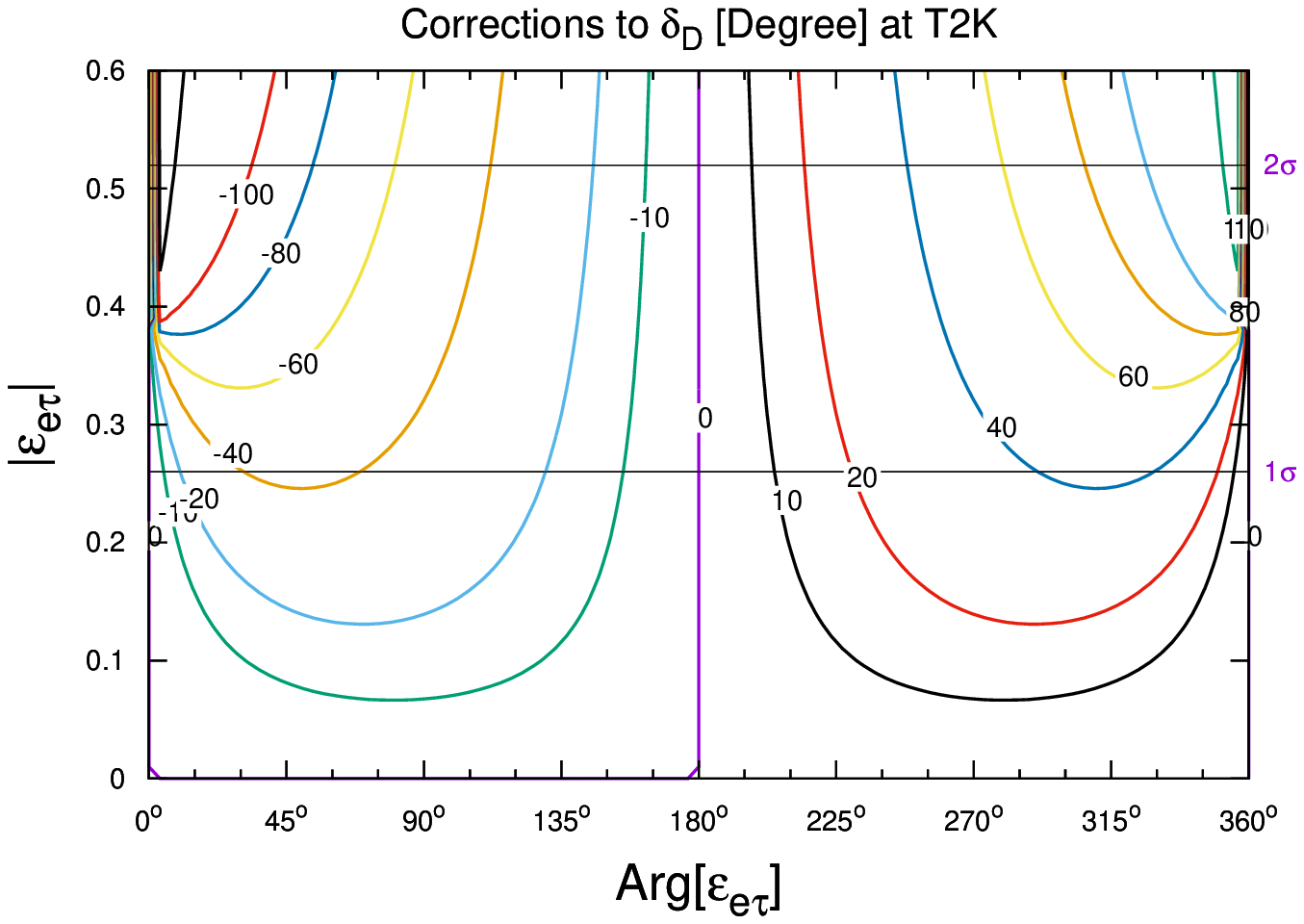}
\includegraphics[width=5cm,height=8cm,angle=-90]{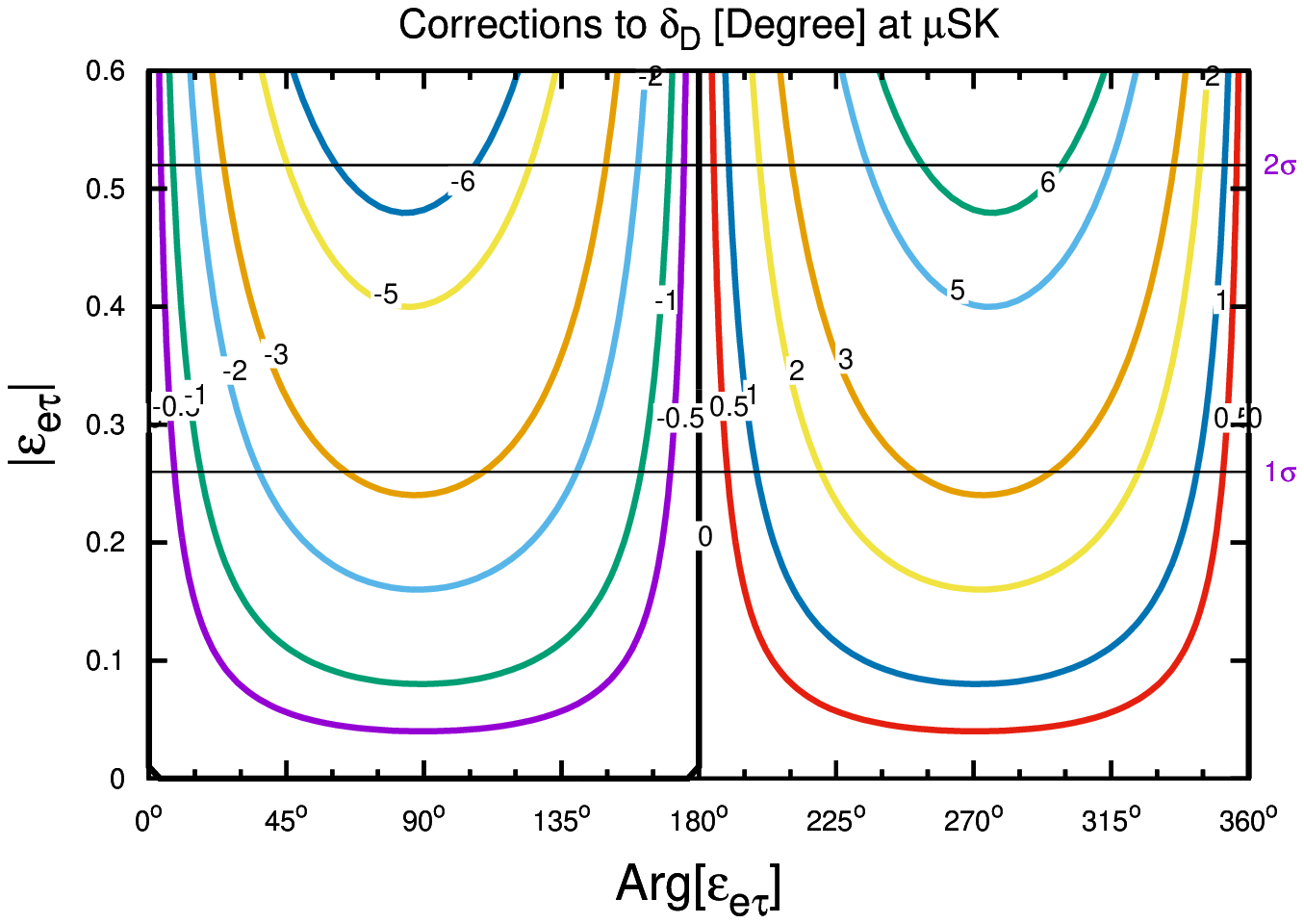}
\includegraphics[width=5cm,height=8cm,angle=-90]{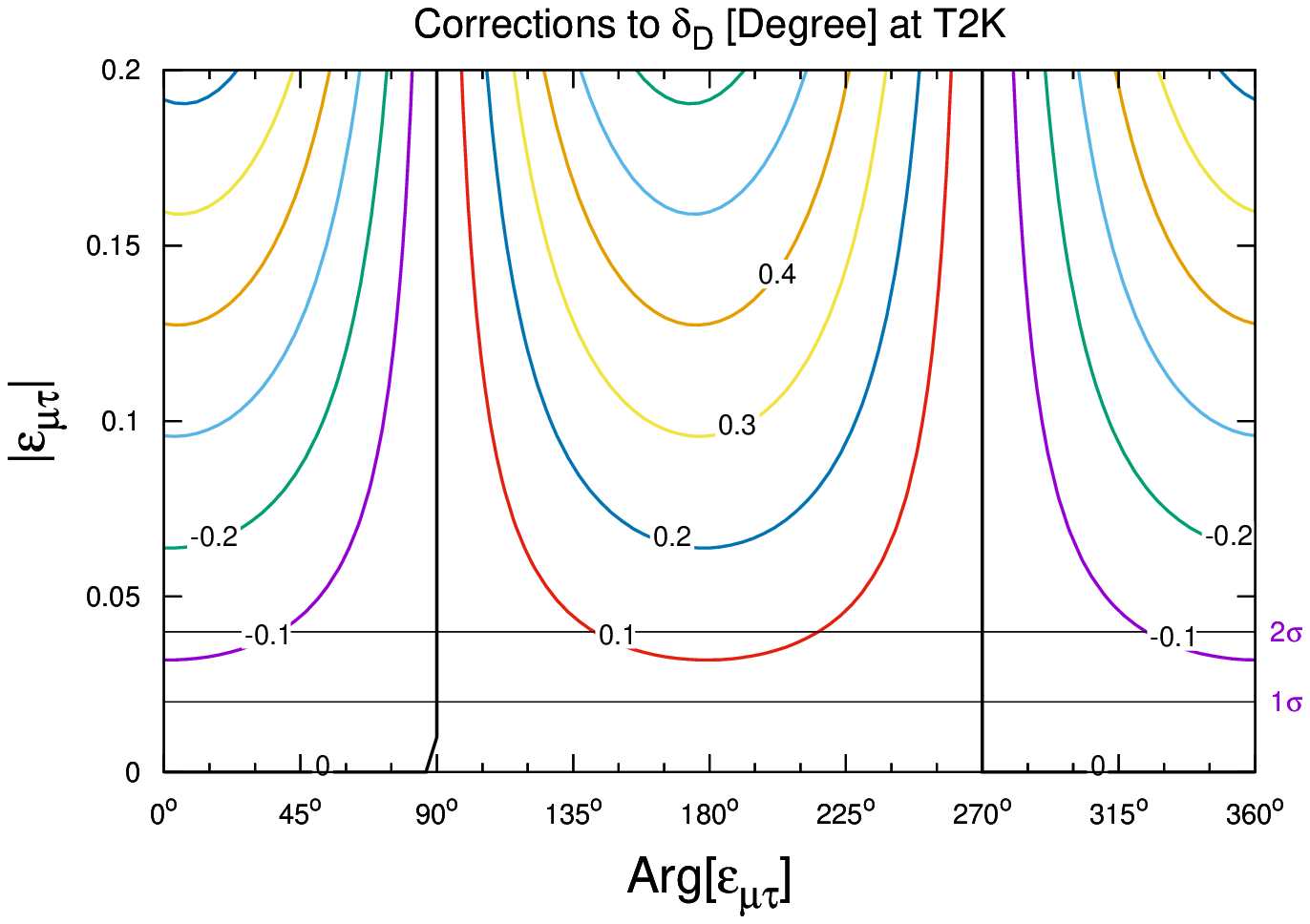}
\includegraphics[width=5cm,height=8cm,angle=-90]{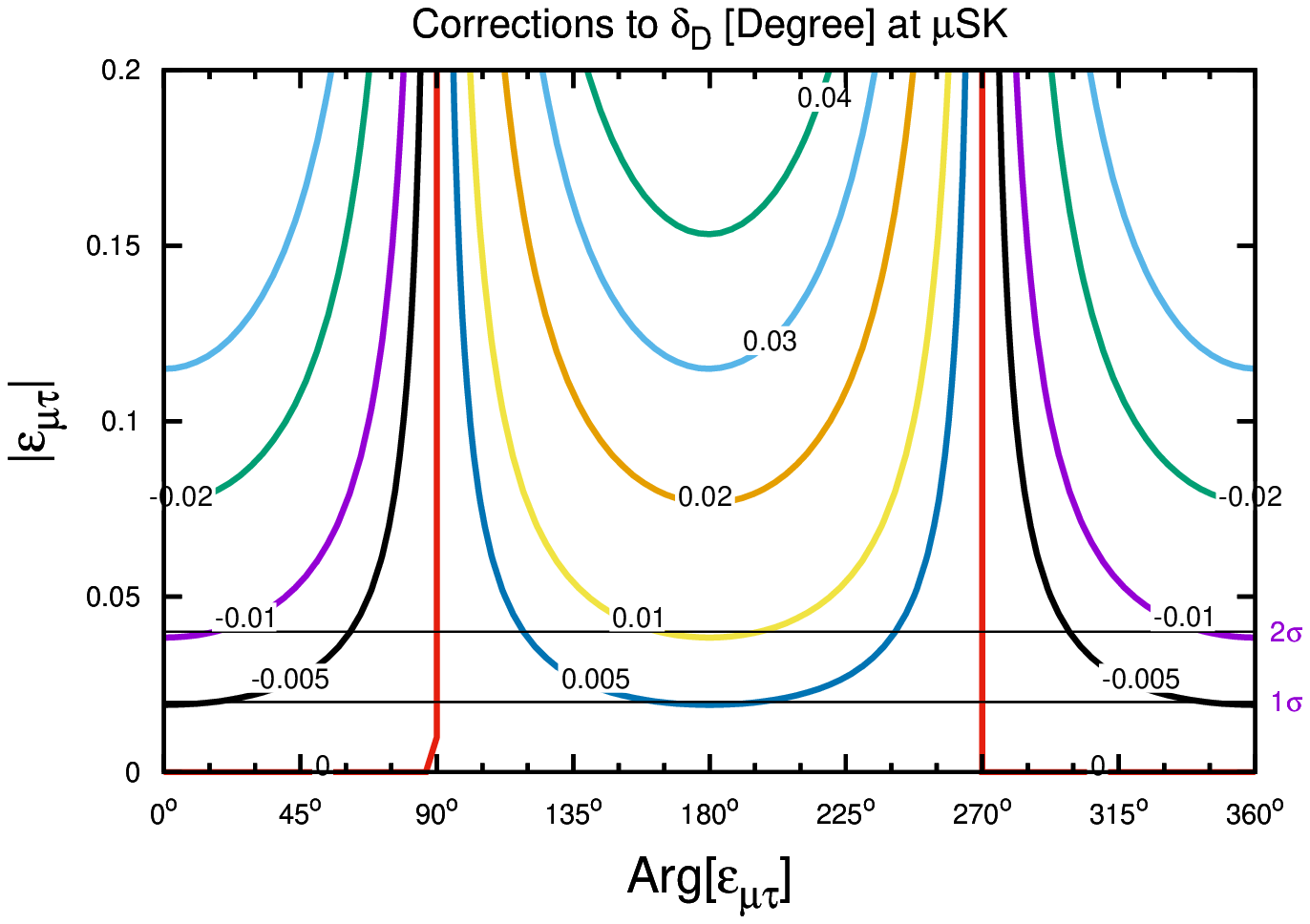}
\caption{The same as in Fig. \ref{fig:rVacuum-NSI} for the matter correction to the leptonic 
Dirac CP phase $\delta_D^m - \delta_D = - \delta_2$.}  
\label{fig:tdCP-NSI}
\end{figure}
%%%%%%%%%%%%%%%%%%%%%%%%%%%%%%%%%%%%%%%%%%%%%%%%%%%%%%%%%%%%%%%%%%%%%%%%%%%

The \gfig{fig:tdCP-NSI} shows the dependence of the correction
to the CP phase $- \delta_2 \approx  \dD^m -  \dD$ on the NSI parameters.
The largest correction comes from $\epsilon_{e \mu}$ 
and $\epsilon_{e \tau}$. With the $1 \sigma$ interval of
$\epsilon_{e \mu}$ we obtain $\delta_2 \approx \pm 30^{\circ}$ at T2K
and $\delta_2 = \pm 2.2^{\circ}$ at $\mu$DAR.
Similar numbers are obtained for $\epsilon_{e \tau}$: $\delta_2 \sim  \pm 40^{\circ}$ and 
$\delta_2 = \pm 3^{\circ}$, respectively.
Contrary to the situation for $\rV $, the parameter 
$\epsilon_{\tau \tau} - \epsilon_{\mu \mu}$ has larger effect on $\delta_D^m - \dD$
than $\epsilon_{\mu \tau}$ does.

The attraction points in the
$\epsilon_{e \mu}$ and   $\epsilon_{e \tau}$ planes  at 
$|\epsilon_{e \mu}| \approx -0.35$, $\phi_{e \mu} = \pi$ 
and $\epsilon_{e \tau} \approx 0.4$, $\phi_{e \tau} = 0, 2\pi$ correspond to 
$\rV = 0$ for $E_\nu = 600\,\mbox{MeV}$,
according to (\ref{eq:te-zero}).
(For $\mu$DAR that would appear at much larger, already excluded,  values of NSI parameters.)
Around the attraction points the correction $\delta_2$ to the CP phase $\delta_D$ can take any value.

From the observational point of view the deviation $r_V - 1$ produces an additional 
degeneracy  with $\delta_D$. 
The factor $r_V$ in the interference term can be absorbed 
in redefinition of $\delta_D$. Indeed, the interference term is proportional to 
$r_V \cos(\delta_D + \phi_{32}^m)$ and  its   
variation   
\begin{equation}
(r_V - 1) \cos(\delta_D + \phi_{32}^m)   
- \Delta \delta_D r_V \sin(\delta_D + \phi_{32}^m)
\end{equation}
vanishes when 
\begin{equation}
\Delta \delta_D = \cot(\delta_D + \phi_{32}^m) 
~\frac{r_V - 1}{r_V}. 
\end{equation}
At the first oscillation maximum,  $\phi_{32}^m \approx \pi/2$,  we obtain 
\begin{equation}
\Delta \delta_D = - \tan \delta_D ~\frac{r_V - 1}{r_V}. 
\end{equation}
Near the maximal CP violation $\delta_D =  3\pi/2$ the correction can be strongly enhanced.

%%%%%%%%%%%%%%%%%%%%%%%%%%%%%%%%%%%%%%%%%%%%%%%%%%%%%%%%%%%%%%%%%
\subsection{Corrections to the total oscillation probability}
\label{sec:Pme}
%%%%%%%%%%%%%%%%%%%%%%%%%%%%%%%%%%%%%%%%%%%%%%%%%%%%%%%%%%%%%%%%%%%%

The complete form of the oscillation probability $P_{\mu e}$ is given by the  expression  
(\ref{eq:Pme-SI}) with oscillation parameters in matter determined in
\gsec{sec:NSI-param}.
According to (\ref{eq:Pme-SI}) the following quantities are modified in matter: 
$r_V$, $\delta_D^m \approx  \delta_D - \delta_2$ 
which are the parameters of the 1-2 sector, 
and  $\theta^m_{13}$, $\phi_{31}^m$, $\phi_{32}^m$ - the parameters of the 1-3 sector. 
We will neglect the corrections due to $\kappa$ 
as well as due to the matter effect on the 2-3 mixing which is of the order $\lambda^4$.
In what follows we will consider the NSI corrections due to the modification of these
oscillation parameters in order, comparing the 
oscillation probabilities with ($P_{\mu e}^{NSI}$) and without ($P_{\mu e}$) the NSI effects: 
\begin{equation}
\frac{\delta P_{\mu e}}{P_{\mu e}} \equiv 
\frac{P_{\mu e}^{NSI} - P_{\mu e}}{P_{\mu e}} .  
\end{equation}

1. The violation of vacuum mimicking gives two contributions: 
from $r_V - 1$ and $\delta_2$, both determined by $\te_{12} \sim \lambda$. 
They affect the interference term $P^I_{\mu e}$. The violation parameter $r_V$ also modifies the 
``solar'' probability $P^S_{\mu e}$. However, since $P_{\mu e}^{S}$ itself is very small,
the correction to the total probability via $P_{\mu e}^{S}$ can be neglected. 

The correction due to  $r_V - 1$ is given by 
\begin{equation}
\frac{\delta P_{\mu e}^I}{P_{\mu e}^{I}} = 
r_V - 1. 
\end{equation}
Since $P_{\mu e}^I \approx (0.25 - 0.30) P_{\mu e}$, the correction can be estimated as 
$(0.25 - 0.30) (r_V - 1)$. According to the computations in \gsec{sec:leading}, this can
lead up to $(10 - 15)\%$ correction for $\epsilon_{e \mu}$ and $(15 - 22)\%$ for
$\epsilon_{e \tau}$ at T2K in the $1\sigma$ allowed intervals. In comparison, the
corrections are about 12 times
smaller at $\mu$SK. They strongly depend on the phases of $\epsilon_{\alpha \beta}$ and the
phase $\delta_D$.

The correction due to $\delta_2$ is given by 
\begin{equation}
  \frac{\delta P_{\mu e}^I}{P_{\mu e}^{I}}
=
  \delta_2 \tan (\delta_D - \delta_2  + \phi_{32}^m ) 
\end{equation}
with $\delta_2$ defined in (\ref{eq:delta2}) being roughly proportional
$\mathbb I(\te_{12})$. Then the  correction to the total probability 
is suppressed by a factor of $\tan (\delta_D - \delta_2 + \phi_{32}^m ) \approx 1/4 \sim 1/3$.
As a result, the contribution from $\delta_2$ 
is somewhat smaller than that from $(r_V - 1)$, reaching $(5 - 10) \%$ at T2K.

2. The correction due to the NSI matter effect on the 1-3 mixing. 
Using Eqs. (\ref{eq:delta13mix}) and (\ref{eq:add13ns}),
we can write the correction to the mixing angle in matter $\theta_{13}^{m, SI}$ with
standard interaction as  
\begin{equation} 
\delta \theta_{13}^m = \theta_{13}^m - \theta_{13}^{m, SI} = 
x_{31} \left[\cos \delta_3 |\Sr \Cr + \te_{13}| - \Sr \Cr \right] .
\label{eq:delta13m}
\end{equation}

Thus, $\delta \theta_{13}^m$ depends on $(\te_{12}, \te_{13}) \sim \lambda$ ({\it i.e.},  
on different combination of the NSI parameters). According to (\ref{eq:delta13m}),
$\delta \theta_{13}^m \sim B s_{13} r_\Delta$ with
$B \sim  4 - 6$ at T2K and $B \approx 1$ at $\mu$SK.

The corrections to the 1-3 mixing can contribute to the ``atmospheric'' probability as 
\begin{equation}
\frac{\delta P_{\mu e}^A}{P_{\mu e}^{A}} \approx 
4 \cot 2 \theta_{13} \delta \theta_{13}^m \sim B r_\Delta . 
\label{eq:13correction}
\end{equation}
For the interference term the relative correction is 2 times smaller and the
contribution to the total probability is further suppressed by a factor $1/4 \sim 1/3$.
The total correction to $P_{\mu e}$ due to $\delta \theta_{13}^m$ can be as large as $(10 \sim 15)\%$. 

3. The correction via the 1-3 oscillation phase. The modification of the 3-1 phase due to NSI is
\begin{equation}
\delta \phi_{13}^m = \phi_{13}^{m, NSI} - \phi_{13}^m = 
(\Delta H_{31}^{NSI} - \Delta H_{31}) \frac{L}{2}.  
\end{equation}
Using \geqn{eq:13splitting} we obtain 
\begin{equation}
\Delta H_{31} - \Delta H_{31}^0 = \frac{1}{2} V (2\te_{33} - \te_{11} - \te_{22}) 
+ \frac{1}{2} [\Delta H_{21} - \Delta H_{21}^0],  
\label{eq:13-splsum}
\end{equation}
where $\Delta H_{21}$ is defined in (\ref{eq:delta2}), 
$\Delta H_{31}^0$,   $\Delta H_{21}^0$ are splittings between the eigenvalues 
without NSI ($\epsilon = 0$).
In turn, being in general of the order $V \epsilon$,
see (\ref{eq:delta2}),
$\Delta H_{21} - \Delta H_{21}^0 \approx 
- V \cos 2\theta_{12} \te_{11}$ below the 1-2 resonance and 
$\Delta H_{21} - \Delta H_{21}^0 \rightarrow V \te_{11}$ above the resonance. 
In the second case there is a cancellation between the last term in \geqn{eq:13-splsum}
with $V \te_{11}$ in the first term there. 
Thus, the phase difference becomes
\begin{equation}
\delta \phi_{13}^m \approx  h \frac{V L}{4} \sim r_\Delta,   
\end{equation}
where the coefficient $h$ is smaller than 1  and can be suppressed at high energies. 
Notice that $\delta \phi_{13}^m$ depends not only on $\te_{11} = O(1)$ but also on 
other diagonal elements  $\te_{22}$, $\te_{11} \sim \lambda^2$. 
So,  $\delta \phi_{13}^m \sim r_\Delta$.

%%%%%%%%%%%%%%%%%%%%%%%%%%%%%%%%%%%%%%%%%%%%%%%%%%%%%%%%%%%%%%%%%%%%%%%%%%%%%%%%%%%%%%%%%%
\begin{figure}[h!]
\centering
\includegraphics[width=5cm,height=8cm,angle=-90]{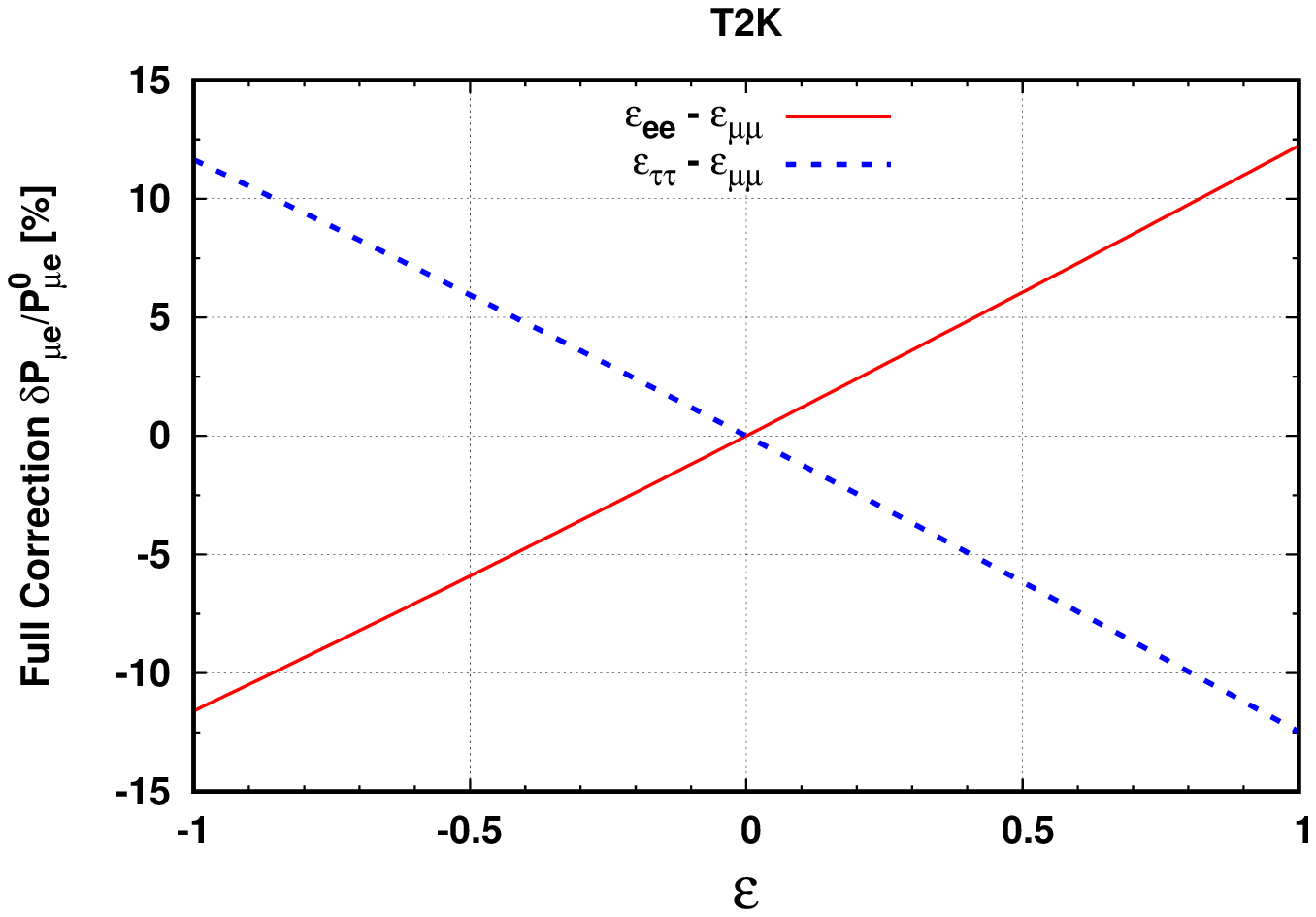}
\includegraphics[width=5cm,height=8cm,angle=-90]{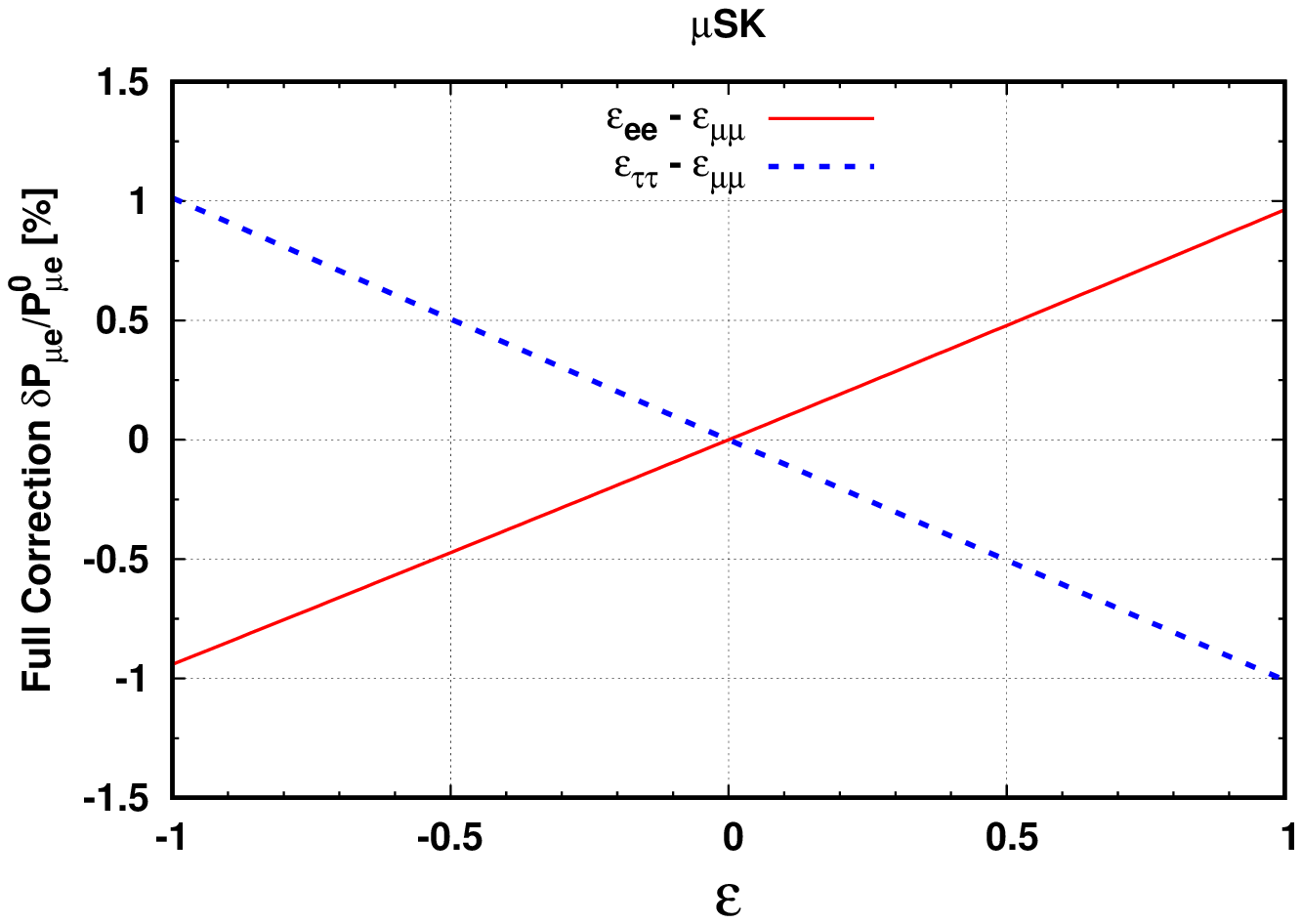}
\includegraphics[width=5cm,height=8cm,angle=-90]{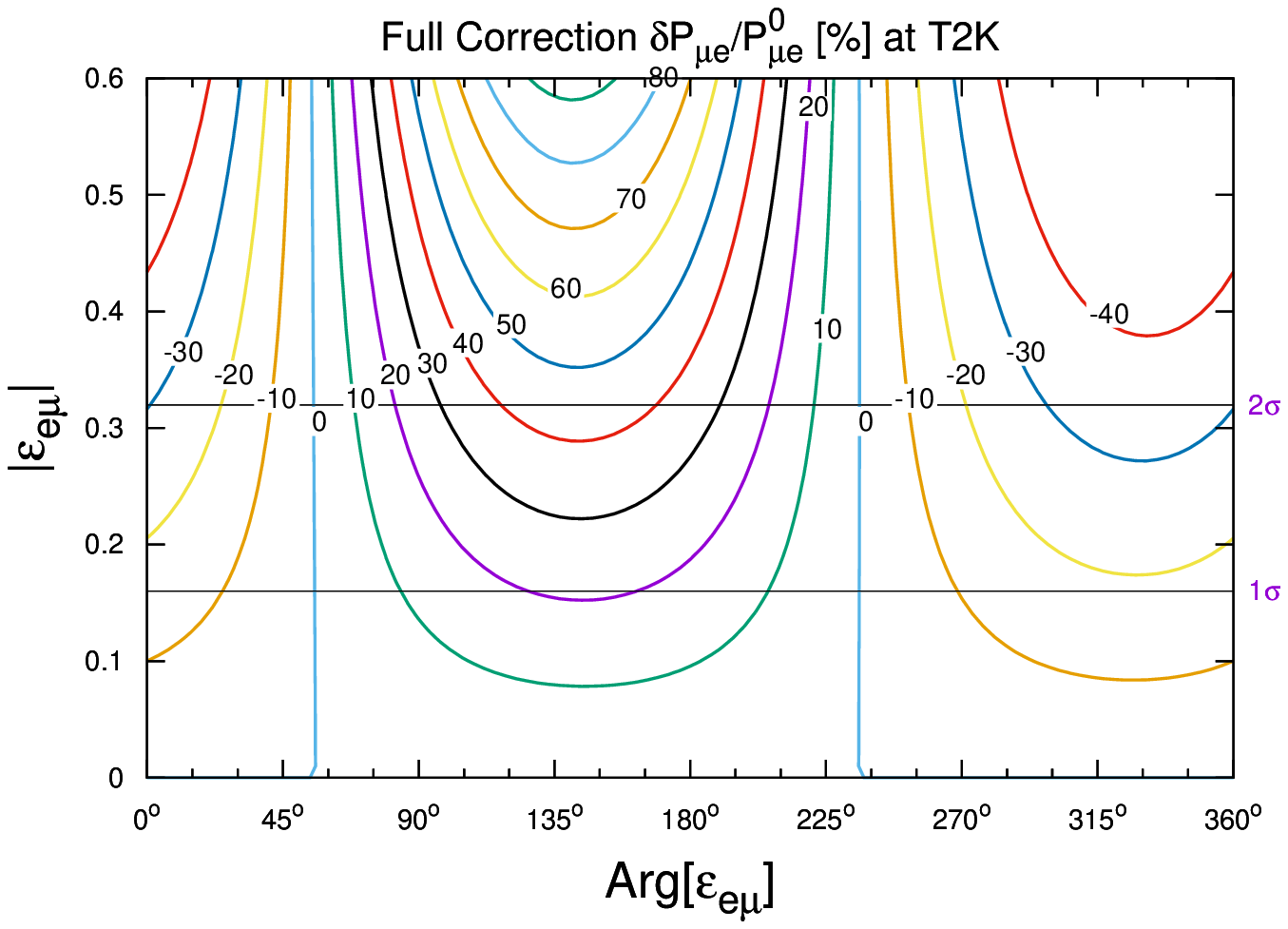}
\includegraphics[width=5cm,height=8cm,angle=-90]{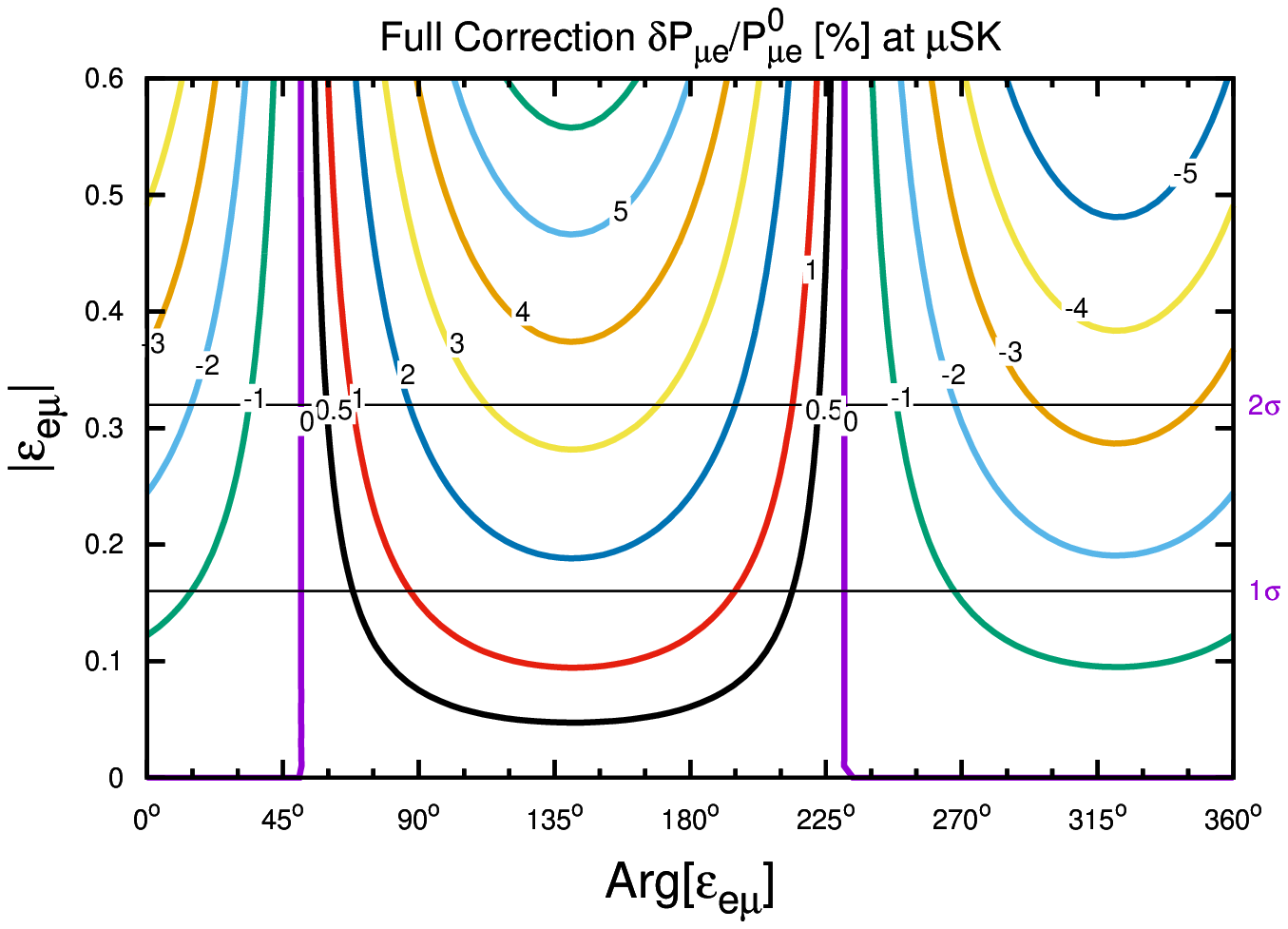}
\includegraphics[width=5cm,height=8cm,angle=-90]{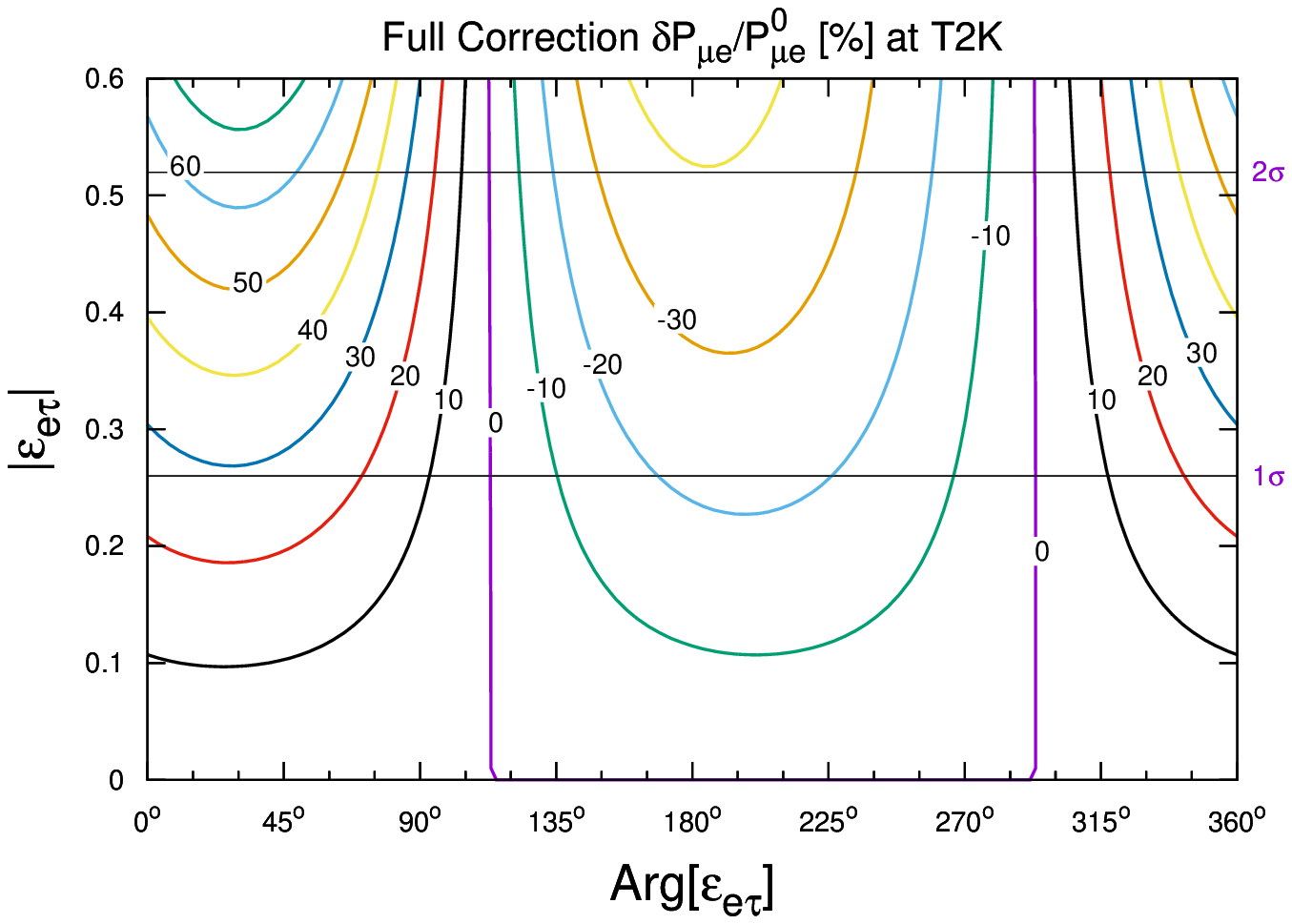}
\includegraphics[width=5cm,height=8cm,angle=-90]{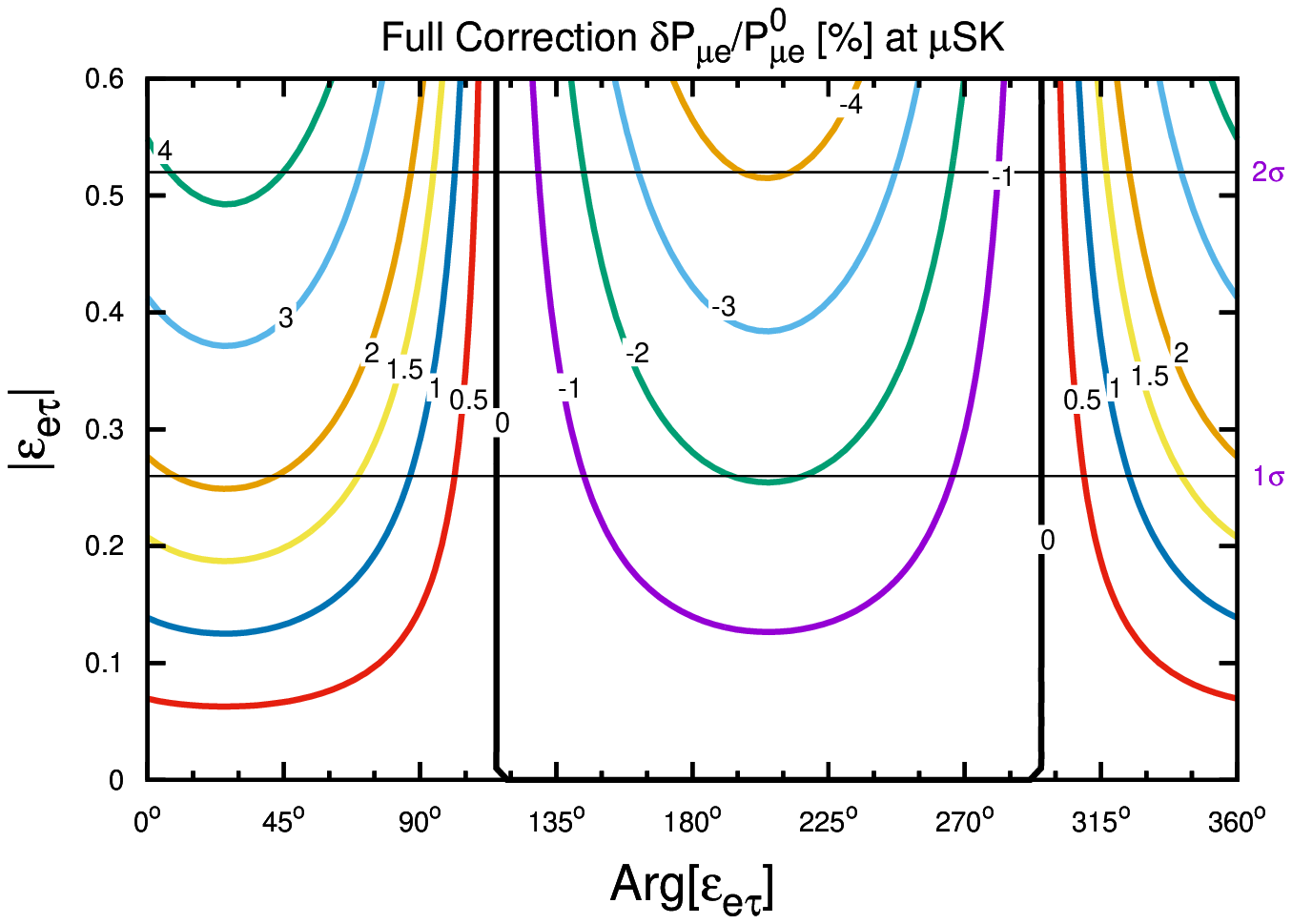}
\includegraphics[width=5cm,height=8cm,angle=-90]{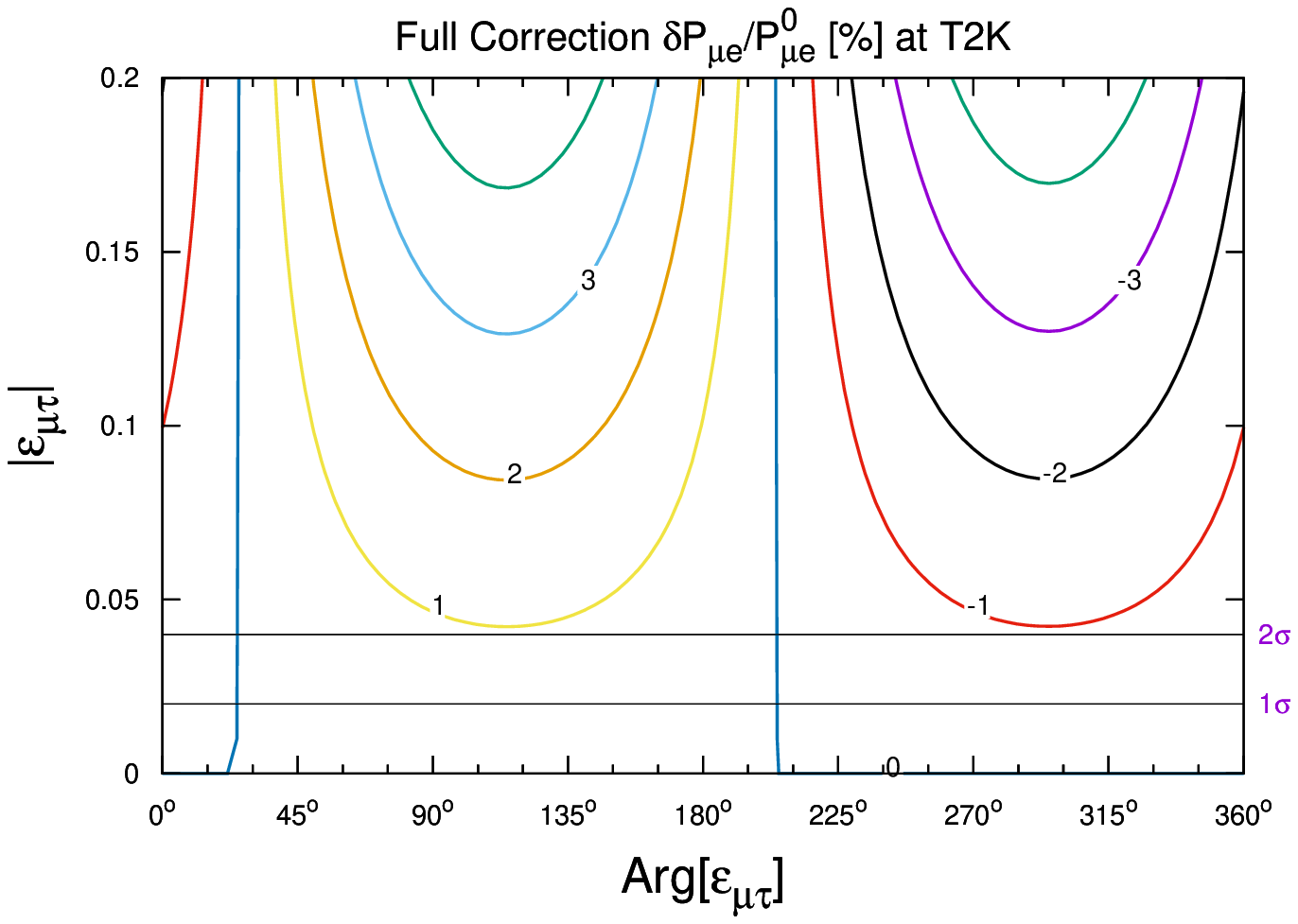}
\includegraphics[width=5cm,height=8cm,angle=-90]{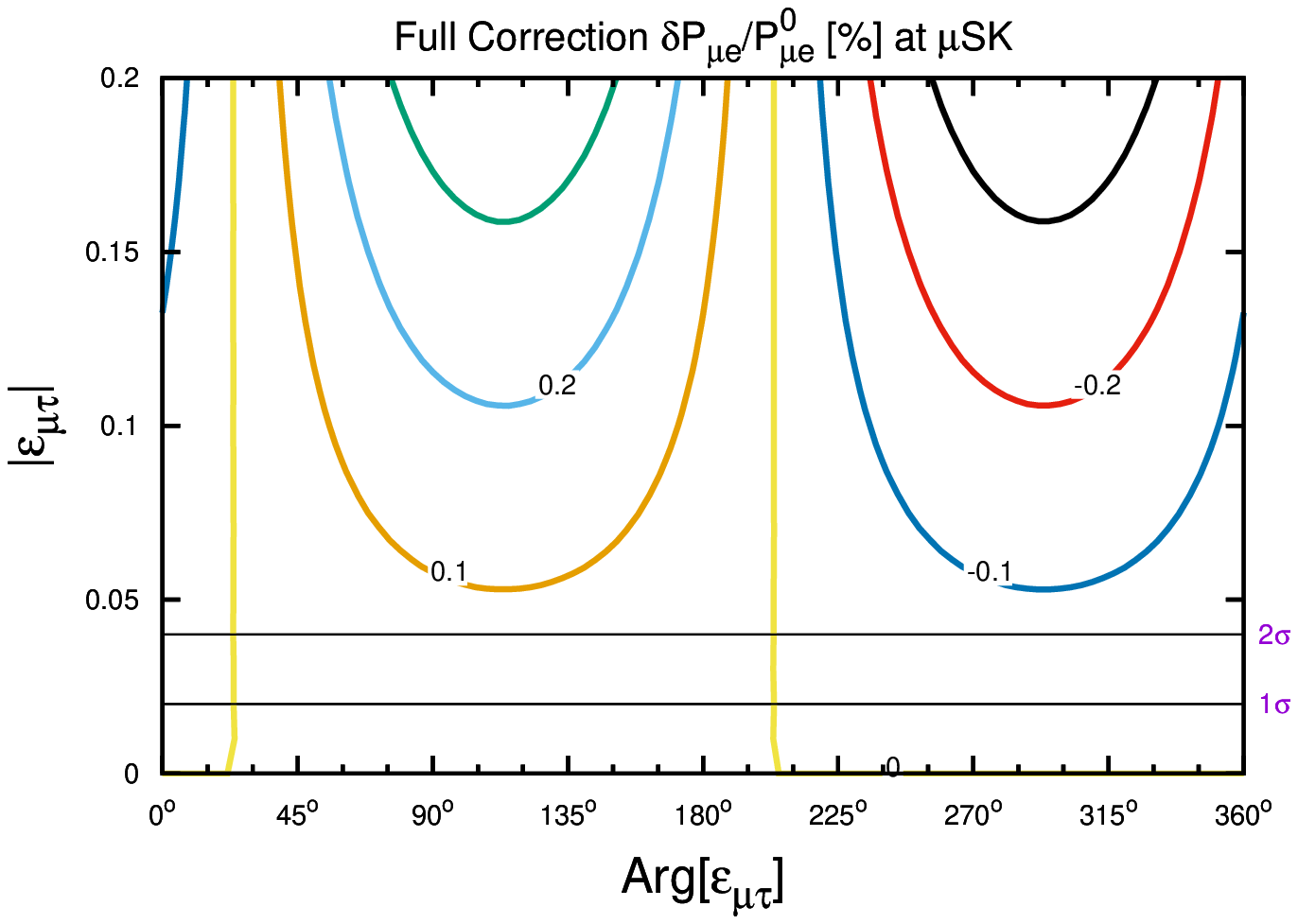}
\caption{The dependence of the corrections to total oscillation probability 
$P_{\mu e}$ on values of  NSI parameters:
         (a) $\epsilon_{ee} - \epsilon_{\mu \mu}$ and $\epsilon_{\tau \tau} - \epsilon_{\mu \mu}$;
         (b) $\epsilon_{e \mu}$;
         (c) $\epsilon_{e \tau}$;
         (d) $\epsilon_{\mu \tau}$ with $\delta_D = -90^\circ$ and $\rho_m = 3 \mbox{g/cm}^3$.
         The left panels are for $E_\nu = 600 \, \mbox{MeV}$ and $L = 295\,\mbox{km}$ 
at \tk while the right panels are
         for $E_\nu = 50 \, \mbox{MeV}$ and $L = 23\,\mbox{km}$ at $\mu$SK.}
\label{fig:fPme-NSI}
\end{figure}
%%%%%%%%%%%%%%%%%%%%%%%%%%%%%%%%%%%%%%%%%%%%%%%%%%%%%%%%%%%%%%%%%%%%%%%%%%%%%%%%%%%%%%%%%%%%%%%%%%%

The correction to the ``atmospheric'' probability equals 
\begin{equation}
\frac{\delta P_{\mu e}^A}{P_{\mu e}^{A}} \approx 
2 \cot \phi_{13} ~\delta \phi_{13}^m.  
\end{equation}
At the first oscillation maximum an additional suppression comes 
from $\cot \phi_{13} \approx 0$.

The analytic consideration presented here allows us to reproduce the results of
numerical computations and analyse their dependence on the NSI parameters.
In general, the violation of vacuum mimicking gives substantial or even dominant contribution to the 
probability change.
The largest  corrections come from $r_V - 1$, $\delta_2$, and $\delta \theta_{13}^m$. 
Their relative contributions depend on the phases of $\epsilon$'s. For 
${\rm Arg}[\epsilon_{e\mu}] \sim 0$ or $\pi$ the violation of vacuum mimicking gives
the main contribution, which is however suppressed at ${\rm Arg}[\epsilon_{e\mu}] \sim \pi/2,~3\pi/2$. 
If the NSI parameters are such that $\te_{12} = 0$, the main contribution to
$P_{\mu e}$ comes from the 1-3 mixing determined by $\te_{13}$. 
And {\it vice versa} if $\te_{13} = 0$, the effect of 1-3 mixing is suppressed and 
contributions from $r_V - 1$ and $\delta_2$ dominate, {\it etc}. 

These features are in agreement with our numerical computations. With all the contributions
combined, the total effect of NSI parameters is shown in \gfig{fig:fPme-NSI}.
Although the NSI parameter $\epsilon_{ee} - \epsilon_{\mu \mu}$ does  not contribute to
the deviation from vacuum mimicking (see \gfig{fig:rVacuum-NSI})  
or to the leptonic Dirac CP phase
(see \gfig{fig:tdCP-NSI}), it affects the total oscillation 
probability $P_{\mu e}$, mainly due to the correction $\delta \theta^m_{13}$ in \geqn{eq:13correction},
which is suppressed by $s_{13}$. 
The main contribution to the corrections comes from  
$\epsilon_{e \mu}$ and $\epsilon_{e \tau}$, reaching $20\%$ and $30\%$, respectively,
for the $1\sigma$ upper bounds.
The contribution from ($\epsilon_{ee} - \epsilon_{\mu \mu}$,
$\epsilon_{\tau \tau} - \epsilon_{\mu \mu}$, $\epsilon_{\mu \tau}$) 
can be as large as $(5 - 10)\%$.

%%%%%%%%%%%%%%%%%%%%%%%%%%%%%%%%%%%%%%%%%%%%%%%%%%%%%%%%%%%%%%%%%%%%
\section{CP Measurement in the Presence of Non-Standard Interactions}
\label{sec:CP}
%%%%%%%%%%%%%%%%%%%%%%%%%%%%%%%%%%%%%%%%%%%%%%%%%%%%%%%%%%%%%%%%%%%%%

In this section 
we use T2K and $\mu$SK as examples to illustrate the NSI effect on the
measurement of the genuine Dirac CP phase $\dD$.
To estimate the CP sensitivity we have simulated
the pseudo-data at T2K and $\mu$DAR for presently favored ``true'' value $\dD = 3\pi/2$.
We take the following values of the neutrino oscillation parameters
\begin{subequations}
\begin{eqnarray}
&&
  \sin^2 2 \theta_{13} = 0.089 \pm 0.005 \,,
\quad
  \sin^2 2 \theta_{12} = 0.857 \pm 0.024 \,,
\quad
  \sin^2 2 \theta_{23} = 0.5 \pm 0.055 \,,
\qquad
\\
&&
  \Delta m^2_{21} = (7.5 \pm 0.2) \times 10^{-5} \mbox{eV}^2 \,,
\quad
  \Delta m^2_{31} = (2.4 \pm 0.1) \times 10^{-3} \mbox{eV}^2 \,,
\end{eqnarray}
\label{eq:prior-osc}
\end{subequations}
to be consistent with the setup in \cite{TNT2K}.
The pseudo-data is simulated in the absence of NSI. 

We use the following $\chi^2$ function 
\begin{equation}
  \chi^2
\equiv
  \chi^2_{stat}
+ \chi^2_{sys}
+ \chi^2_{prior} \, 
\end{equation}
to fit the pseudo-data, with or without NSI. Here $\chi^2_{stat}$  
corresponds to  statistical errors:   $\chi^2_{stat} 
\equiv \sum_i (N^{data}_i - N^{fit}_i)^2/N^{data}_i$ while  
$\chi^2_{sys} = \sum_i [(f_i - 1)/\delta f]^2$
accounts for the systematical uncertainty 
in the neutrino fluxes from the J-PARC and $\mu$DAR neutrino sources.
The J-PARC flux has 5\% uncertainty for both neutrino 
and anti-neutrino modes, also for the
$\mu$DAR flux we take  about ${5\%}$ \footnote{If the near detector $\mu$Near 
\cite{TNT2K-nonunitarity}
is also implemented, the $\mu$DAR flux uncertainty can be significantly reduced.} 
uncertainty, see \cite{TNT2K} for details. Finally, the prior part
$\chi^2_{prior}$ takes into account our current knowledge
of the neutrino oscillation parameters in vacuum \geqn{eq:prior-osc} 
 and the prior constraints \geqn{eq:globalfit} on the NSI parameters.
The later is included only when we fit the pseudo-data with the NSI parameters.

In simulations we used both the  neutrino and antineutrino channels.  
Recall that the standard oscillation probability 
\geqn{eq:Pme-SI} at the first oscillation maximum corresponding to
$\phi_{31}^m = \pi/2$ and $\phi_{21}^m \approx r_\Delta \times \pi/2$
can be approximated as 
\begin{equation}
  P_{\mu e}
\approx
  \Sa^2 \sin^2 2 \Tr^m
\mp \Cr^m \sin 2 \Tr^m \sin 2 \Ta \sin 2 \Ts   \frac{\pi}{2} r_\Delta
\sin \dD \,,
\label{eq:Pme-peak}
\end{equation}
where $\pm$ stand for neutrinos and anti-neutrinos, respectively.
The first term in \geqn{eq:Pme-peak} is larger than the second term by a factor of $\Sr/ r_\Delta \sim 5$.
This magnifies the effect of the uncertainties in the first term.
The largest one appears in the 2-3 mixing, $s_{23}$.   
Around the maximal mixing, $\sin 2 \Ta \approx 1$, the variations of $\Sa^2$ are enhanced 
with respect to the variations of $\sin 2 \Ta$:  
$\Delta(\Sa^2) \approx \Delta(\sin 2 \Ta) \times (\sin 2 \Ta / 2 \cos 2 \Ta)
\gg \Delta(\sin 2 \Ta)$. 
Consequently, a small uncertainty in $\sin 2 \Ta$ propagates to a large uncertainty in $s_{23}$.
It effectively reduces the sensitivity to $\delta_D$ which originates from  
the second term in \geqn{eq:Pme-peak}.
Running T2K in both neutrino and anti-neutrino modes can avoid this problem. With both
$P_{\mu e}$ and $P_{\bar \mu \bar e}$ measured, the difference between them is purely
the CP violating term that is proportional to $\sin \dD$. 

The results in this section are simulated by the NuPro package \cite{NuPro}.

\subsection{The CP sensitivity at T2K in the presence of NSI}
\label{sec:T2K}
%%%%%%%%%%%%%%%%%%%%%%%%%%%%%%%%%%%%%%%%%%%%%%%%%%%%%%%%%%%%%%%%%%%%%%%%%%

In simulations for T2K we use $7.8 \times 10^{21}$ 
proton on target (POT) for the J-PARC flux, corresponding
to 342 events in the neutrino mode or 83 events
in the anti-neutrino mode for $\delta_D = 3\pi/2$
\cite{T2K1409}. 
As discussed above,
to reduce uncertainties in the determination of $\dD$ at T2K, it is necessary to run the
experiment in both neutrino and anti-neutrino modes. However, 
in both production and detection, the cross sections for
anti-neutrinos are smaller than those for neutrinos.
To collect comparable number of events, and hence balance the statistical uncertainty in both modes,
the antineutrino mode should run longer. 
Therefore, in simulations we use 2 years of running in the neutrino mode 
and 4 years in the antineutrino mode. Correspondingly, 114 neutrino events
and 56 antineutrino events are expected.

%%%%%%%%%%%%%%%%%%%%%%%%%%%%%%%%%%%%%%%%%%%%%%%%%%%%%%%%%%%%%%%%%%%%%%%%
\begin{figure}[h!]
\centering
\includegraphics[height=0.67\textwidth,angle=-90]{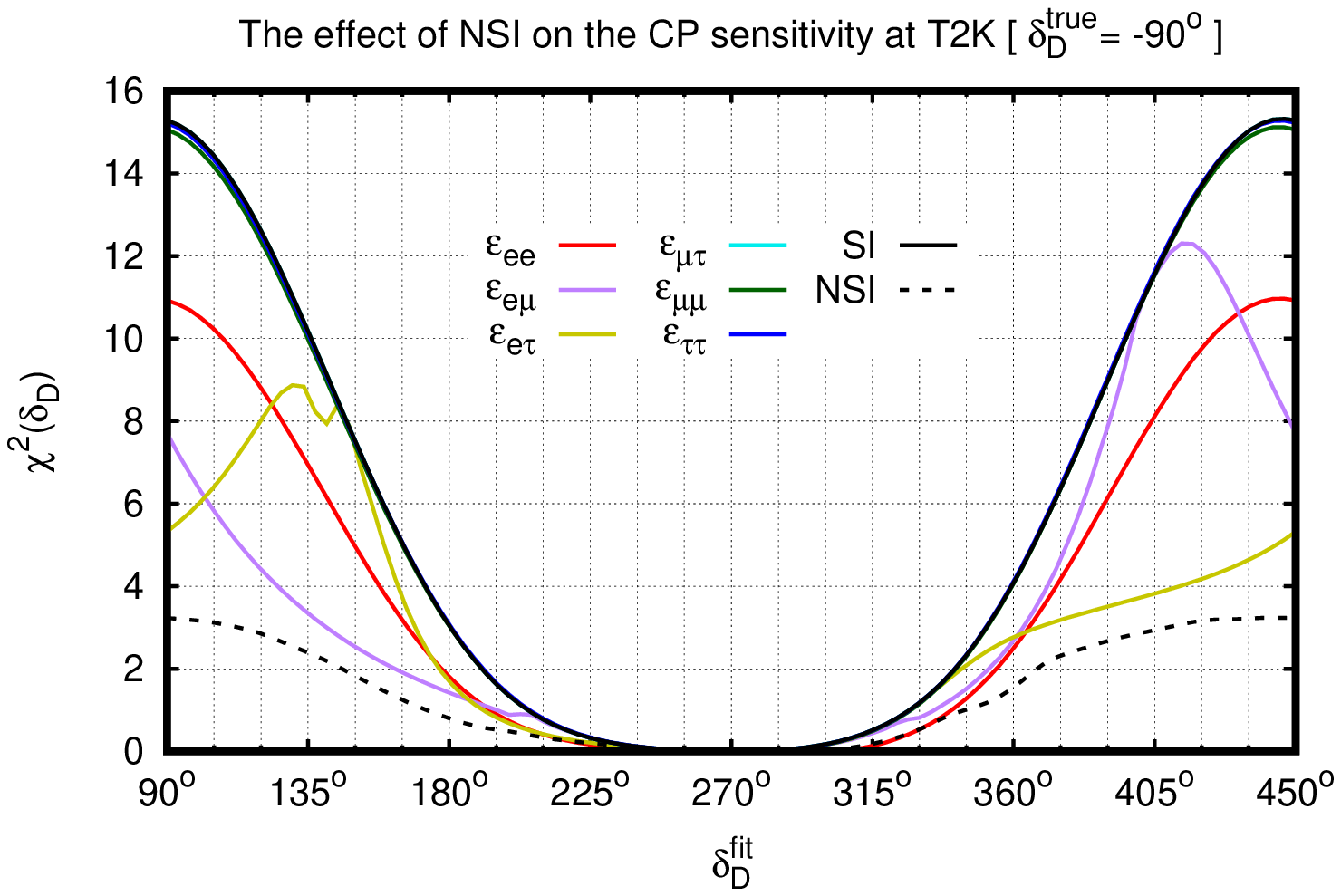}
\caption{The effect of NSI on the CP sensitivity at T2K with 2 years of running 
for the neutrino mode and 4 years for the antineutrino mode,
with the prior constraints (\ref{eq:globalfit}).
The pseudo-data with standard interaction and $\dD^{true} = 3\pi/2$ is
fit with standard interaction (SI) as well as with  individual NSI parameters
($\epsilon_{ee}$, $\epsilon_{e \mu}$, $\epsilon_{e \tau}$, 
$\epsilon_{\mu \tau}$, $\epsilon_{\mu \mu}$,
$\epsilon_{\tau \tau}$) and full set of NSI parameters (dashed line).}
\label{fig:chi2-T2K}
\end{figure}
%%%%%%%%%%%%%%%%%%%%%%%%%%%%%%%%%%%%%%%%%%%%%%%%%%%%%%%%%%%%%%%%%%%%%%%%

We compute the minimal $\chi^2$ value as a function of the fit value of the phase 
$\delta_D^{fit}$, with and without the NSI parameters. The results are shown in \gfig{fig:chi2-T2K}. 
The solid black line in \gfig{fig:chi2-T2K} presents the CP sensitivity at T2K without NSI. 
We can see that this line is nearly symmetric with respect to 
$\delta_D^{fit} = \delta_D^{true} = 3\pi/2$ which  is related to  
the degeneracy between $\dD$ and $\pi - \dD$ due to the
$\sin \delta_D$ dependence in the oscillation probability 
(\ref{eq:Pme-peak}). 
According to \gfig{fig:chi2-T2K}, the true value $\dD^{true} = 3 \pi/2$
can be distinguished from $\dD^{fit} = 0~(\pi)$ 
by $\chi^2 = 3$ ($1.7 \sigma$) and from $\dD^{fit} = \pi/2$ by $\chi^2 = 15$ ($3.8 \sigma$). 
The results can be easily scaled for higher statistics at T2HK or T2K-II.

The sensitivity to $\dD$ can be significantly undermined by NSI.
In \gfig{fig:chi2-T2K} we show the dependence of  
$\chi^2$ on the fit value $\dD^{fit}$ for 
different NSI parameters separately, although only two combinations of the three diagonal 
elements ($\epsilon_{ee}$,
$\epsilon_{\mu \mu}$, and $\epsilon_{\tau \tau}$) affect oscillations,  
as we used in \gsec{sec:matter}.
The color curves in \gfig{fig:chi2-T2K} show that the CP sensitivity at T2K is
significantly reduced by $\epsilon_{ee}$, $\epsilon_{e \mu}$, and $\epsilon_{e \tau}$, but
almost not affected by $\epsilon_{\mu \mu}$, $\epsilon_{\mu \tau}$, $\epsilon_{\tau \tau}$.
This is because $\epsilon_{\tau \tau} - \epsilon_{\mu \mu}$ and
$\epsilon_{\mu \tau}$ are  constrained much stronger according to  \geqn{eq:globalfit} and  also
their contributions  to $\te_{12}$ are suppressed by $s_{13}$, see \geqn{eq:tilde-epsilon-explicit}.
Consequently, they have small contribution to the violation of vacuum mimicking via
$r_V$ \geqn{eq:rV} and to the Dirac CP phase via $\delta_2$ \geqn{eq:delta2}.
Although $\epsilon_{ee} - \epsilon_{\mu \mu}$ cannot contribute
to the leading order corrections via the Dirac CP phase $\delta^m_D$ and $r_V$,
see \geqn{eq:vacuum-NSI}, its effect on the oscillation probability
via $\delta \theta^m_{13}$ \geqn{eq:delta13m} as a function of $\te_{13}$
\geqn{eq:tilde-epsilon-explicit} can still be significant
due to weak prior constraint \geqn{eq:globalfit}. For $\epsilon_{e \mu}$ and $\epsilon_{e \tau}$, the prior constraints are 
stronger but their contributions are not suppressed in $\te_{12}$, see \geqn{eq:tilde-epsilon-explicit}. 

The non-vanishing $\epsilon_{ee}$, $\epsilon_{e \mu}$, $\epsilon_{e \tau}$
reduce the distinguishability between
$\dD^{fit} = 0~(\pi)$ and $\dD^{true} = 3\pi/2$ from $\chi^2 = 3$ to $1.7$ ($1.3\sigma$).
The difference between the two maximal CP values $\dD = \pm \pi/2$ decreases
from $\chi^2 \approx 15$ (roughly $4\sigma$) to $\chi^2 \approx 5.5$ ($2.3 \sigma$). If 
all NSI parameters are present, the CP sensitivity at T2K can be further reduced
as is shown by the dashed black line
in \gfig{fig:chi2-T2K}. The distinguishability of 
$\dD^{fit} = \pi/2$ from $\dD^{true} = 3\pi/2$
is reduced by roughly a factor of 5 to $\chi^2 \approx 3.2$ ($1.8 \sigma$)
and for $\dD^{fit} = 0~(\pi)$ it becomes $\chi^2 \approx 0.8$ ($0.9 \sigma$).
The value of $\chi^2$ for $\delta^{fit}_D = 0~(\pi)$
and the corresponding best-fit values of the NSI parameters are shown in
\gtab{tab:chi2-0dD}.

%%%%%%%%%%%%%%%%%%%%%%%%%%%%%%%%%%%%%%%%%%%%%%%%%%%%%%%%%%%%%%%%%%%%%%%
\subsection{Improving the CP sensitivity with neutrinos from $\mu$DAR}
\label{sec:muDAR}
%%%%%%%%%%%%%%%%%%%%%%%%%%%%%%%%%%%%%%%%%%%%%%%%%%%%%%%%%%%%%%%%%%%%%%%%%

Due to smaller energies, $\sim 50\, \mbox{MeV}$,  as compared with 
the peak energies, $\sim 600\,\mbox{MeV}$, of the J-PARC flux,
neutrinos of the $\mu$DAR flux will experience 12 times smaller 
matter effect than in T2(H)K.
In simulation we use the characteristics proposed for 
$\mu$SK \cite{TNT2K} which includes the existing Super-K detector and 
a $9\,\mbox{mA}$ cyclotron to produce $\bar \nu_\mu$ from $\mu$DAR. The
optimal distance between the $\mu$DAR source and the SK detector is around $23\,\mbox{km}$.
For 6 years of running, a total number  $1.1\times 10^{25}$ POT
can be collected, corresponding to 212  $\bar \nu_\mu \rightarrow \bar \nu_e$
oscillation events in the SK detector for $\dD^{true} = 3\pi/2$.

%%%%%%%%%%%%%%%%%ffff7%%%%%%%%%%%%%%%%%%%%%%%%%%%%%%%%%%%%%%%%%%%%%%%%%%%%%%%%%%%%%%%%%%%%%%%%%
\begin{figure}[h!]
\centering
\includegraphics[height=0.7\textwidth,angle=-90]{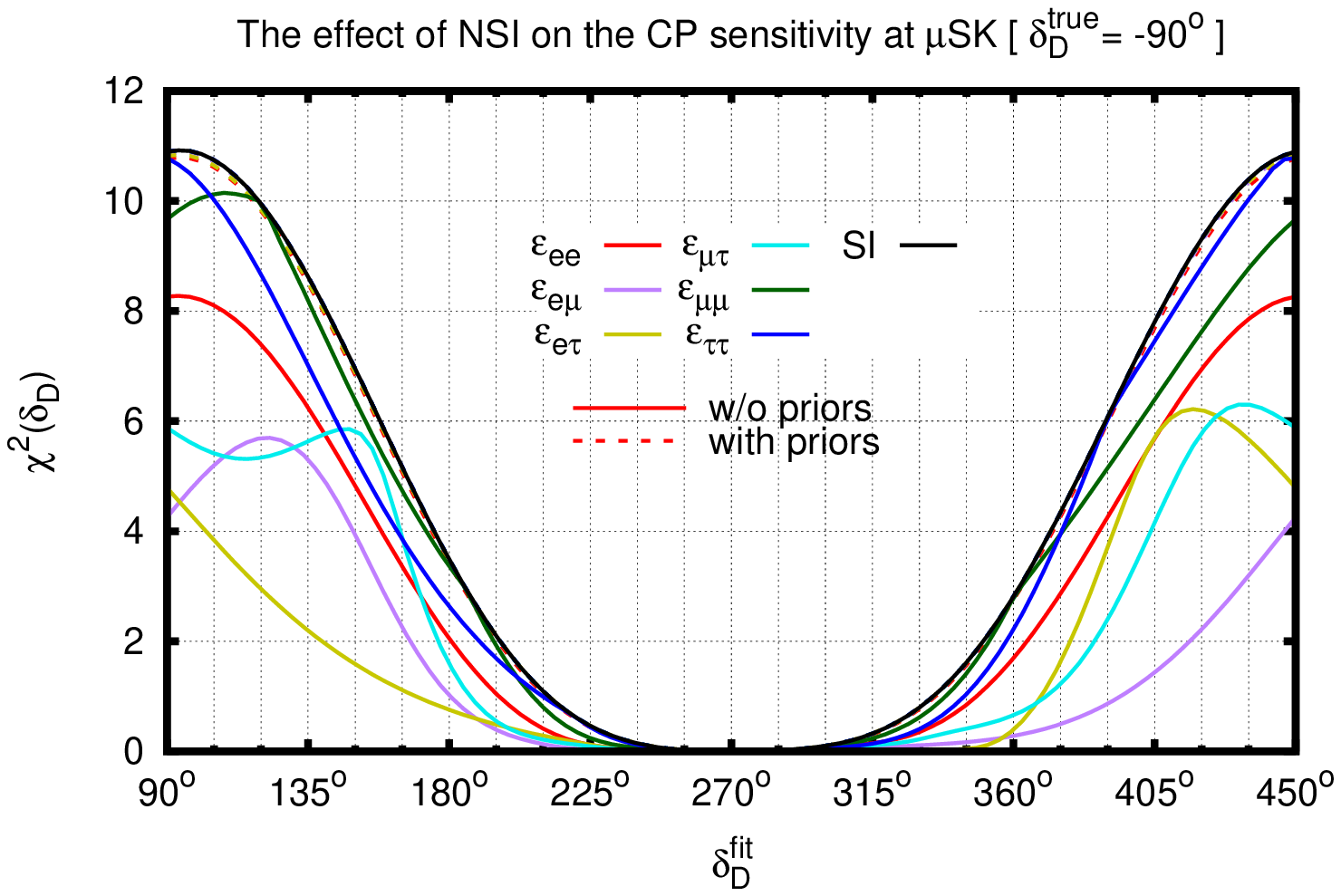}
\caption{The effect of NSI on the CP sensitivity at $\mu$SK with 
6 years of running in the antineutrino mode, without (solid lines)
and with (dashed line) the prior constraints (\ref{eq:globalfit}).}
\label{fig:chi2-muSK}
\end{figure}
%%%%%%%%%%%%%%%%%%%%%%%%%%%%%%%%%%%%%%%%%%%%%%%%%%%%%%%%%%%%%%%%%%%%%%%%%%%%%%%%%%%%

In \gfig{fig:chi2-muSK}, we show the CP sensitivity at $\mu$SK. 
Without NSI the true value $\dD^{true} = 3\pi/2$ 
can be distinguished from $\dD^{fit} = 0 \mbox{ or } \pi$ 
by $\chi^2 = 2.8$ ($1.7\sigma$) and from $\dD^{fit} = \pi/2$ by $\chi^2 = 11$ ($3.3\sigma$). 
Notice that the curve is not symmetric with respect to $\dD =  3\pi/2$
due to the interplay between
the $\sin \dD$ and $\cos \dD$ dependences in the oscillation probability
$P_{\bar \mu \bar e}$.

The matter effects at $\mu$SK are an order of magnitude smaller than at T2K, and consequently 
the sensitivity to $\dD$ is less affected by NSI, in comparison with \gfig{fig:chi2-T2K}. With the
prior constraint \geqn{eq:globalfit} imposed, the sensitivity to $\dD$ at $\mu$SK is almost unaffected.  
The sensitivity to $\dD$ is significantly reduced by the NSI parameters only if no prior constraint is imposed.
Even in this case the sensitivity to $\dD$ is not reduced to almost zero by the NSI parameters ($\epsilon_{ee}$,
$\epsilon_{e \mu}$, $\epsilon_{e \tau}$) , which is the case at T2K.

\begin{figure}[h!]
\centering
\includegraphics[height=0.47\textwidth,angle=-90]{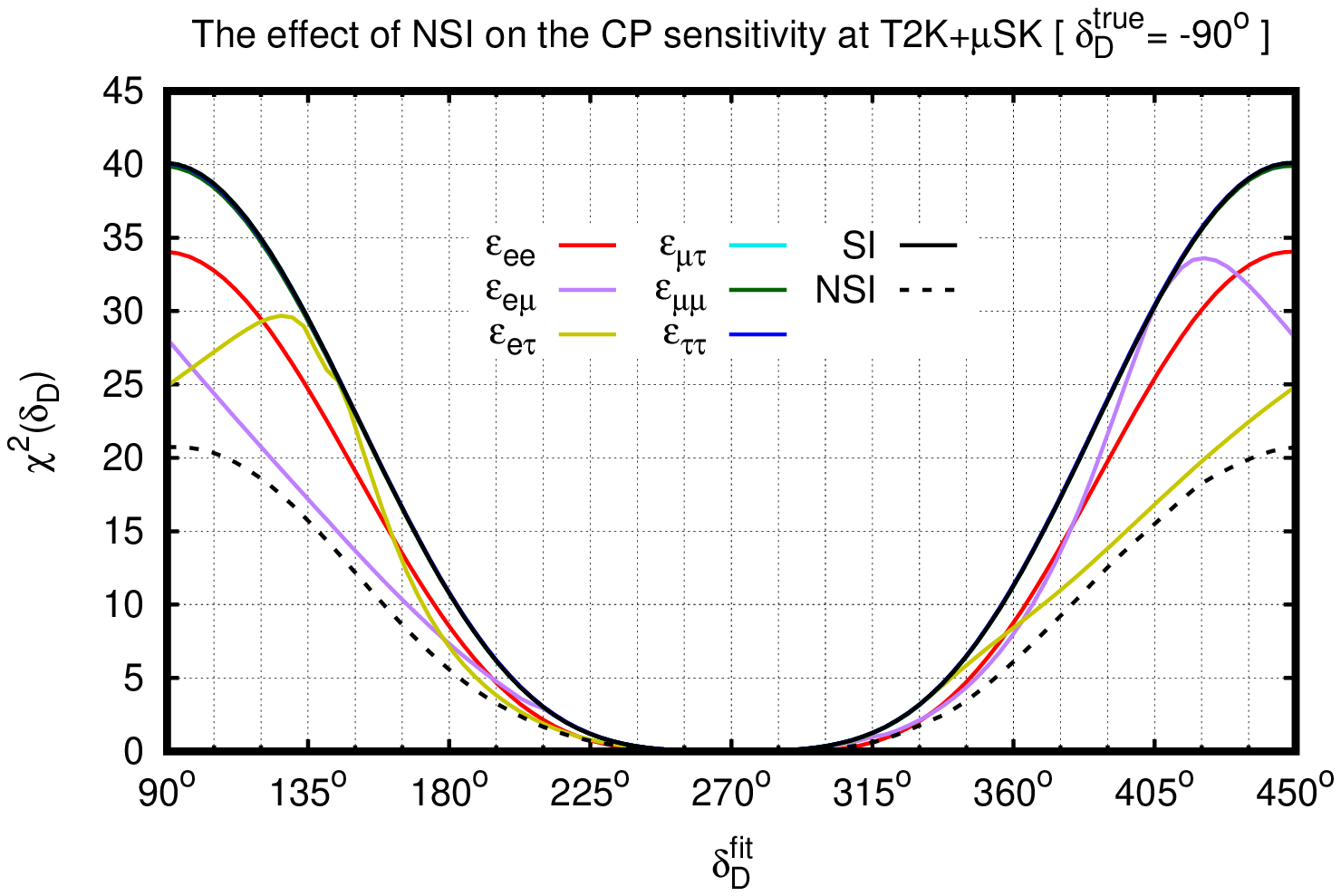}
\includegraphics[height=0.47\textwidth,angle=-90]{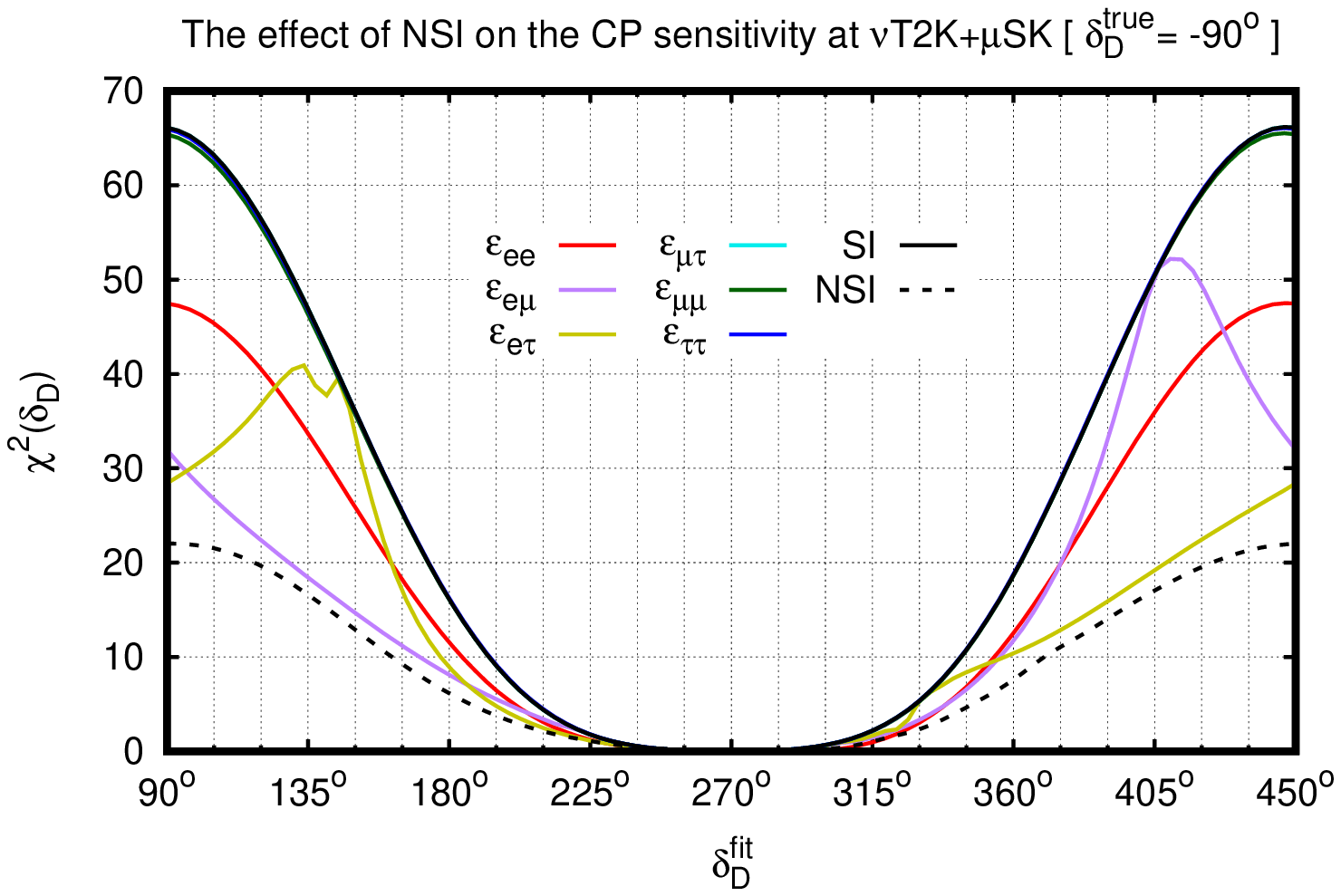}
\caption{The effect of NSI on the CP sensitivity at TNT2K with the prior constraints (\ref{eq:globalfit}).
For comparison, the same configuration
         of T2K as in Fig.\ref{fig:chi2-T2K}, 2 years of running in the neutrino mode and 4 years
         for anti-neutrino, has adopted in the left panel (T2K) while the right takes
         6 years of running in the neutrino mode ($\nu$T2K).
         For both panels, the $\mu$SK experiment runs for 6 years in the anti-neutrino mode.
The pseudo-data with standard interaction and $\dD^{true} = - 90^\circ$ is
fit with standard interaction (SI), individual NSI parameters
($\epsilon_{ee}$, $\epsilon_{e \mu}$, $\epsilon_{e \tau}$, $\epsilon_{\mu \tau}$, $\epsilon_{\mu \mu}$,
$\epsilon_{\tau \tau}$), or full set of NSI parameters (dashed line).}
\label{fig:chi2-TNT2K}
\end{figure}

Let us consider the sensitivity to $\dD$ of the combined measurements at TNT2K \cite{TNT2K}.
While T2K unavoidably measures both $\dD$ and
NSI, $\mu$SK mainly provides a determination of the phase $\dD$. 
The \gfig{fig:chi2-TNT2K} shows how $\mu$SK can improve 
the sensitivity to $\dD$, in comparison with \gfig{fig:chi2-T2K} from T2K only. With the prior constraints
\geqn{eq:globalfit} imposed, the significance of distinguishing $\dD^{true} = 3\pi/2$
from $\dD^{fit} = \pi/2$ can reach $5\sigma$ and the distinguishability from
vanishing CP values, $\dD^{} = 0 \mbox{ or } \pi$, can be about $2.5 \sigma$ for individual NSI parameters. 
The numbers become $4.6\sigma$ and $2.2\sigma$ with the full set of NSI parameters, in comparison
with $1.8 \sigma$ and $0.9 \sigma$ at T2K as well as $3.3 \sigma$ and $1.7 \sigma$ at $\mu$SK.
Combining the T2K and $\mu$SK results will substantially improve the sensitivity to $\dD$,
in comparison to the sensitivity of T2K or $\mu$SK alone.

With a $\mu$DAR antineutrino source built near the Super-K/Hyper-K detectors,
\tk can devote all its exposure time to the neutrino run. This can
significantly increase the event number.
We can obtain a factor of 3 increase in the number of neutrino events 
and a factor of 4 increase in the number of antineutrino events in comparison to those at T2K. 
In addition, the wide energy spectrum of the $\mu$DAR antineutrino will
break the degeneracy between $\dD$ and $\pi - \dD$ by providing $\cos \dD$ dependence of $P_{\mu e}$, 
especially for non-maximal CP phase, $\dD \neq \pi - \dD$.
If the CP phase happens to be $\dD = \pm \pi /2$, $\mu$SK can also reduce
the uncertainty around $\dD^{true} = 3\pi/2$ \cite{TNT2K}. With mainly $\sin \dD$
dependence at T2K, the CP uncertainty $\Delta(\dD) \propto 1/\cos \dD$ diverges
for maximal CP violation. This can be significantly improved due to the $\cos \dD$ dependence
provided by the $\mu$DAR flux.

%%%%%%%%%%%%%%%%%%%%%%%%%%%%%%%%%%%%%%%%%%%%%%%%%%%%%%%%%%%%%%%%%%%%%%%%%%%%%%%%%%%%%%%%%%%%%%%%
%%%%%%%%%%%%%%%%%%%%%%%%%%%%%%%%%%%%%%%%%%%%%%%%%%%%%%%%%%%%%%%%%%%%%%%%%%%%%%%%%%%%%%%%%%%

In the right panel of \gfig{fig:chi2-TNT2K} 
we show the CP sensitivity of the combined measurement with T2K running 6 years
purely in the neutrino mode ($\nu$T2K)
and $\mu$SK running 6 year purely in the antineutrino mode.
The latter CP sensitivity is higher than the sensitivity of the combination of T2K (neutrino and antineutrino) 
and $\mu$SK. Without NSI, the distinguishability of $\dD^{fit} = \pi/2$ from $\dD^{true} = 3\pi/2$
increases from around $6\,\sigma$ to around $8\,\sigma$. After including individual NSI parameter,
it increases from $5\,\sigma$ to $5.4\,\sigma$ with prior constraints.
The distinguishability of $\dD^{fit} = 0 \mbox{ or } \pi$ from $\dD^{true} = 3\pi/2$ increases
from $3.3\,\sigma$ to $4\,\sigma$ for SI and from $2.6\,\sigma$ to $2.8\,\sigma$ for individual NSI parameter with 
prior constraints. Thus the  combination of $\nu$T2K (fully neutrino running) 
and $\mu$SK, is better than splitting the T2K run among neutrino and antineutrino modes.
Comparing the solid black curve for SI and the dashed one for the full set of NSI parameters,
we find that the distinguishability of $\dD^{fit} = \pi/2$ 
from $\dD^{true} = 3\pi/2$ decreases from
$\chi^2 \approx 40$ to $21$ at T2K (2 + 4 years) + $\mu$SK, 
while it decreases from around 65 to $22$ 
for $\nu$T2K + $\mu$SK. In both cases 
the CP violation can be established  at more  than $4.5\sigma$. Although the CP sensitivity
with NSI is roughly the same for T2K+$\mu$SK and $\nu$T2K+$\mu$SK, 
it can be significantly increased at $\nu$T2K
if there is no NSI.

%%%%%%%%%%%%%%%%%%%%%%%%%%%%%%%%%%%%%%%%%%%%%%%%%%%%%%%%%%%%%%%%%%%%%%%%%%%%%%%%%%
\begin{table}[h]
\setlength{\tabcolsep}{3mm}
\centering
\begin{tabular}{c||cc|cc|cc|cc}
  $\dD^{true} = -90^\circ \mbox{ vs } \dD^{fit} = 0^\circ$ & \multicolumn{2}{c|}{T2K} & \multicolumn{2}{c|}{$\mu$SK} & \multicolumn{2}{c|}{T2K+$\mu$SK} & \multicolumn{2}{c}{$\nu$T2K+$\mu$SK} \\
  Event Numbers & \multicolumn{2}{c|}{$\gpurple{\bf 114 \nu + 56 \bar \nu}$} & \multicolumn{2}{c|}{$\gpurple{\bf 212 \bar \nu}$} & \multicolumn{2}{c|}{$\gpurple{\bf 57 \nu + 268 \bar \nu}$} & \multicolumn{2}{c}{$\gpurple{\bf 342 \nu + 212 \bar \nu}$} \\
\hline \hline
% $\chi^2$ for \gred{\bf SI} \& \gblue{\bf NSI} & \gred{\bf 2.43} & \gblue{\bf 1.21} & \gred{\bf 4.13} & \gblue{\bf 2.75} & \gred{\bf 8.20} & \gblue{\bf 5.33} & \gred{\bf 13.6} & \gblue{\bf 6.60} \\
  $\chi^2$ for \gred{\bf SI} \& \gblue{\bf NSI} & \gred{\bf 4.08} & \gblue{\bf 1.54} & \gred{\bf 2.81} & \gblue{\bf 2.75} & \gred{\bf 11.3} & \gblue{\bf 6.10} & \gred{\bf 18.7} & \gblue{\bf 7.59} \\
\hline
% $\chi^2$                & \gblue{\bf 1.58} & -- & \gblue{\bf 4.09}  & -- & \gblue{\bf 6.81}  & -- & \gblue{\bf 10.1} & -- \\
% $\epsilon^{\mbox{best fit}}_{ee}$         & 0.60  & 0.55 & 0.47  & 0.47 & 0.69  & 0.65 & 0.90 & 0.69 \\
  $\chi^2$                & \gblue{\bf 2.50} & -- & \gblue{\bf 2.75}  & -- & \gblue{\bf 8.77}  & -- & \gblue{\bf 12.5} & -- \\
  $\epsilon^{\mbox{bf}}_{ee}$         & 0.69  & 0.57 & 0.48  & 0.47 & 0.81  & 0.63 & 1.07 & 0.70 \\
\hline
% $\chi^2$                & \gblue{\bf 2.42}  & -- & \gblue{\bf 4.13}  & -- & \gblue{\bf 8.20}  & -- & \gblue{\bf 13.6} & -- \\
% $\epsilon^{\mbox{best fit}}_{\mu \mu}$    &-0.02  & -0.01 &-0.02  & -0.01 &-0.02  & -0.01 &-0.02 & 0.09 \\
  $\chi^2$                & \gblue{\bf 4.06}  & -- & \gblue{\bf 2.81}  & -- & \gblue{\bf 11.3}  & -- & \gblue{\bf 18.6} & -- \\
  $\epsilon^{\mbox{bf}}_{\mu \mu}$    &-0.02  & -0.01 &-0.02  & -0.01 &-0.02  & -0.01 &-0.02 & -0.01 \\
\hline
% $\chi^2$                & \gblue{\bf 2.43}  & -- & \gblue{\bf 4.13}  & -- & \gblue{\bf 8.20}  & -- & \gblue{\bf 13.6} & -- \\
% $\epsilon^{\mbox{best fit}}_{\tau \tau}$  & 0.02  & 0.01 & 0.02  & 0.01 & 0.02  & 0.01 & 0.01 & 0.10 \\
  $\chi^2$                & \gblue{\bf 4.11}  & -- & \gblue{\bf 2.81}  & -- & \gblue{\bf 11.3}  & -- & \gblue{\bf 18.7} & -- \\
  $\epsilon^{\mbox{bf}}_{\tau \tau}$  & 0.02  & 0.01 & 0.02  & 0.01 & 0.02  & 0.01 & 0.01 & 0.01 \\
\hline
% $\chi^2$                & \gblue{\bf 1.76}  & -- & \gblue{\bf 4.11}  & -- & \gblue{\bf 6.54}  & -- & \gblue{\bf 9.12} & -- \\
% $\epsilon^{\mbox{best fit}}_{e \mu}$      & 0.10  & 0.04 & 0.02  & 0.01 & 0.15  & 0.11 & 0.19 & 0.10 \\
  $\chi^2$                & \gblue{\bf 2.68}  & -- & \gblue{\bf 2.81}  & -- & \gblue{\bf 8.01}  & -- & \gblue{\bf 11.7} & -- \\
  $\epsilon^{\mbox{bf}}_{e \mu}$      & 0.12  & 0.07 & 0.01  & 0.01 & 0.19  & 0.11 & 0.23 & 0.10 \\
\hline
% $\chi^2$                & \gblue{\bf 1.94}  & -- & \gblue{\bf 4.13}  & -- & \gblue{\bf 7.02}  & -- & \gblue{\bf 9.17} & -- \\
% $\epsilon^{\mbox{best fit}}_{e \tau}$     & 0.17  & 0.11 & 0.01  & 0.01 & 0.27  & 0.24 & 0.43 & 0.16 \\
  $\chi^2$                & \gblue{\bf 2.77}  & -- & \gblue{\bf 2.81}  & -- & \gblue{\bf 8.42}  & -- & \gblue{\bf 10.4} & -- \\
  $\epsilon^{\mbox{bf}}_{e \tau}$     & 0.25  & 0.14 & 0.01  & 0.01 & 0.37  & 0.21 & 0.51 & 0.30 \\
\hline
% $\chi^2$                & \gblue{\bf 2.43}  & -- & \gblue{\bf 4.13}  & -- & \gblue{\bf 8.20}  & -- & \gblue{\bf 13.6} & -- \\
% $\epsilon^{\mbox{best fit}}_{\mu \tau}$   & 0     & 0 & 0     & 0 & 0     & 0 & 0 & 0
  $\chi^2$                & \gblue{\bf 4.08}  & -- & \gblue{\bf 2.81}  & -- & \gblue{\bf 11.3}  & -- & \gblue{\bf 18.7} & -- \\
  $\epsilon^{\mbox{bf}}_{\mu \tau}$   & 0     & 0 & 0     & 0 & 0     & 0 & 0 & 0
\end{tabular}
\caption{The $\chi^2_{min}$ and \mbox{best fit} values of the NSI parameters when using $\delta^{fit}_D = 0^\circ$
         to fit the pseudo-data generated with SI and $\delta^{true}_D = -90^\circ$, 
         under the prior constraints (\ref{eq:globalfit}). In each experimental
         configuration T2K, $\mu$SK, T2K+$\mu$SK, and $\nu$T2K+$\mu$SK, the $\chi^2$ for \gred{\bf SI} comes
         from the fit with standard interaction, \gblue{\bf NSI} with either individual or the full set of ($\epsilon_{ee}$, 
$\epsilon_{e \mu}$, $\epsilon_{e \tau}$, $\epsilon_{\mu \tau}$, $\epsilon_{\mu \mu}$,
$\epsilon_{\tau \tau}$).}
\label{tab:chi2-0dD}
\end{table}
%%%%%%%%%%%%%%%%%%%%%%%%%%%%%%%%%%%%%%%%%%%%%%%%%%%%%%%%%%%%%%%%%%%%%%

%

%%%%%%%%%%%%%%%%%%%%%%%%%%%%%%%%%%%%%%%%%%%%%%%%%%%%%%%%%%%%%%%%%%%%%%
\section{Conclusions}
\label{sec:conclusion}
%%%%%%%%%%%%%%%%%%%%%%%%%%%%%%%%%%%%%%%%%%%%%%%%%%%%%%%%%%%%%%%%%%%%

We explored the CP violation and the matter effects in neutrino oscillation
in the presence of standard and non-standard interactions at low energies and relatively small baselines.
This experimental setup is realized in T2K, the experiments based on $\mu$DAR, {\it etc}.     
A simple analytic formalism is elaborated which describes  
the NSI effects and their dependence on the parameters 
of NSI interactions. The vacuum mimicking and its violation in the 1-2 sector
as well as the use of the separation basis play central roles in the formalism. 

In the case of standard interactions due to vacuum mimicking the matter affects the oscillation
probability $P_{\mu e}$ mainly via the correction to the 1-3 mixing. 
We find that matter changes the probability by
about $13\%$ at T2K and $1\%$ at $\mu$DAR. 
Also we show that vacuum mimicking provides a simple explanation 
of the fact that the usual formula for $3\nu-$ oscillation probability
gives a very good description at energies around the 1-2 resonance.

In the presence of NSI, the vacuum mimicking in the 1-2 sector
is strongly broken. The breaking shows up  
in a very specific way and is related to the single NSI 
parameter $\te_{12}$ in the separation basis. 
The same parameter $\te_{12}$ controls the additional contribution to the effective CP 
phase in matter. 
The 1-3 mixing is modified by another parameter $\te_{13}$ while
the 1-3 oscillation phase is corrected by the diagonal elements 
$\te_{11}$, $\te_{22}$ and $\te_{33}$. 

We show that the total probability $P_{\mu e}$ is mainly affected by 
the violation of vacuum mimicking parametrized by ($r_V - 1$) and $\delta_2$ as well as 
by the correction to the 1-3 mixing,  $\delta \theta_{13}^m$.
The relative effects of the violation of vacuum mimicking and $\delta \theta^m_{13}$ on
$P_{\mu e}$ depend on the phases of $\te_{e \mu}$ or/and $\te_{e \tau}$.
Within the $1\sigma$ intervals, the correction to $P_{\mu e}$ due to NSI  can reach $(20 - 30)\%$ 
at T2K and $2\%$ at $\mu$DAR. The corrections to the CP phase can be as large as $(40 - 50)^\circ$ 
at T2K and $3^\circ$ at $\mu$DAR. 

We apply our analytic formalism to the CP phase measurement at low energies.
The standard interaction leads to vacuum mimicking in the solar amplitude and keeps 
the Dirac CP phase $\dD$ unaffected.  
On the contrary, NSI can introduce significant deviation from vacuum mimicking, and consequently, 
modify $\dD$ to practically any value at T2K. 
With prior constraints \geqn{eq:globalfit} on the size of NSI parameters, 
the CP sensitivity in terms of $\chi^2(\delta_D)$ at T2K can be reduced by
almost a factor of 2$\sim$3. The effect of NSI can be even larger at
NO$\nu$A and DUNE with higher neutrino energies.

We show that TNT2K -- the combination of T2K and the new component $\mu$SK can resolve the $\dD$--NSI
degeneracy.
The $\mu$SK component uses the SK detector to study the oscillations of the antineutrinos
from a $\mu$DAR source.  Since the $\mu$DAR flux has 10 times lower
energy, the effect of NSI at $\mu$SK is much smaller than at T2K.
While T2K  measures both the genuine CP phase $\dD$ and NSI, $\mu$SK
can provide clean determination of $\dD$ simultaneously. 
The sensitivity to $\delta_D$ which can be achieved by this combination of T2K and $\mu$SK is much higher
than the sensitivity of T2K or $\mu$SK alone. The TNT2K configuration can guarantee high sensitivity
to $\dD$ in the presence of NSI.

\section{Acknowledgements}
SFG would like to thank Manfred Lindner
for various discussions as well as Hong-Jian He and Center for High Energy Physics at
Tsinghua University for hospitality during the preparation of this paper.
A.S. is grateful to E. K. Akhmedov for useful discussions. 

\bibliographystyle{unsrt}
\bibliography{nuMatter}
\nocite{*}

\end{document}